\documentclass[%
reprint,
superscriptaddress,
showpacs,
amsmath,amssymb,
aps,
pre,
showkeys
]{revtex4-1}

\usepackage{graphicx}
\usepackage{subcaption}
\usepackage{dcolumn}
\usepackage{bm}
\usepackage{mathtools}
\DeclarePairedDelimiter{\ceil}{\lceil}{\rceil}

\usepackage{color}

\begin{document}

\preprint{APS/123-QED}

\title{Distribution of volume, microvoid percolation, \\and packing density in globular proteins}

\author{Jenny Farmer}
\affiliation{Department of Physics and Optical Science, University of North Carolina at Charlotte, Charlotte, NC 28223, USA}
\affiliation{Department of Bioinformatics and Genomics, University of North Carolina at Charlotte, Charlotte, NC 28223, USA}
\author{Sheridan B. Green}
\affiliation{Department of Physics, Yale University, New Haven, CT 06520, USA}
\author{Donald J. Jacobs}
\email{djacobs1@uncc.edu}
\affiliation{Department of Physics and Optical Science, University of North Carolina at Charlotte, Charlotte, NC 28223, USA}
\affiliation{Center for Biomedical Engineering and Science, University of North Carolina at Charlotte, Charlotte, NC 28223, USA}

\date{\today}

\begin{abstract}
A fast and accurate grid-based method with low memory requirement is presented to calculate volume characteristics in molecular systems. The distribution of volume and packing density is characterized in globular proteins, where void space is decomposed into microvoid volume and cavities based on a spherical test probe with variable radius. A scan over test probe radius is mapped onto a site percolation problem for microvoid volume. Finite-size scaling is applied to determine critical exponents, which are found to be consistent with connectivity percolation exponents in three dimensions. Disparate results in the literature regarding packing density in the core of a protein compared to on its surface, and with respect to protein size, is elucidated in terms of microvoid volume within a unified implicit-solvent model. By parameterizing the model to match the results of explicit-solvent models that agree with experimental data, we verify that packing density within globular proteins is spatially uniform and independent of protein size.

\end{abstract}

\maketitle


\section{Introduction}\label{sec:intro}

The volume of space occupied by and surrounding a protein is an important property that affects protein stability and function. In particular, globular proteins are generally densely packed in the native state \citep{Liang2001,Richards1974,Tsai1999}, yet ramified interior void spaces are present due to imperfect packing. Local volume information is important for pressure unfolding \citep{Cioni2006,Frye1998,Roche2012}, as well as structure prediction and drug design \citep{Perot2010,Raunest2011,Petrek2006,Brezovsky2013}. Boundary volume, related to accessible surface area, correlates with hydrophobic transfer energies, which are a driving force in conformational changes \citep{Baldwin2013,Chandler2005,Reynolds1974}.  Furthermore, clefts and pockets that form on a protein surface are distinguishing characteristics that are effectual in predicting catalytic areas and binding sites \citep{Winter2011,McCallum2000,Liang1998,Weisel2007,Schmidtke2010}. 

Volume decomposition strategies frequently employ a spherical test probe for purposes of classification, where the results depend on the test probe radius. For example, a cavity defines a region in space that is large enough to hold a probe of a certain size. As probe size changes, different types of volume contributions merge or separate. It is common to use a probe radius of $1.4$ \AA{} to represent a water molecule in order to infer functional mechanisms from local volume characteristics \citep{Chen2015,Voloshin2011,Gaines2018,Shaytan2009,Richards1977,Fleming2000}. Identifying how cavities change in size within a protein for different probe sizes is useful in quantifying how various types of local volumes cluster as probe radius is continuously varied as an independent variable.

Many computational methods have been implemented to calculate protein volume, void space, surface area, as well as to identify clefts and channels \citep{Raunest2011,Petrek2006,Liang1998,Weisel2007,Chen2015,Voss2010,Ho1990,Kleywegt1994,Levitt1992,Laskowski1995}. For example, the Voronoi diagram \citep{Voronoi1908} was applied to protein packing by Richard and Finney in the 1970s \citep{Richards1974, Finney1970}. Alpha shapes \citep{Edelsbrunner1983}, closely related to the Delaunay triangulation \citep{Delaunay1934}, are based on the Voronoi diagram \citep{Voronoi1908}, where a volume is assigned to each atom based on its proximity to neighboring atoms. Algorithms employing methods from computational geometry, such as the alpha shape, continue to be developed \citep{Li2013,Albou2009,Zhou2014}. Alternatively, grid-based methods \citep{Raunest2011,Weisel2007,Voss2010,Levitt1992,Kleywegt1994} discretize volume onto a regular three-dimensional lattice, making the calculated volumes approximate and sensitive to the orientation of a protein relative to the grid \citep{Liang1998,Chen2015}. With a time complexity that is linear in protein size \citep{Petrek2006,Weisel2007,Chen2015,Mach2011}, grid-based methods have been shown to have a competitive advantage over analytical methods for large macromolecular structures \citep{Voss2010,Mach2011}. 

In this paper, we report a novel grid-based algorithm with several distinctive features that provide important advantages over existing methods.  First, we incorporate compressive elasticity into the test probe as a means to identify cavities and to mitigate artifacts from the discrete lattice. Second, scalability and performance are optimized using a generalized cluster labeling technique, similar to the original Hoshen-Kopelman (HK) method \citep{Hoshen1976}, to characterize volume in a protein as a multi-component percolation problem. Third, a variable-sized spherical test probe is utilized to classify clusters of different volume types, where a direct connection to percolation theory is made by mapping the probe to a site percolation probability. Fourth, we model the effect of protein dynamics by averaging results over a narrow distribution of test probe sizes. Finally, after partitioning the volume into distinct types, we assign contributions to each volume type on a per atom basis, enabling the accurate calculation of packing density.

A survey of the literature shows that a variety of methods that calculate local packing density yield inconsistent results, although most methods report at least marginally higher packing in the core \citep{Tsai1999,Fleming2000,Gerstein1995,Esque2010,Dill1990,Rother2003}. The nature of packing density remains an open problem, in large part due to inconsistent operational definitions. Conceptually, packing density is a measure of the percentage of the protein volume that is comprised of the atoms themselves, not including the empty space between the atoms. Voronoi tessellation should provide an accurate method for computing packing density. Unfortunately, the Voronoi method has difficulties near the surface of a protein when the solvent is not explicitly modeled. In past works, these technical problems were avoided by limiting the analysis to buried residues.

Attempts to extend applicability to surface residues have been handled in a variety of ways, such as by imposing boundary conditions to model a solvation shell surrounding the protein \citep{Liang2001,Rycroft2009}, by excluding certain intractable surface volumes from the calculations \citep{Tsai1999,Andersson1998,Gerstein1996}, or by strategically placing water molecules around the protein \citep{Chakravarty2002,Angelov2002}. A different approach, called occluded surface packing (OSP) \citep{Fleming2000}, computes packing density by extending lines from each atom perpendicular to its atomic surface, until the lines either intersect with another atomic surface or reach a length equal to the diameter of a water molecule. The lengths of these lines are used to determine the packing density, which is typically less than the corresponding Voronoi packing estimate \citep{Fleming2000}. Yet another approach, by \citet{Liang2001}, applies a separate definition for surface packing density that considers pockets along the protein surface. 

The various implicit solvent models differ in their predictions of local packing density for residues that are exposed to solvent. A straightforward way to resolve this problem is to perform a molecular dynamics (MD) simulation of a protein with explicit solvent at the all-atom level and apply Voronoi methods \citep{Gerstein1995,Esque2010}. Notwithstanding subtle technical problems of its own, this direct approach arguably provides the most accurate representation of the microenvironment that surrounds a protein but suffers from being too computationally expensive to use in high-throughput applications in structural bioinformatics and drug design. Our implicit model is parameterized to match the results from these all-atom MD simulations.

This paper is outlined as follows. In Section \ref{sec:method}, we define four distinct, mutually exclusive types of volume and describe the algorithm for assigning each grid point's volume type and cluster connectedness. In Section \ref{sec:volume_chars}, we benchmark the accuracy and speed of the method and present results for volume characteristics over a dataset of 108 globular proteins that are diverse in size and fold structure. Microvoid clusters are analyzed in the context of percolation theory. By employing finite size scaling, we show universal behavior in cluster size scaling for microvoid volume across a diverse set of globular proteins. In Section \ref{sec:pd}, we quantify packing density characteristics in terms of partial volumes. Local packing density depends on how much of a residue is exposed to solvent and how partial volume is partitioned within an implicit solvent model. Within our unified framework, we elucidate how different trends appear depending on the operational definition of partial volume. We conclude in Section \ref{sec:discussion} by highlighting the advantages of our method and discussing future work and applications. 

\section{Method}\label{sec:method}
\subsection{Definitions of volume types}\label{subsec:vol_types}

Different definitions of volume types are possible because the decomposition of protein volume is not unique. For example, solvent-accessible volume and solvent-excluded volume are two commonly invoked terms with varying definitions throughout the literature \citep{Chen2015,Richmond1984,Connolly1983,Connolly1984}. Molecular volume and accessible volume are commonly defined by the prescription given by \citet{Richards1977}, which involves using a spherical probe of radius  that rolls around the entire surface of a protein. The contact surface and accessible surface, first calculated by \citet{Lee1971}, are two distinct areas carved out by this process. The contact surface is also called the molecular surface, or Connolly surface, and is defined by the leading edge of the spherical probe. The accessible surface is obtained by extending the contact surface at each point by the probe radius. The molecular volume is the volume enclosed by the contact surface and the accessible volume is enclosed by the accessible surface. 

\begin{figure}
  \centering
    \includegraphics[width=0.48\textwidth]{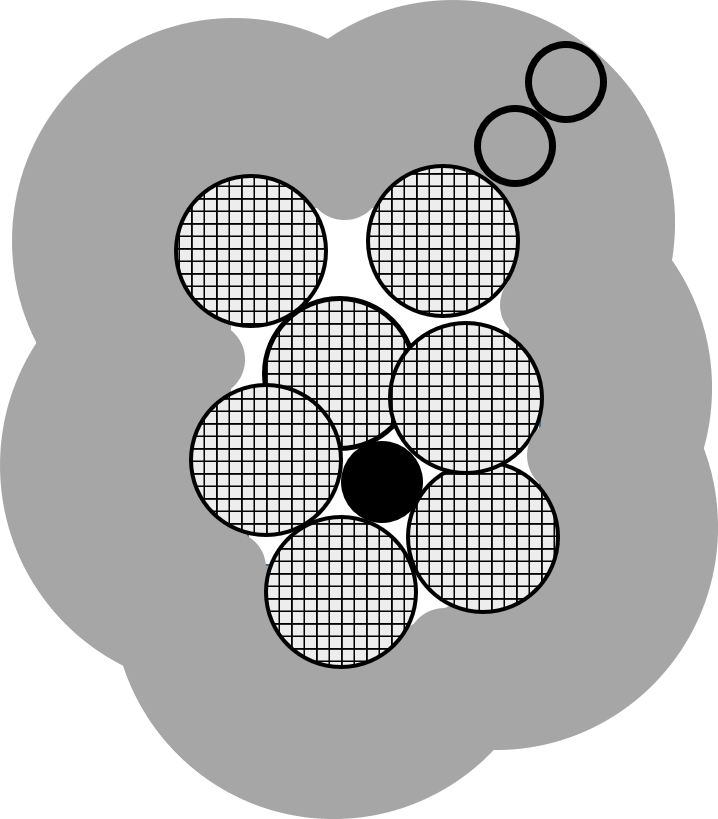}
\caption{Volume space definitions. The vdW volume of the protein atoms is denoted by the cross-hatched circles. The white space represents microvoid, and a single cavity is show in black.  The gray surrounding layer is the solvent extended volume layer, with a default depth of two times the diameter of the probe molecule.}\label{fig:vol_defs}
\end{figure}

In this work, molecular volume is expressed as a sum of microvoid volume, cavity volume, and vdW volume.  Any space within the molecular volume that is not occupied by atoms (i.e. is not part of the vdW volume) is called void space, which we divide into cavity and microvoid volume. Cavity volume consists of regions that are large enough to accommodate the spherical probe, and we denote the remaining volume as microvoid. A schematic of the volume decomposition is shown in Fig. \ref{fig:vol_defs}. Note that the minimum span of any cross section of microvoid volume must be less than the probe diameter. Therefore, clusters of contiguous microvoid volume are highly ramified because large pockets of microvoid volume cannot exist. Microvoid volume, also called "dead space", typically percolates throughout a protein. To our knowledge, this microvoid volume, which represents the vast majority of molecular volume, was not previously calculated as a quantity of interest. Note that, unlike the vdW volume, the clustering of microvoid and cavity volumes will depend on probe size. 

In addition to the molecular volume, we include an extended volume layer beyond the molecular surface. The surface of this extended volume layer extends past the contact surface up to length $L_s$, forming an outer shell surrounding the protein. All grid points that are located within the extended surface are included in the volume calculation. Setting $L_s=4R$ allows for two layers of probes to fit inside the extended volume layer, which is useful for studying solvation properties of residues by encapsulating the solvation shell. Setting $L_s= R$ recovers the accessible volume defined by Richards. We also parameterize $L_s$ by comparing to all-atom MD simulation results that indicate uniform packing density throughout a protein. In this case, a function $L_s (R)$ is constructed to ensure that the average packing density on the surface of a protein is the same as within its core.

We define boundary volume as the final volume type. Boundary volume consists of the volume in the extended layer as well as all cavity volume that has a connecting path to the extended volume layer. For example, a cylindrical channel through a protein would be part of boundary volume provided that a spherical probe can traverse through channel. However, if the probe size is too large to enter the channel on both ends, then the channel forms a separate microvoid region. When the probe radius reflects the size of a water molecule, boundary volume identifies the pathways where water can penetrate into the protein. In summary, we define four volume types: vdW, cavity, microvoid, and boundary. Adding these four partitions reconstitutes the molecular volume augmented by the extended volume surrounding a protein.

\subsection{Volume connectedness}\label{subsec:connectedness}

We employ a modified three-dimensional version of the Hoshen-Kopelman (HK) method \citep{Hoshen1976}. The HK method is widely used to identify which points on a grid form connected clusters, utilizing the union-find algorithm that is common to many computer science applications \citep{Teuler2000,Al-Futaisi2003,Hoshen1998,Frijters2015,Hoshen1997,Khaimov1989,Moreira1996}. The HK method is an efficient cluster-labeling technique that assigns a common label to a set of neighboring grid points. The time complexity is linear in the number of grid points and the method requires only a single scan through the lattice to identify each distinct cluster \citep{Al-Futaisi2003,Hoshen1998}. We map an all-atom protein structure onto a cubic lattice with lattice spacing $a$. The lattice is traversed by processing successive pairs of planes of grid points along their common normal direction. The leading plane is denoted $F$ for front and the trailing plane is denoted $B$ for back, where $B$ lags $F$ by one lattice constant in the dimension normal to the planes. The size of these planes is set such that the edges of the $B$- and $F$-planes reside just outside the extended volume. The initial $B$-plane is placed just outside of the extended volume of the protein such that all grid points in this plane are initialized as boundary volume. Similarly, all grid points on the edges of each plane are initialized as boundary volume. The first cluster label is used to represent boundary volume. We note that there can only be one boundary volume cluster based on our definition and boundary conditions. No other initialization is required. After the HK scan is complete, the protein fits within a box with side lengths $\{Lx,\ Ly,\ Lz\}$. All grid points on each face of this box share the first cluster label. The HK clustering label technique merges neighboring grid points with the same volume type into clusters with unique cluster labels. Note that every grid point inside this box is assigned to one of the four volume types defined above. In the case of boundary volume regions, a point is discarded if it is outside the extended volume that is set by the $L_s$ distance cutoff.

During the HK scan, each grid point within the $F$-plane is systematically evaluated row by row. First, the volume type of a grid point must be identified. If a grid point is within the vdW radius of an atom, it is identified as vdW volume. Otherwise, it is marked as void space. To further distinguish between cavity and microvoid volume requires a compressive energy test applied to the test probe, which is described in the next subsection. Each grid point is identified as one of three volume types: vdW, microvoid or cavity. Note that boundary volume represents a specific cluster of grid points that are marked as cavity, which is automatically identified based on the boundary conditions described above. Second, an unlabeled grid point that is in question within the $F$-plane is assigned a cluster label based on its neighboring grid points where volume types have already been identified within the $F$-plane and $B$-plane. In particular, the minimum cluster label among all its neighboring grid points with the same volume type is assigned to this unlabeled grid point, followed by the merging of cluster labels according to the HK method. Otherwise, if no previously-scanned neighbors of the same volume type are present, a new cluster label is introduced and the grid point in question is assigned to this cluster. Third, when the $F$-plane is completely scanned, no further history of the previously scanned $B$-plane is needed. The $F$-plane then serves as the new $B$-plane and the current $B$-plane is recycled as the new $F$-plane in an alternating fashion. 

The HK cluster labeling technique identifies contiguous space defined by connected grid points of the same volume type. Two grid points of the same volume type are connected if they are nearest, second-nearest or third-nearest neighbors on the cubic lattice. Altogether, a grid point can be connected to up to 26 neighbors. We considered models with lower connectivity by limiting connections between grid points to be nearest neighbors only or nearest and second-nearest neighbors only. All the models showed qualitatively similar results. However, the greater connectivity model that we use in all of the results presented here was found to approximate continuous space better based on this model obtaining lower uncertainties in volume predictions across random orientations of the protein when it is embedded on the cubic lattice. 

During the HK scan, the size of each uniquely labeled cluster using a canonical label, is accumulated on the fly using counters, which can subsequently combine as different clusters merge and canonical labels condense. Thus, the memory requirement to store grid points and their associated characteristics is two-dimensional, where we define the $z$-axis to be normal to the $B$- and $F$-planes such that $L_x L_y$ gives the cross-sectional area of the planes. After the HK scan is complete, the volume for each individual cluster is recorded using histograms for cluster statistics. To minimize memory requirements, we can minimize the cross-sectional area by selecting a special direction for the $z$-axis that coincides with the eigenvector of the largest eigenvalue of the protein's moment of inertia tensor. Reducing memory also increases speed when grid points no longer need to be processed. However, there is roughly an average 10\% reduction in memory and increase in speed when embedding a large set of diverse globular proteins on the lattice, each optimized to have an orientation with minimum cross-sectional area. Although a particular $z$-axis direction can provide greater runtime and memory savings for long, cigar-shaped proteins, in this work all of our results are based on the averaging of over 300 uniformly random orientations per protein calculation.  

In tandem with tracking cluster volumes for all void types, we calculate partial volumes by assigning the volume and type of each grid cell to its nearest-neighbor atom. As each grid point is classified, it is recorded by volume type in cumulative counters for the atom whose vdW shell is closest to that grid point. Thus, if the distance between a grid point $g$ and atom $a$ from center to center is $r_{ga}$, the distance that the grid point is away from the atom's vdW shell is $d_{ga}= r_{ga}-R_\textrm{vdW}$, where $R_\textrm{vdW}$ is the vdW radius of the atom. As such, each atom is ascribed four partial volumes (vdW, cavity, microvoid, and boundary) to describe its immediate local environment.  The total vdW volume is non-additive because the spherical balls representing vdW interactions generally overlap among atom pairs that are covalently bonded. Note that the vdW volume is weakly dependent on conformation whereas cavity, microvoid and boundary volume all strongly depend on the protein conformation. For example, partial boundary volume is a measure of how exposed an atom is to solvent. The partial boundary volume of a residue is obtained by summing all partial boundary volumes from all atoms within the residue. At the residue level, partial boundary volume correlates well with solvent accessible surface area (SASA) when compared to DSSP and NACCESS \citep{Touw2015,Hubbard1993}. Partial volumes (of various types) can also be assigned to the backbone or sidechain portions of residues, where this distinction has been shown to be critically important for implicit solvation models \citep{Auton2005}.

\subsection{Mitigation of discrete lattice artifacts}\label{subsec:lattice_art}

Classifying void space requires checking if a test probe can fit into a region of space without overlapping with any neighboring atom's vdW shell. If the boundary of a test probe centered at a grid point does not overlap with the boundary of any atom, then all grid points that are enclosed by the test probe are classified as cavity. Importantly, a grid point can be part of cavity while being arbitrarily close to the boundary of an atom. Clearly, it is not sufficient to check if a grid point is greater than a minimum distance from all nearby atoms to classify it as cavity or microvoid. To take advantage of the efficiency in the HK method where only the $B$- and $F$-planes are stored in memory at any given time, each grid point that appears in the $F$-plane must be identified on the fly as microvoid or cavity. Therefore, a procedure is needed that can displace the location of the test probe (defined by its geometric center) from the initial grid point position to a position that no longer overlaps with neighboring atoms while keeping the initial grid point within the span of the test probe. We employ a compressive energy test that is capable of performing this procedure.  

The compressive energy test is constructed through a pseudo-potential energy function of the form 

\begin{equation}
    V(r) = \begin{cases}
                \frac{1}{2}k(r-b)^2 &\text{for $r<b$} \\
                0 &\text{for $r>b$}
            \end{cases},
\end{equation}
where the length $b=R+R_\textrm{vdW}$, with $R$ the probe radius and $R_\textrm{vdW}$ the atom type's vdW radius. This pairwise potential energy is calculated for all atoms that create a compressive force on the test probe. The same spring constant is used for all atom types. In addition, we apply a geometric constraint that requires $r_{pg}<R$, where $r_{pg}$ is the distance between the test probe and the grid point in question. The model parameters are adjusted to maintain distance tolerances to within $0.0125$ \AA{}. This means that the probe boundary and vdW boundary of an atom can overlap by up to $0.0125$ \AA{} before being classified as a clash. Compared to the lattice constant a, the relative percent error tolerated in overlap is $1.25/a$. Hence, for {0.5 \AA{}}, {0.3 \AA{}}, and {0.1 \AA{}} grid sizes, the relative errors are respectively 2.5\%, 4.2\%, and 12.5\%. Alternatively, the relative error can be fixed at $\varepsilon$, which can be satisfied by setting distance tolerances to be within $\varepsilon a$. However, a fixed distance tolerance is employed because it is more consistent with the inherent uncertainty present due to the soft vdW interactions and protein dynamics that are unrelated to grid size. Said another way, uncertainties due to microscopic interactions between the atoms reduces the need for exact geometric results. Therefore, we set a maximum energy threshold of $E_\textrm{max}=\frac{1}{2}k(0.0125\textrm{ \AA{}})^2$. Since units are arbitrary, we set $k=2$. 

\begin{figure*}
  \centering
    \includegraphics[width=0.8\textwidth]{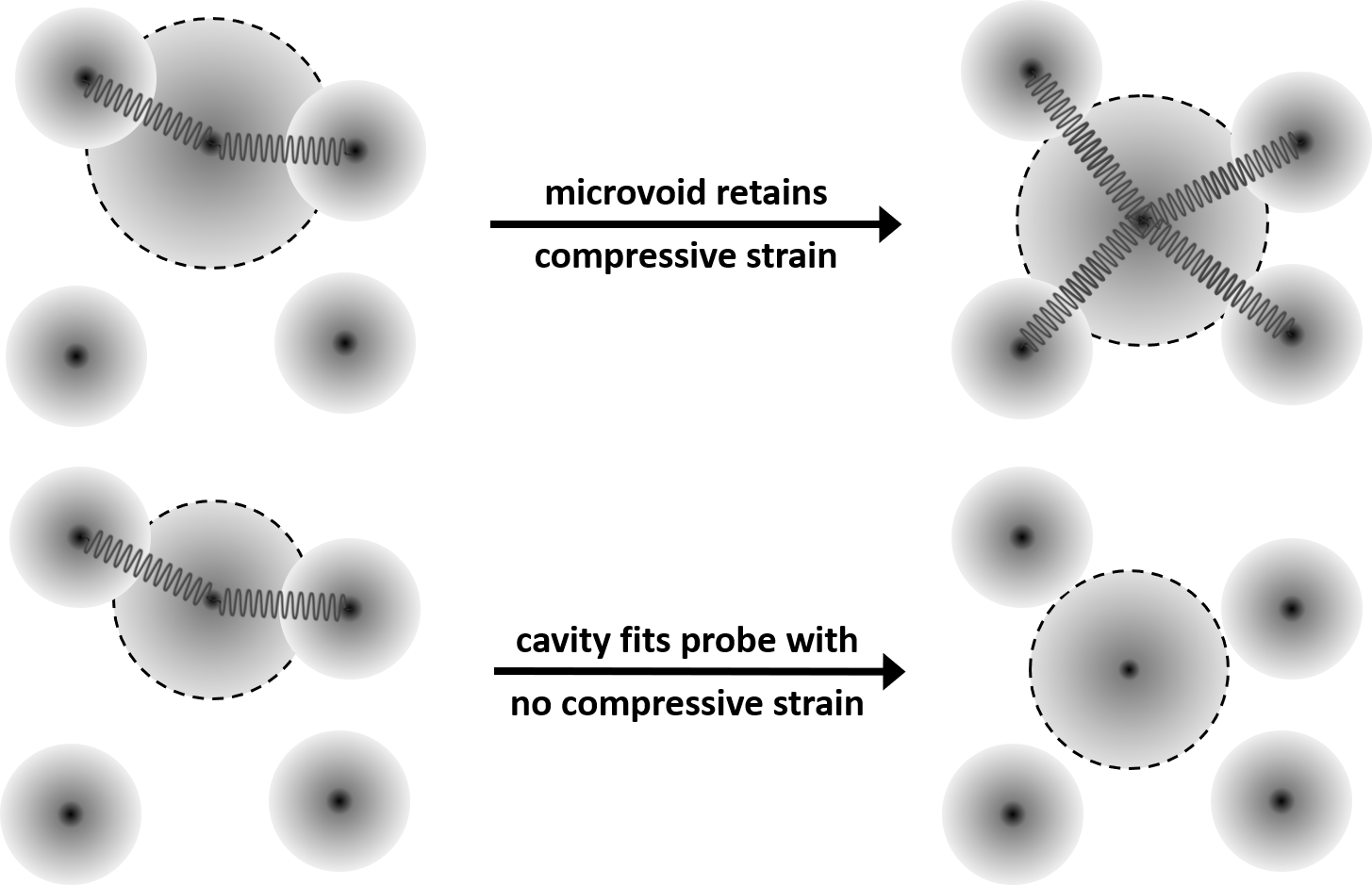}
\caption{Visual demonstration of the compressive elasticity model. Top: Compressed springs cannot be fully relaxed and maintain nonzero strain energy. Therefore, the initial grid point is classified as microvoid. Bottom: After relaxing the compressive energy for a smaller probe, the lack of remaining strain energy implies that the probe is able to fit within the available space. Thus, the initial grid point is classified as cavity.}\label{fig:spring_model}
\end{figure*}

Determining if a grid point is microvoid or cavity requires numerical relaxation to minimize the compressive energy on the test probe of radius $R$ that is initially placed at a reference grid point. This minimized energy is denoted as $E(R,n|k)$, where $n$ is the number of atoms within the local environment surrounding the reference grid point that can possibly clash with the probe and $k$ is the number of iterations applied in the minimization process before termination. Note that $n$ counts all atoms that could potentially clash with the test probe, but the actual number of clashes, if any, will be far less than $n$. This is because as the test probe moves in a certain direction, clashes are more likely to occur with distant atoms positioned in that direction, while distant atoms positioned in the opposite direction will become too far away to create a compressive force. Operationally, $n$ gives the array size needed to account for all possible clashes the test probe could encounter as it moves away from the reference grid point. If $E(R,n|k)>E_\textrm{max}$, then the grid point is counted as microvoid; otherwise, the point contributes to cavity because the probe does not overlap with any atom's vdW radius by more than the distance threshold. A schematic for compressive elasticity is shown in Fig. \ref{fig:spring_model}. 

The compressive energy minimization process is summarized by five key properties. First, a proof by construction shows that a test probe cannot clash with $n \le 3$. When a test probe has compressive energy, it typically clashes with several atoms. Consequently, when $0<E(R,n|k)<E_\textrm{max}$, the overlap errors per atom are typically much lower than the distance tolerance. Second, it is only possible to misidentify a true cavity region as microvoid but not vice versa. This misidentification can occur if the relaxation method fails to locate the true energy minimum, denoted as $E(R,n|\infty)$. Specifically, $E(R,n|\infty)<E_\textrm{max}<E(R,n|k)$ is the erroneous case where the grid point is a true cavity but is classified as a microvoid. Third, we employ conservative convergence thresholds to abort the energy minimization process when the rate of energy decrease becomes slow while the gap in energy above $E_\textrm{max}$ remains high. That is, once $E(R,n|k)- E(R,n|k+1) \ll E(R,n|k+1)-E_\textrm{max}$, we terminate the energy minimization. The error cases are quite rare and only occur when convergence is slow and $E(R,n|\infty) \lesssim E_\textrm{max}$. Fourth, a larger test probe will result in more grid points being identified as microvoid because more neighboring atoms will typically contribute to compressive energy. Additional compressive springs slow down the energy minimization process. Fifth, when a grid point is identified as microvoid using a small probe, it will remain a microvoid for any larger probe because $E(R_2,n_2|\infty) \geq E(R_1,n_1|\infty)$ when $R_2>R_1$ where it is also the case that $n_2 \geq n_1$. 

Taking these properties into account leads to an algorithm that, for each grid point, begins by minimizing the compressive energy for a small probe size and iteratively considers larger sizes until the probe size of interest is reached. We find that a schedule for probe sizes that is linear in the volume of the probe works best. However, we determine the number of different probe sizes in the schedule to be $N_\textrm{probe} = \ceil{\frac{R}{0.5\textrm{ \AA{}}}}$, where $R$ is the probe radius of interest. For example, if we wish to consider a 4 \AA{} probe, we employ a total of 8 different probes whose volumes are equally spaced. As we iterate over the probe sizes, if the energy relaxation clearly shows strain energy, then the grid point must be microvoid and the remaining, larger probe sizes need not be checked. However, if a grid point cannot be identified as microvoid due to lack of convergence or is identified as cavity for a probe size iteration, then the algorithm continues on to the next probe size. This process is repeated until the probe size of interest is reached. Note that misclassification rates are reduced by using a higher energy threshold for smaller probes in the loop. On the final probe size, more energy minimization iterations are allotted to reach convergence if necessary. Otherwise, the energy minimization procedure is the same.

In summary, an intrinsic problem to iterative relaxation methods is to decide upon a convergence threshold. An incorrect identification as microvoid indicates that the system had not yet reached its minimum energy when a decision was made and $E(R,n|\infty)<E_\textrm{max}$. But to maintain speed, some errors can be tolerated to avoid large number of iterations. Therefore, we tested the method by repeatedly making predictions on the same set of proteins with the same orientations while varying the maximum number of iterations, threshold criteria, and other parameters that alter the rate of convergence and performance. The most accurate combinations of parameters result in the algorithm becoming rather slow. While evaluating different parameterizations, we located discrepancy cases and checked the reasons for and the severity of the incorrect assignment of the grid point as microvoid volume. We then adjusted the model parameters to arrive at a low non-systematic error rate that is less than one incorrect microvoid assignment per $2.5\times 10^5$ microvoid assignments. Overfitting is not an issue due to the model only having a handful of parameters compared to hundreds of millions of generic local molecular geometries that were tested. Note that no error occurs when a grid point is assigned as cavity. In addition, because we average the calculation over 300 random rotations and the errors virtually never occur at the exact same point in the protein, the overall error rate is less than ${10}^{-6}$ percent in misidentification. A much more substantial source of error is due to the finite resolution of the lattice when discretizing space, which we demonstrate is also acceptably low in Section \ref{sec:volume_chars}.

In this work, we calculate volume characteristics of proteins based on their static structure determined via X-ray crystallography. Although these static structures are commonly employed to characterize the native state of a protein, there are ambient vibrations about the equilibrium conformation. To account for local volume fluctuations as atoms vibrate, we consider a variable test probe size. This requires us to calculate the average volume characteristics of a protein over an ensemble of test probe sizes. These probe sizes are drawn from a Gaussian distribution with a specified mean radius and a standard deviation set to 5\% of this mean radius. With negligible additional computational cost, we randomly rotate the protein about its center of mass before embedding it onto the lattice in tandem with performing the calculation over the ensemble of test probe radii. We find that averaging over a narrow spectrum of probe radii helps to smooth out the discrete nature of the data quite well. By modeling probe sizes with slight variations and using the compressive energy model to capture off-lattice structural variations, we effectively mitigate artifacts arising from geometrical idealizations based on static structure. As a consequence, accurate calculations of volume can be obtained for any grid size provided it is equal to or less than the mean probe radius. 

\section{Volume characteristics}\label{sec:volume_chars}

To test the volume calculations, we apply our method to a subset of structures  selected from the Top 500 dataset of globular proteins, originally published by \citet{Hobohm1994}.  This list was compiled to ensure good structure quality, resolution, and non-redundancy amongst members of the set. We removed all structures from the original Top 500 with any missing chains, residues, or atoms. Additionally, we removed structures with an incomplete biological unit definition and those with fewer than 50 residues. The remaining 108 structures span a range of sizes from 50 to over 800 residues, providing a solid basis for testing and benchmarking the method as well as establishing general trends in volume characteristics across globular proteins. 

\begin{figure*}
 \centering
    \begin{subfigure}{0.32\textwidth}
        \centering
    \includegraphics[width=\textwidth]{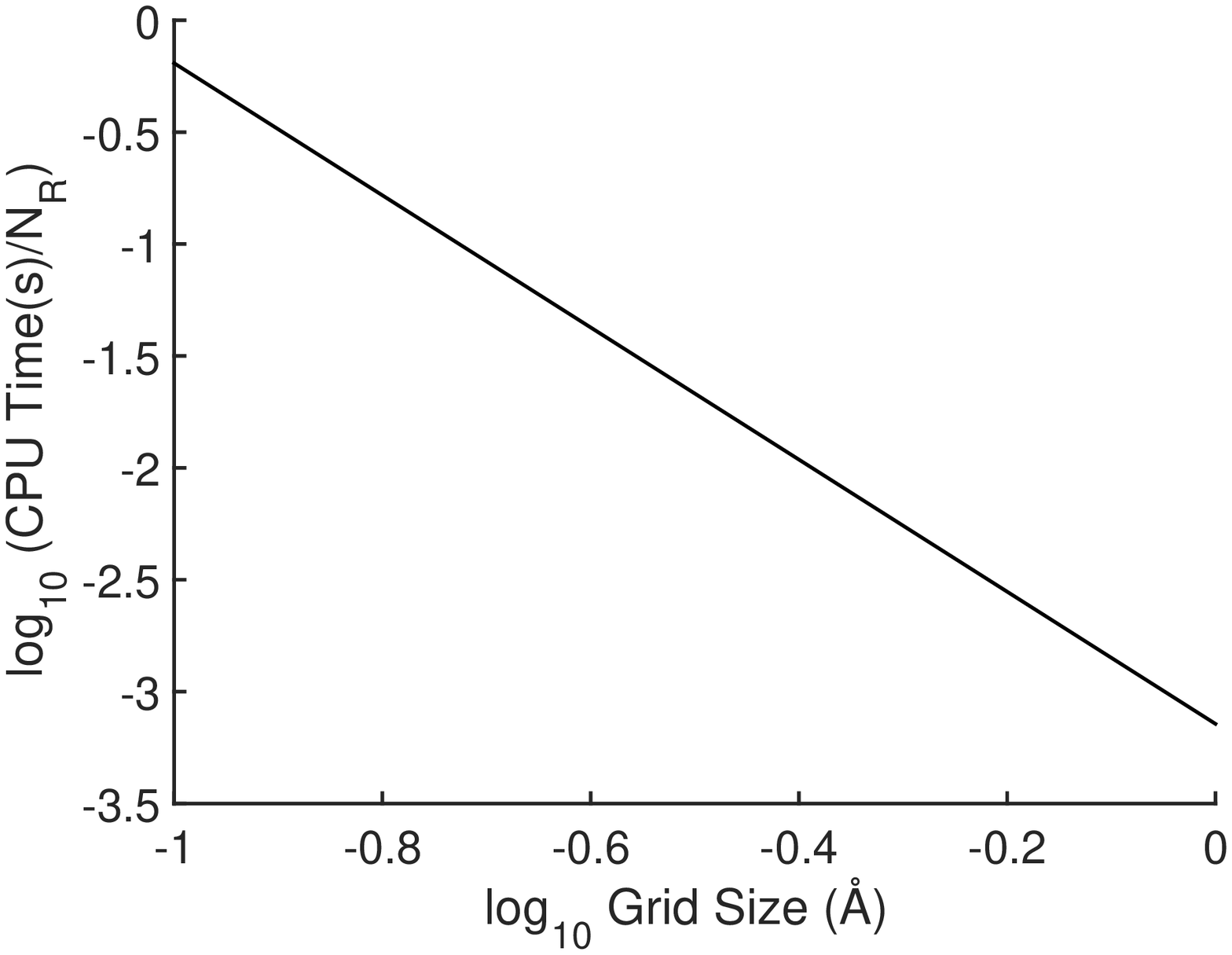}
        \caption{}\label{subfig:bench_grid}
    \end{subfigure}
    \begin{subfigure}{0.32\textwidth}
        \centering
    \includegraphics[width=\textwidth]{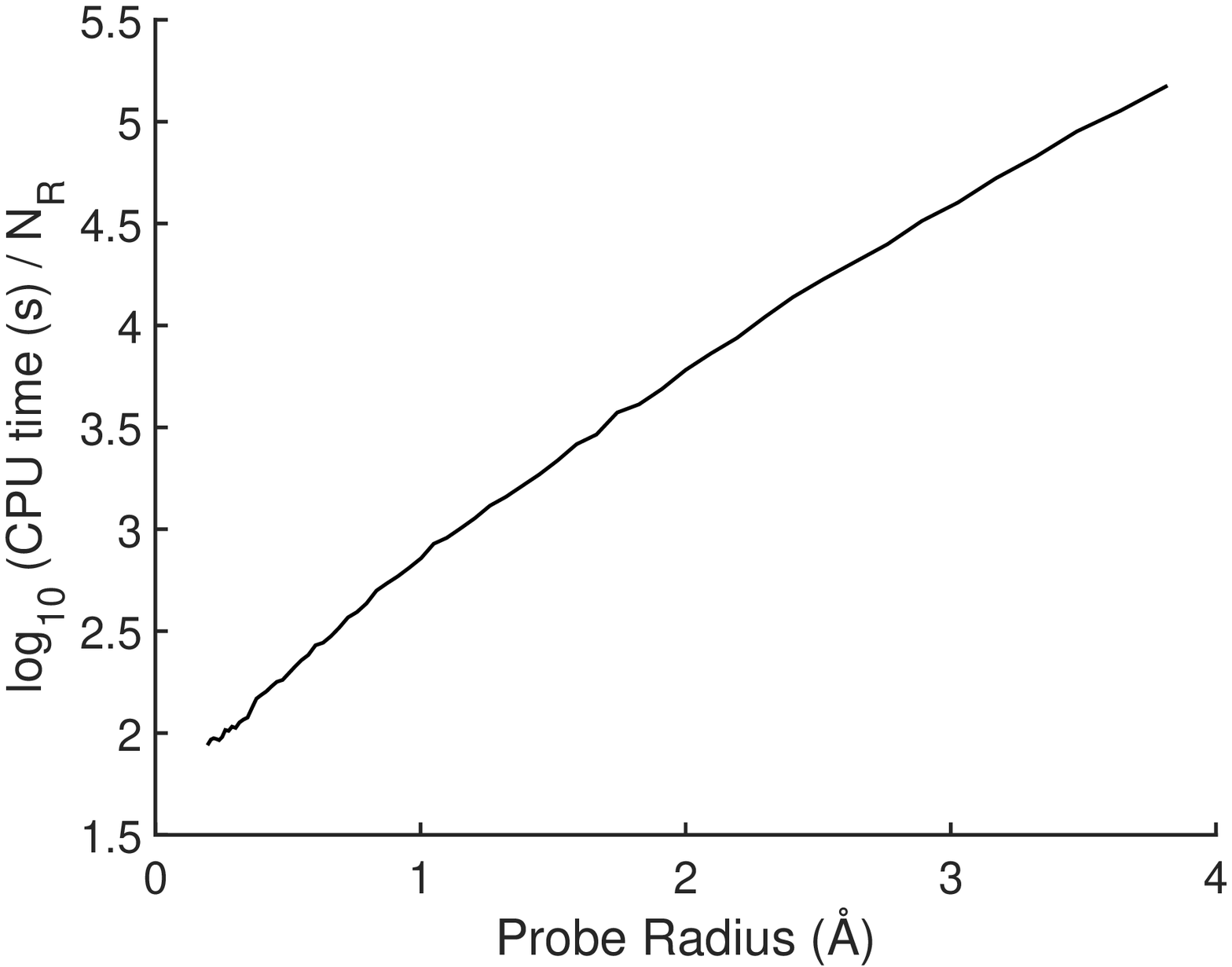}
        \caption{}\label{subfig:bench_probe}
    \end{subfigure}
    \begin{subfigure}{0.32\textwidth}
        \centering
    \includegraphics[width=\textwidth]{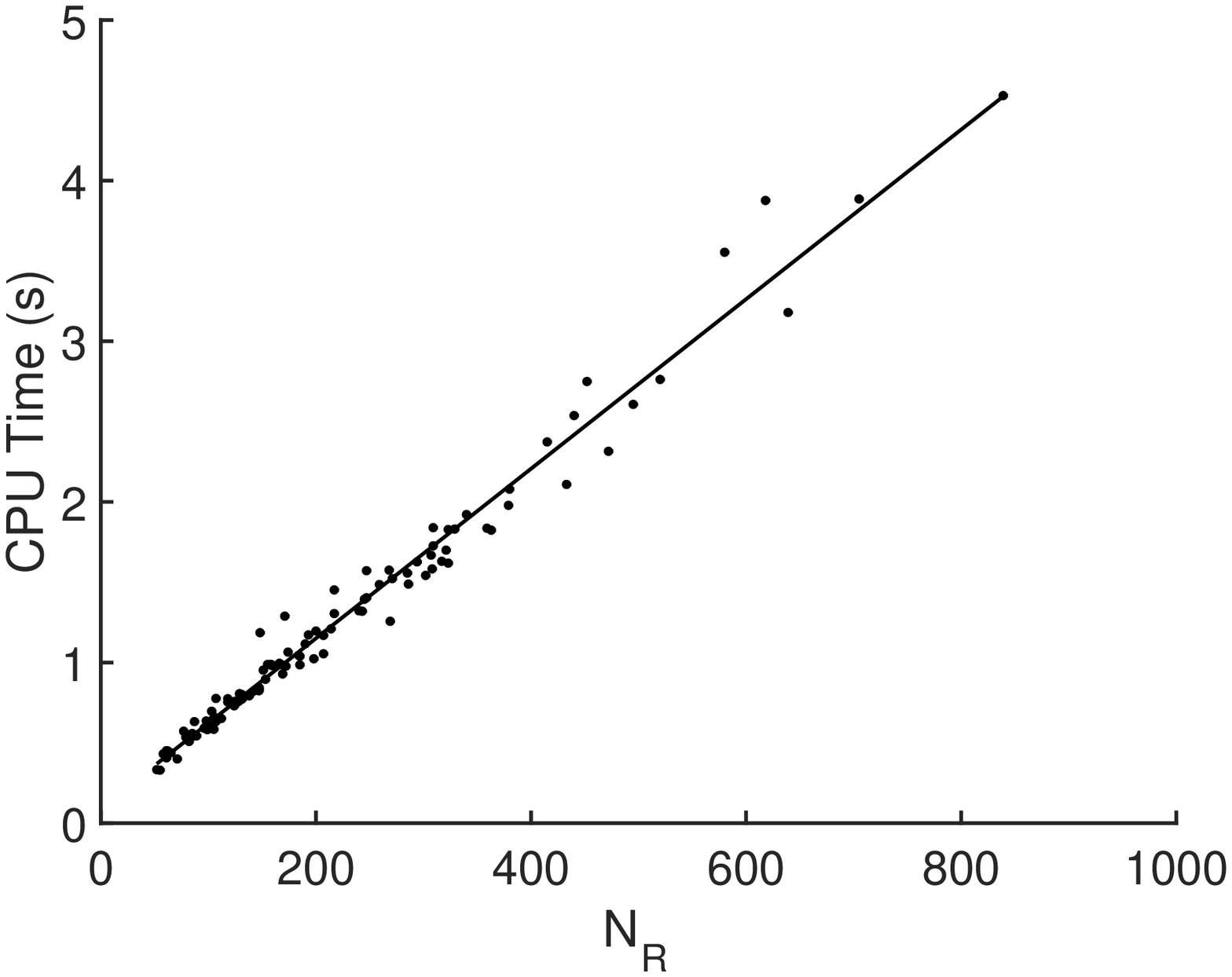}
        \caption{}\label{subfig:bench_length}
    \end{subfigure}
\caption{Benchmark time metric trends by (a) grid size, (b) probe radius, and (c) protein length}
\label{fig:benchmarks}
\end{figure*}

\subsection{Performance and convergence properties}\label{subsec:perf_convergence}

For each protein, 300 rotations in uniformly random directions for lattice embedding were performed for grid sizes $0.1 - 1.0$ \AA{} in linear increments of 0.1 \AA{} and up to 76 logarithmically spaced probe radii ranging from the grid size to 4.0 \AA{}.  Mean values and variances over the ensemble of rotations were calculated for each volume type as a global quantity as well as for partial volumes per atom and residue. The CPU times were clocked on the same high-performance cluster using a single core with an Intel Xeon 2.93 GHz processor for all calculations.  Fig. \ref{fig:benchmarks} summarizes how CPU times vary as a function of grid size, probe radius, and protein length.  The results shown in Fig. \ref{subfig:bench_grid} are for the case of varying grid size with a fixed probe radius of 1.4 \AA{}.  We find that $T_\mathrm{CPU} \propto a^{-2.95}$, indicating that the method is linear in the number of grid points.  The time complexity dependence on test probe radius is more complicated.  Fig. \ref{subfig:bench_probe} shows compute times versus test probe radius with a fixed lattice constant of $a=0.1$ \AA{}.  The nearly linear relationship between $R$ and $\log (T_\mathrm{CPU})$ demonstrates an approximately exponential time complexity for $R$. In both cases, CPU times for each of the 108 proteins were averaged and normalized by the protein's sequence length (e.g. number of residues $N_R$).

A linear scaling of CPU time with protein length is seen in Fig. \ref{subfig:bench_length}, which shows that the method is well suited for large molecular systems. Points are plotted for all 108 proteins for a grid size of 0.1 \AA{} and probe radius of 1.4 \AA{}. Thus, the empirically determined time complexity for processing a protein is given by $T_\textrm{CPU} \sim N_R \frac{e^{cR}}{a^3}$, where c is a constant.

\begin{figure*}
 \centering
    \begin{subfigure}{0.32\textwidth}
        \centering
    \includegraphics[width=\textwidth]{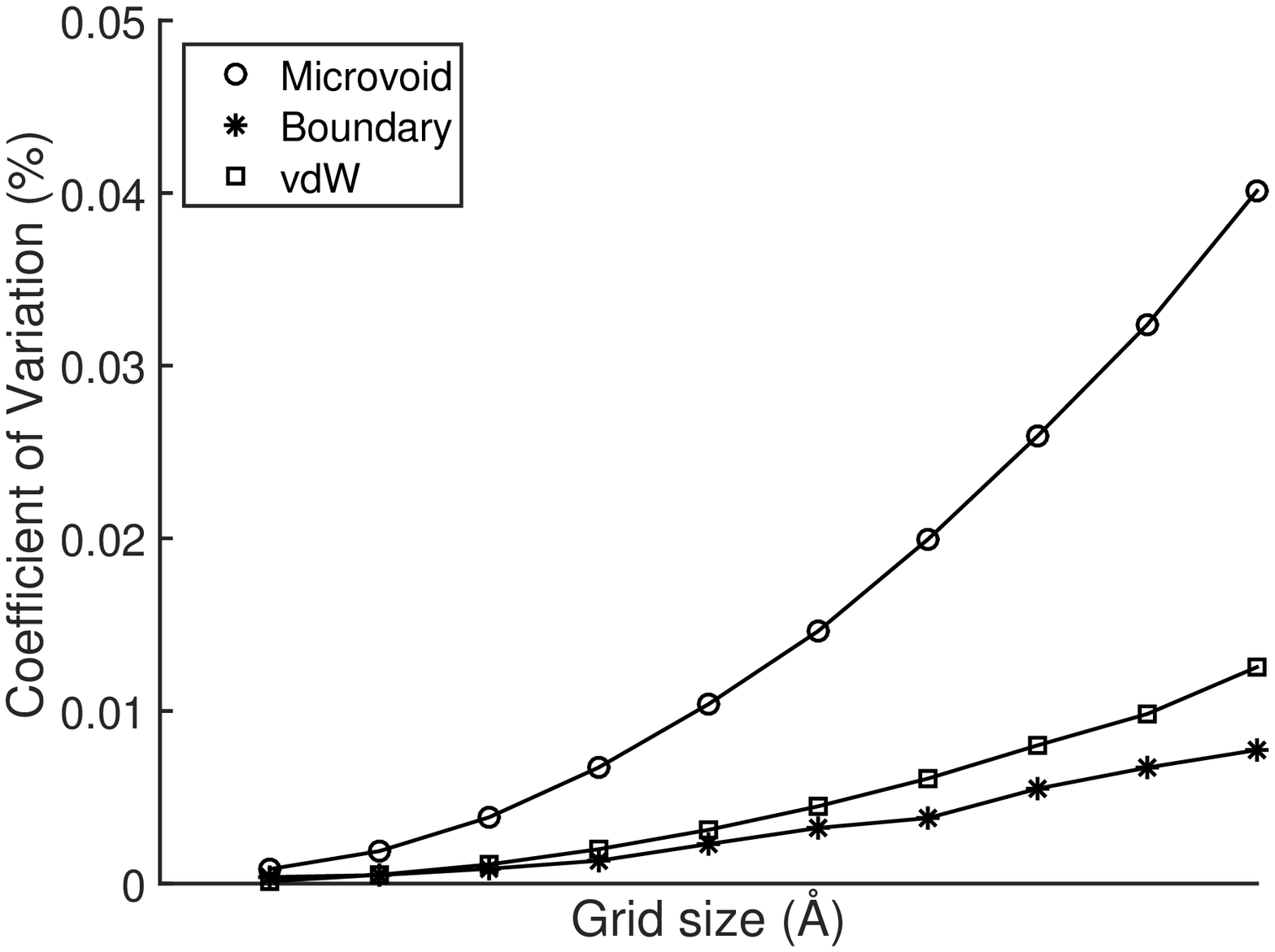}
        \caption{}\label{subfig:convergence_fixed}
    \end{subfigure}
    \begin{subfigure}{0.32\textwidth}
        \centering
    \includegraphics[width=\textwidth]{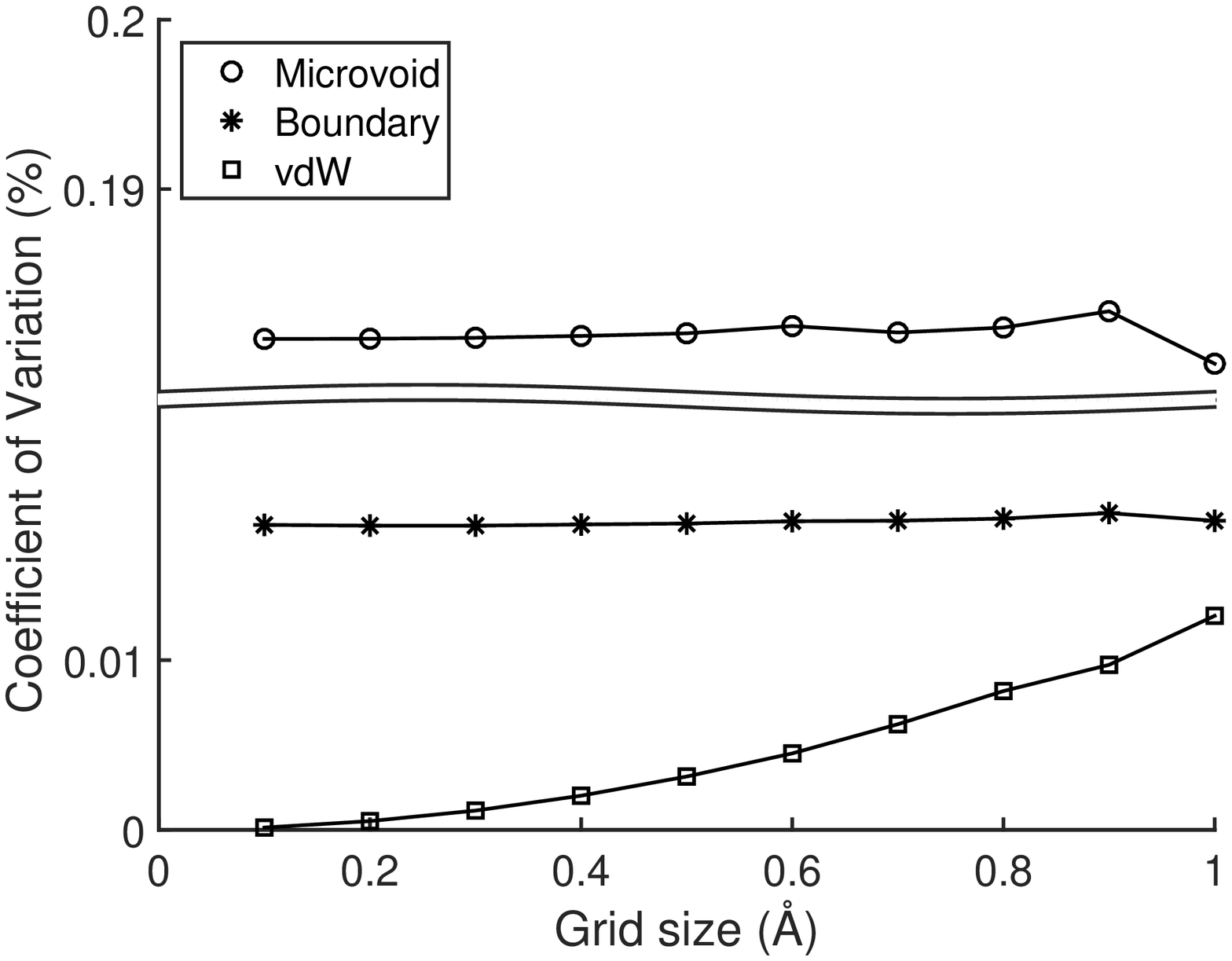}
        \caption{}\label{subfig:convergence_variable}
    \end{subfigure}
    \begin{subfigure}{0.32\textwidth}
        \centering
    \includegraphics[width=\textwidth]{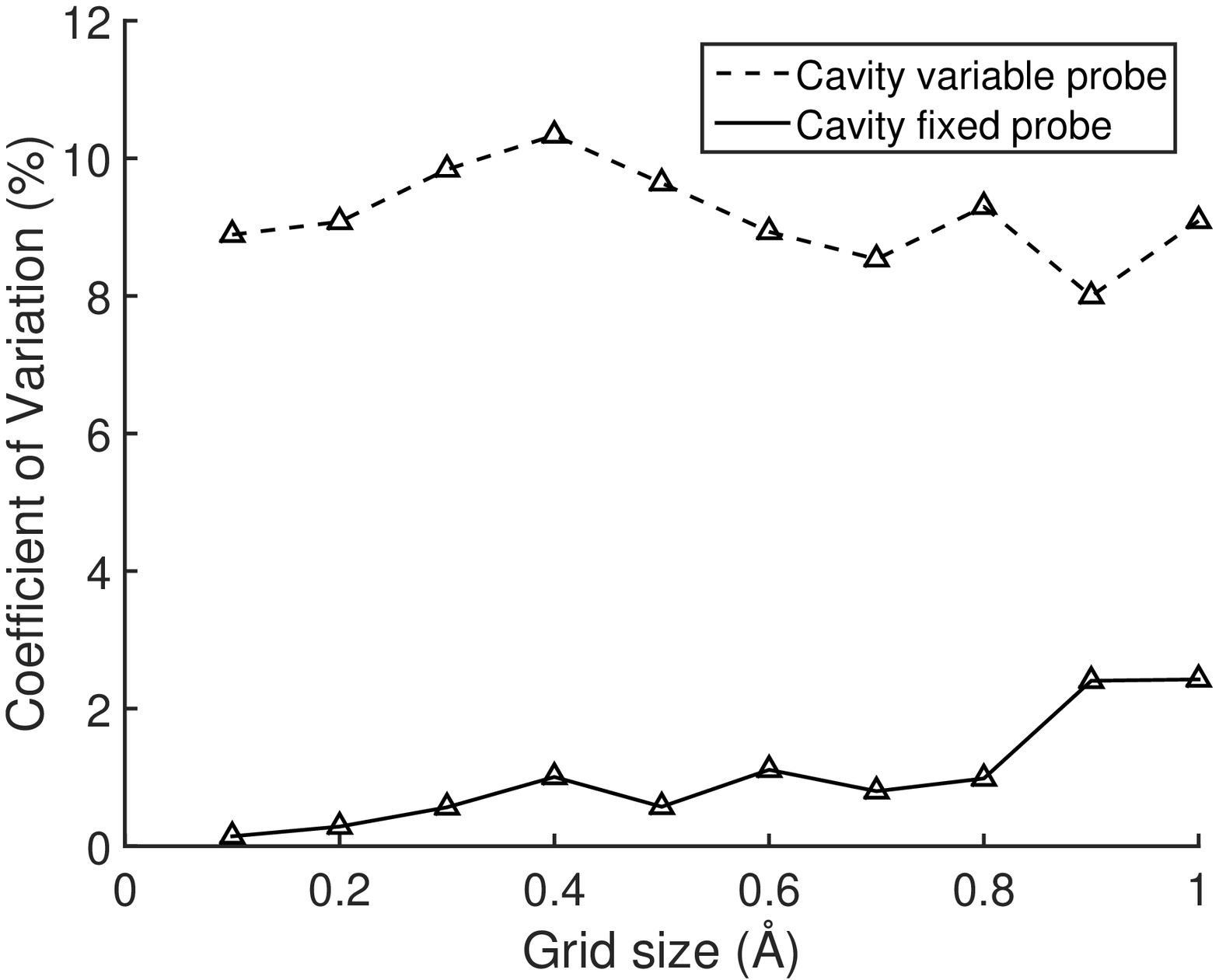}
        \caption{}\label{subfig:convergence_cavity}
    \end{subfigure}
\caption{(a) Convergence with rotations and fixed probe radius. (b) Convergence with rotations and variable probe radius. (c) Convergence for cavity volume.}
\label{fig:convergence}
\end{figure*}

In Fig. \ref{subfig:convergence_fixed} and \ref{subfig:convergence_variable}, the coefficient of variation, defined as the ratio of the standard deviation to the mean and expressed as a percentage, is plotted as a function of grid size for various types of volumes.  Figure \ref{subfig:convergence_fixed} shows the convergence for each volume type with 300 rotations for a fixed probe size of 1.4 \AA{}. The coefficient of variation decreases as grid resolution increases, demonstrating virtually negligible error with rotation, even for relatively large grid sizes. The convergence of the same quantities using the probe-averaging method with a mean radius of 1.4 \AA{} is shown in Fig. \ref{subfig:convergence_variable}. The grid size dependence for the vdW coefficient of variation is essentially the same regardless of fixed or variable probe radius, confirming that vdW volume is independent of probe size. Interestingly, microvoid and boundary volumes have nearly constant coefficients of variation across all grid sizes, indicating that the random variation on the test probe radius introduces much more variance than the resolution effects. The reason for placing variation on the test probe radius is to better model soft vdW interactions and dynamic fluctuations in structure. Since probe size variation is the dominant cause for variations in our ensemble averaging, consideration of multiple rotations is not necessary. Nevertheless, randomizing over different protein orientations on a grid contributes negligible computational cost when performing an ensemble average over probe size.

The variation of cavity volume over random protein orientations for both fixed and varied probe size is shown in Fig. \ref{subfig:convergence_cavity}. As with microvoid and boundary volumes, for fixed probe size, the variations in cavity volume decrease with decreasing grid size. However, the coefficient of variation is considerably larger than that of other volume types, which is explained by noting that cavity volume is small, yet they undergo relatively large percent changes in volume due to sensitivity in connecting to channels that extend to the protein surface. For a small probe radius, the channel opens up, allowing solvent to penetrate into the cavity. In this case, cavity volume becomes boundary volume. Similarly, two small microvoid clusters may merge together to form a cavity.  There is also a greater degree of sensitivity to protein orientation when a probe radius is close to the critical threshold for a cavity to dramatically change size. These effects are enhanced by varying the probe radius, as shown in Fig. \ref{subfig:convergence_cavity}. Again, the variation in probe radius increases the coefficient of variation for cavity volume and appears to dominate over the dependence on grid size.  

Taken together, the plots in Figs. \ref{fig:benchmarks} and \ref{fig:convergence} provide little justification for using grid sizes much smaller than the probe size of interest. The much greater computational cost yields very little gains in accuracy due to intrinsic uncertainties from ambient dynamical fluctuations. Since the probe radius is usually taken to be 1.4 \AA{} when considering volume characteristics in proteins, a grid size of 0.5 \AA{} provides ample accuracy with calculation times that are competitive with the fastest methods reported in the literature.  It is worth noting that methods such as alpha shape are limited to a static view of the protein unless averaging over multiple frames from a MD simulation. If our method were applied to an MD trajectory, we would use a fixed probe radius and not randomize the protein orientation on the grid since the trajectory naturally captures the dynamical fluctuations.

\begin{figure*}
 \centering
    \begin{subfigure}{0.45\textwidth}
        \centering
    \includegraphics[width=\textwidth]{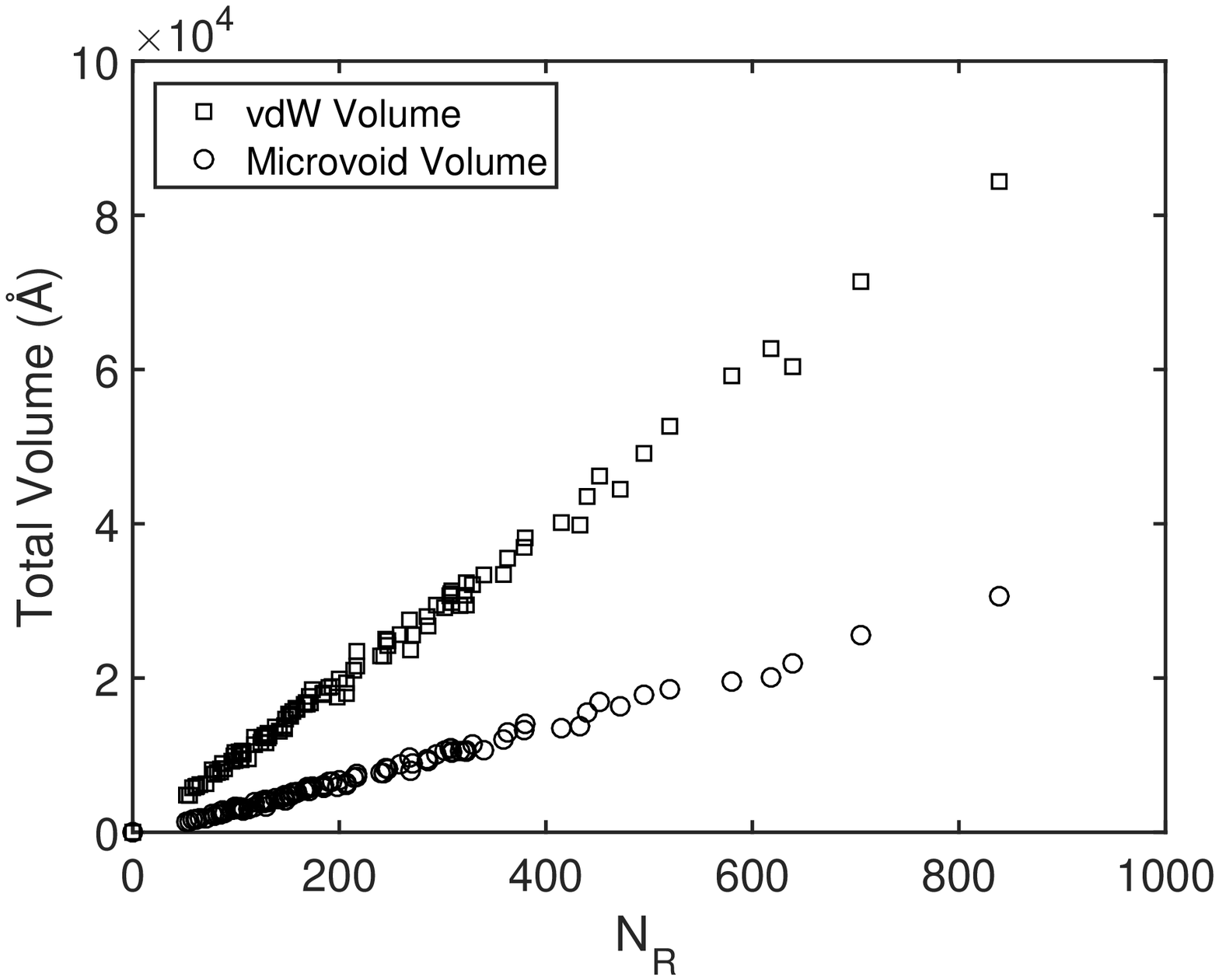}
        \caption{}\label{subfig:vdw_mv_vol}
    \end{subfigure}
    \begin{subfigure}{0.45\textwidth}
        \centering
    \includegraphics[width=\textwidth]{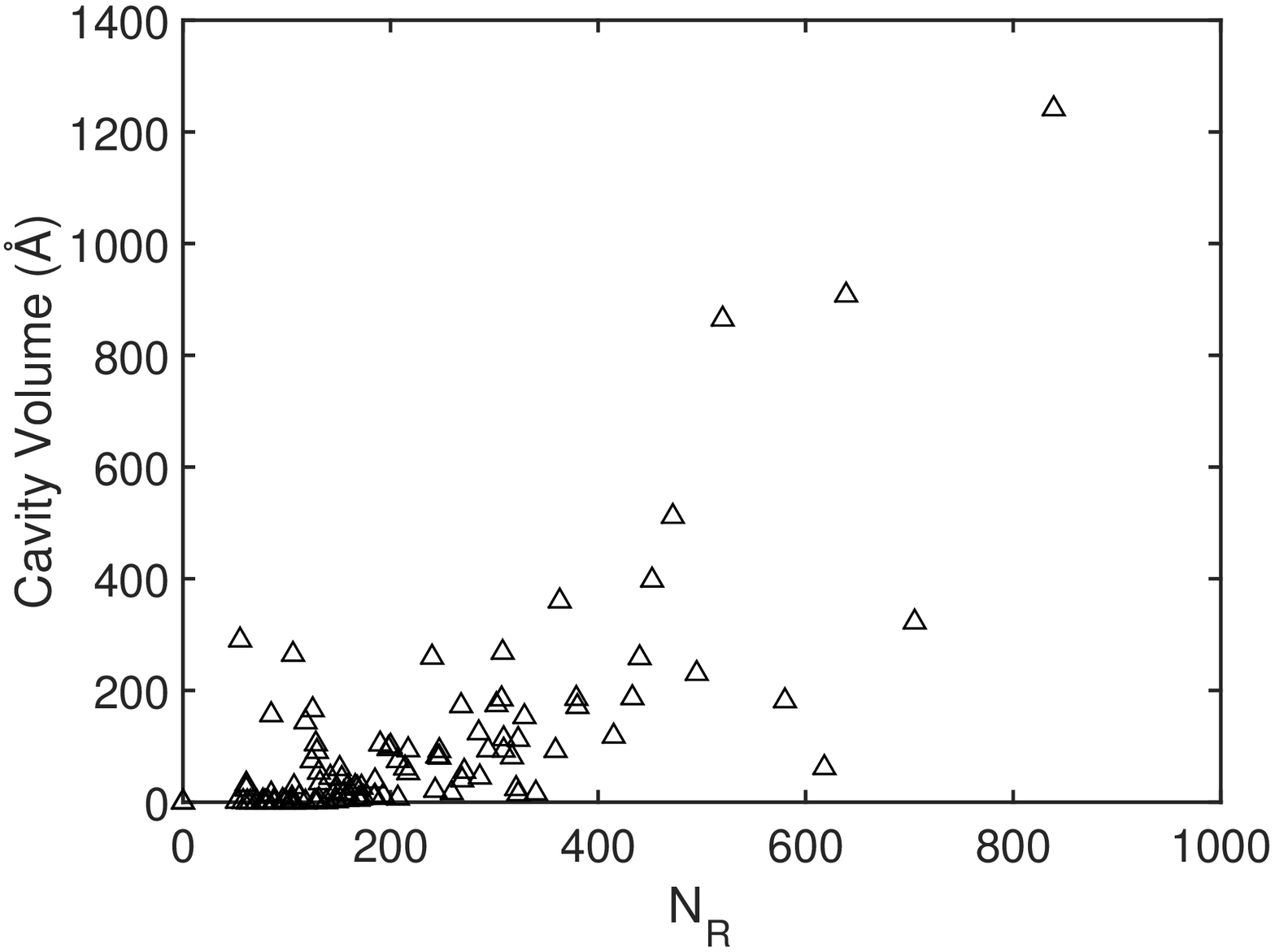}
        \caption{}\label{subfig:cav_vol}
    \end{subfigure}
    \begin{subfigure}{0.45\textwidth}
        \centering
    \includegraphics[width=\textwidth]{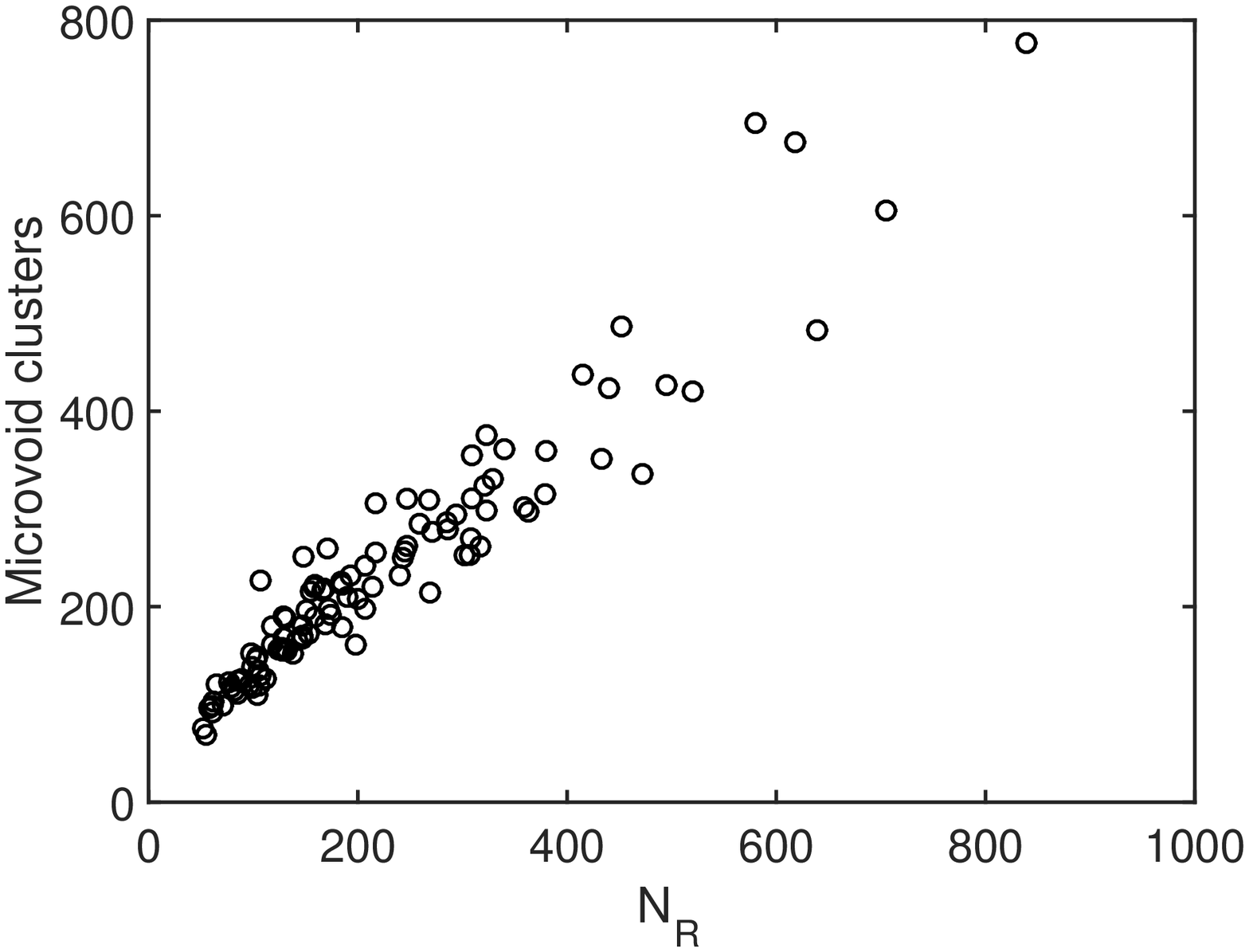}
        \caption{}\label{subfig:mv_clusters}
    \end{subfigure}
    \begin{subfigure}{0.45\textwidth}
        \centering
    \includegraphics[width=\textwidth]{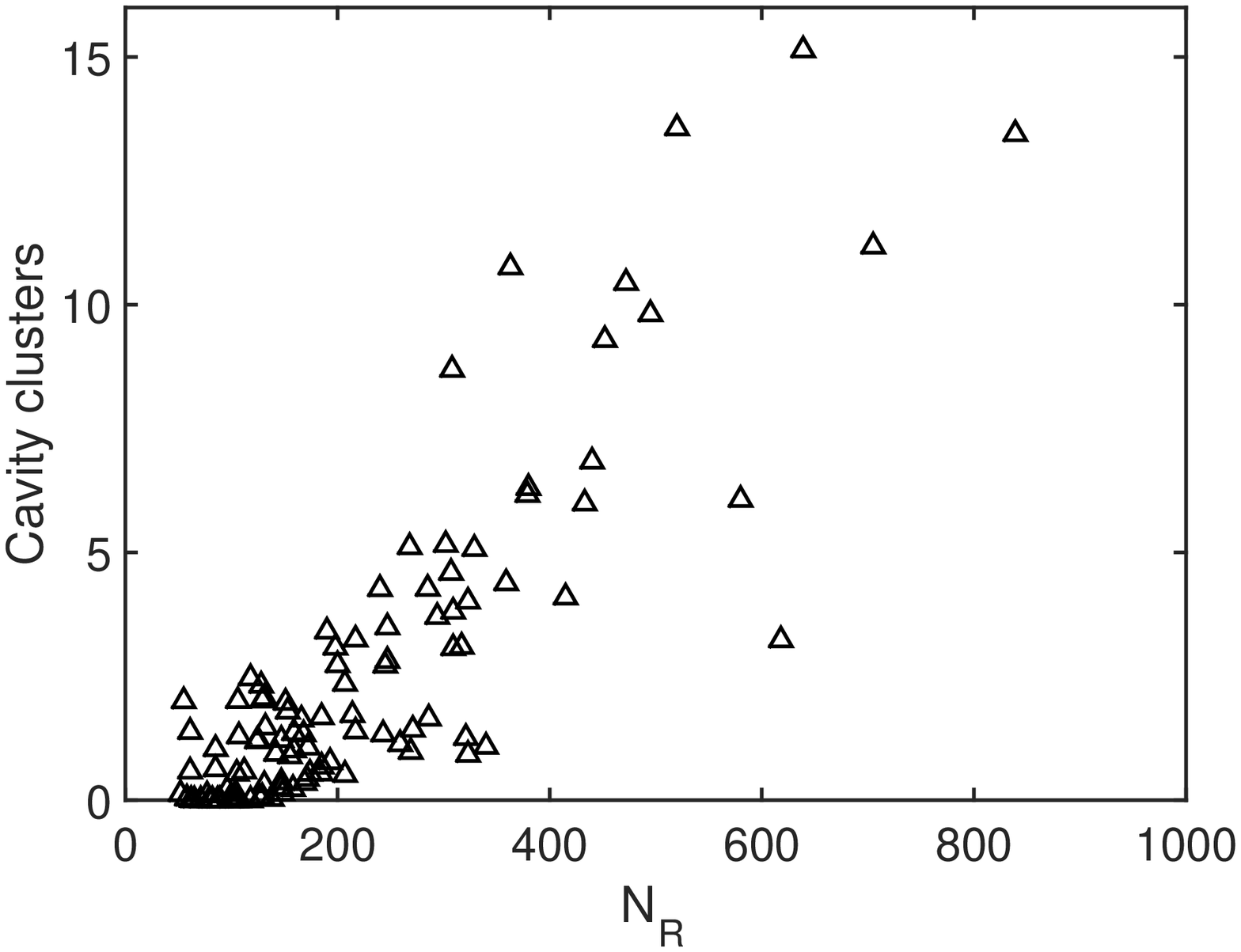}
        \caption{}\label{subfig:cav_clusters}
    \end{subfigure}
\caption{(a) Characteristics as a function of protein length, averaged over 300 rotations for grid size 0.1 and probe radius 1.4 for (a) total vdW and microvoid volume, (b) total cavity volume, (c) number of microvoid clusters, and (d) number of cavity clusters}
\label{fig:volume_characteristics}
\end{figure*}

\subsection{Linear scaling with protein length}\label{subsec:scaling_with_length}

We observe that some quantities depend strongly on protein length. In Fig. \ref{fig:volume_characteristics}, the results are exemplified for the case that grid size and probe radius are 0.1 \AA{} and 1.4 \AA{}, respectively. Note that qualitatively similar correlations are found (not shown) when we replace the number of residues (i.e. protein length) with the number of atoms. In Fig. \ref{subfig:vdw_mv_vol}, we find that vdW volume and microvoid volume scale linearly with protein length. In contrast, cavity volume, shown in Fig. \ref{subfig:cav_vol}, has only a slight tendency to increase with protein length. These trends have been previously established for cavity and vdW volumes using a variety of methods and datasets \citep{Liang2001,Chen2015}. However, to the best of our knowledge, microvoid as a separate quantity has not yet been considered. Figs. \ref{subfig:mv_clusters} and \ref{subfig:cav_clusters} show that the total number of distinct microvoid clusters scales linearly with protein length, and the number of distinct cavities (e.g. cavity volume clusters) also scales approximately linearly.  Although it is intuitive to expect partial volumes to be extensive, like total volume, it is clear that cavity volume is not an extensive quantity, and as noted above is considerably smaller than other volume types. In particular, microvoid volume dominates the total void space. This is because cavity formation is thermodynamically unfavorable \citep{Cioni2006,Hubbard1994,Eriksson1992}. Hence, the presence of a cavity is expected to be of biological importance \citep{Winter2011,Spyrakis2011,Voss2006,Bui2004}. Interestingly, the number of clusters for microvoid and cavity appear to be extensive. These results show that a significant number of globular proteins tend to have greater cavity volumes than a simple linear trend line based on extensive properties alone, which suggests we are observing evolution-driven effects captured in our dataset.

\begin{figure*}
 \centering
    \begin{subfigure}{0.45\textwidth}
        \centering
    \includegraphics[width=\textwidth]{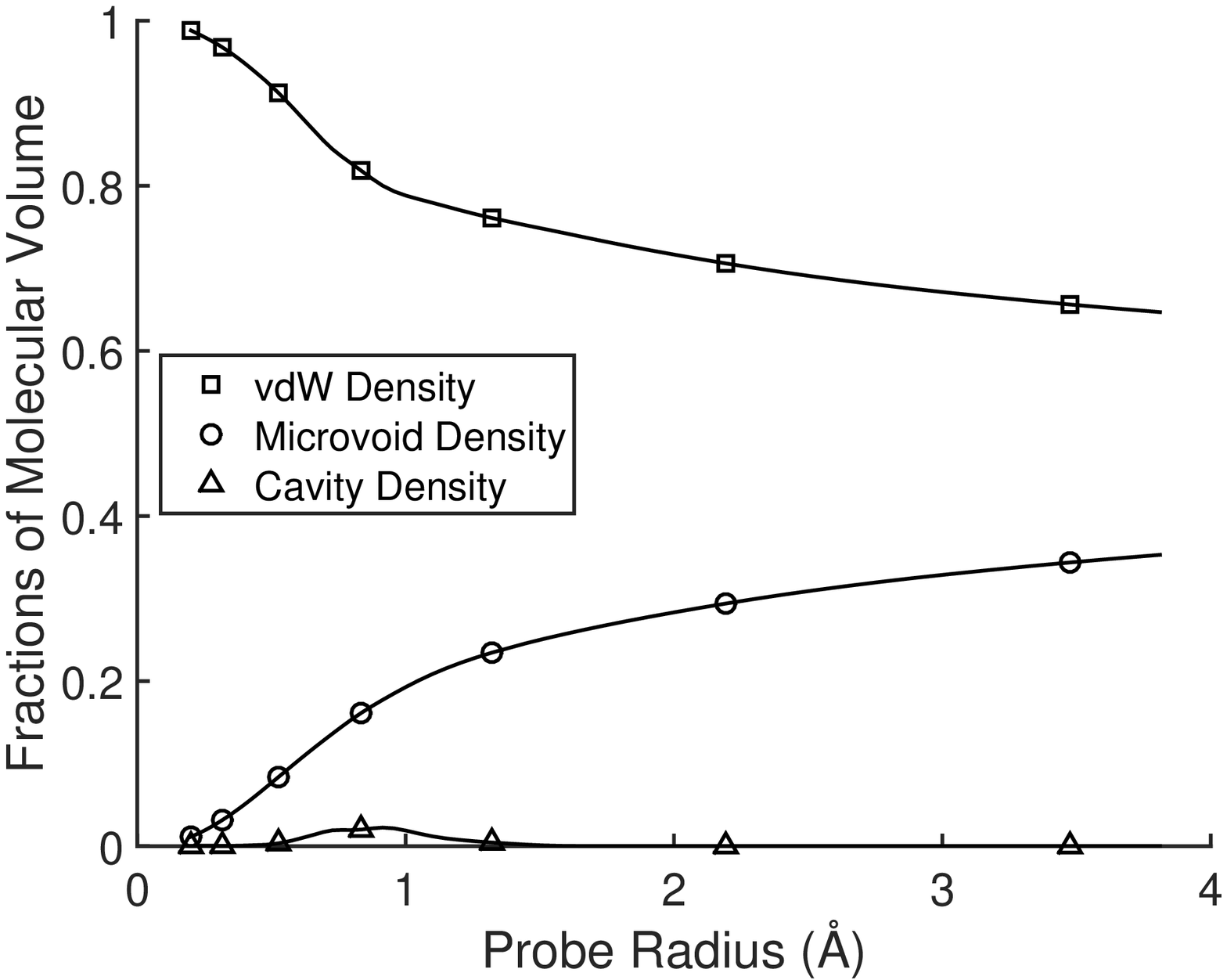}
        \caption{}\label{subfig:volume_densities}
    \end{subfigure}
    \begin{subfigure}{0.45\textwidth}
        \centering
    \includegraphics[width=\textwidth]{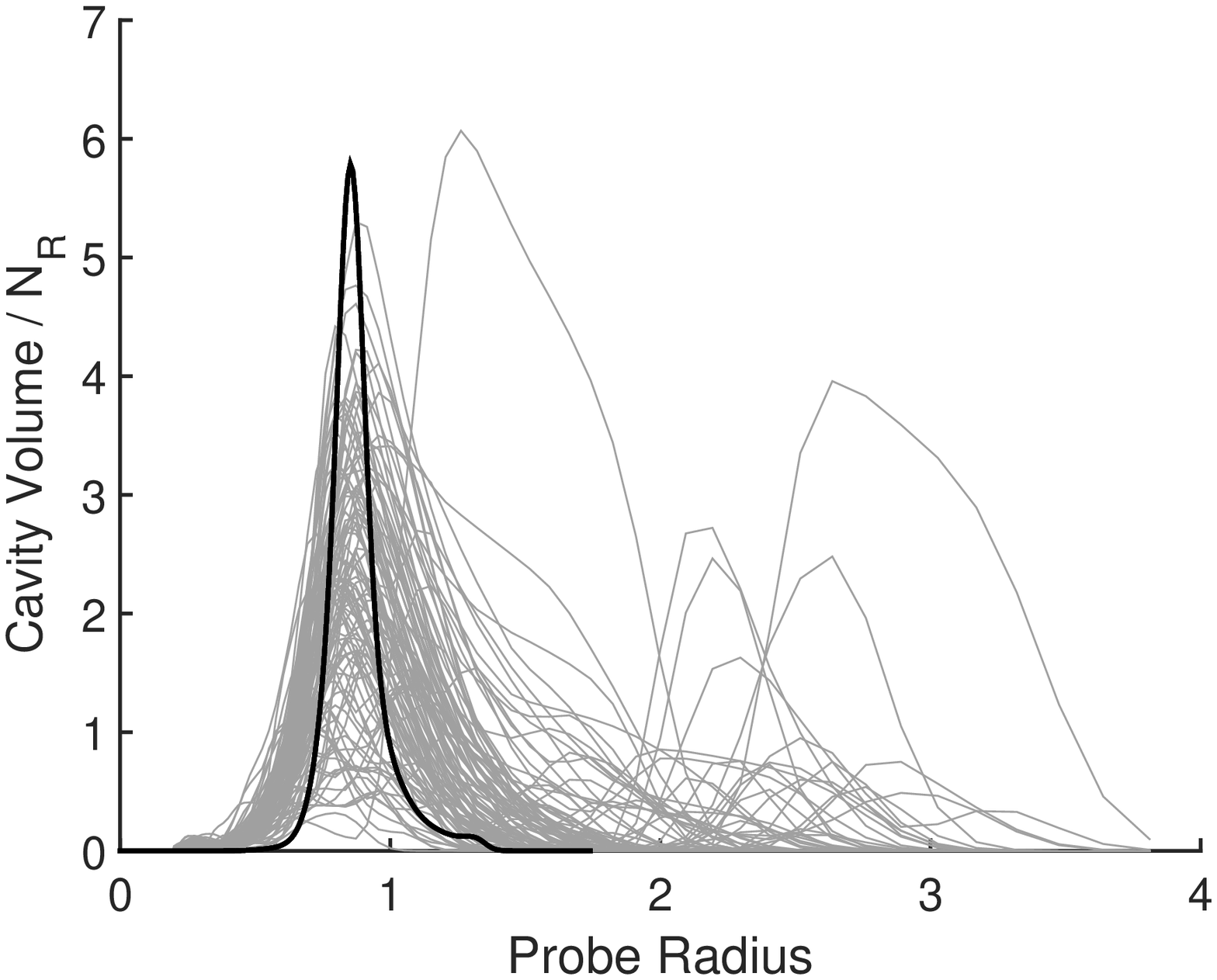}
        \caption{}\label{subfig:cav_probe_pdf}
    \end{subfigure}
\caption{Volume densities as a function of probe radius. (b) Distribution of peak cavity volume as a function of probe radius. Gray lines represent each of the 108 proteins.  Black line represents the distribution of peak probe radii for cavity volumes.}
\label{fig:vol_prob_dependence}
\end{figure*}

\subsection{Volumes as a function of probe size}\label{subsec:vol_vs_probe}

The classifications of cavity, microvoid, and boundary volume are dependent upon the size of the probe.  As the probe size approaches zero, the probe can access virtually all space outside of the van der Waals radii of the protein atoms.  Thus, in this limit, cavity and microvoid volumes shrink to zero and the volume is decomposed into either vdW or boundary volume.  As the size of the probe increases, crevices become unreachable by the probe external to the molecular surface, thus creating void space.  The probe-size dependence of vdW, microvoid, and cavity volumes expressed as fractions of the molecular volume is shown in Fig. \ref{subfig:volume_densities}. All of these volumes are averaged over 300 rotations for 108 proteins with error bars that are no larger than the symbols representing the volume type. For small probe size, the molecular volume is nearly equal to the vdW volume, whereas for large probe size, the relative fraction of vdW volume decreases while microvoid volume contributes much more to the molecular volume.  Note that the vdW volume divided by the molecular volume is a common definition for packing density and the results we obtain in Fig. \ref{subfig:volume_densities} agree with earlier studies \citep{Liang2001,Richards1974,Fleming2000}.

The cavity volume, on the other hand, accounts for a tiny fraction of the molecular volume for most proteins in our dataset.  The peak in cavity volume that occurs around 0.8 \AA{}, as shown in Fig. \ref{subfig:volume_densities}, may potentially be a universal characteristic for globular proteins. This result does not depend on grid size provided it is less than 0.8 \AA{} to capture this resolution. To study this further, a magnified view of the cavity volume versus probe radius is plotted for each of the 108 proteins as separate curves in Fig. \ref{subfig:cav_probe_pdf}. In addition, we overlay the probability density \citep{Farmer2018} for the probe radii at maximum cavity volume for each of the 108 proteins. The observed noteworthy consistency in the probe radius where the maximum cavity volume occurs across proteins is quite interesting and we posit the following explanation. Because globular proteins tend to pack well, the size of void space within local atomic environments is determined mainly by vdW interactions. A small probe radius allows the boundary volume to more easily penetrate the protein, substantially reducing cavity volume. A large probe radius creates more microvoid volume, which also substantially reduces cavity volume. A small range in probe size naturally appears where the cavity volume should be maximized. However, this geometrical and physical explanation does not appear to have a biological interpretation. It is worth mentioning that we observe a slight shift in the peak probe size that maximizes cavity volume when using alternative vdW radii definitions, as discussed below.

Since thermodynamic forces tend to limit cavity formation, the presence of a large cavity is a signature of an important functional characteristic of a protein. We note that, for several proteins in our dataset, very large peaks in total cavity volume are found for probe radii above 2 \AA{} (see Fig. \ref{subfig:cav_probe_pdf}). For each of these cases, we identify a single large pocket within the protein that is connected to solvent accessible tunnels with a much smaller opening (e.g. a lollypop shape) that will prevent large molecules from entering. In particular, PDB:1QFM is a protein in the dataset that is not plotted individually on Fig. \ref{subfig:cav_probe_pdf} for visual clarity, as it contains a large cavity over 8000 \AA{}$^3$ large with an opening of approximately 4 \AA{} in diameter that out-scales the axes \citep{Fulop1998}. These results highlight the value of scanning proteins with multiple probe sizes as a means of gaining a more complete picture of potentially important functional characteristics. This also emphasizes the importance of the local volume characteristics that will be discussed in Section \ref{sec:pd}.

\begin{figure*}
 \centering
    \begin{subfigure}{0.32\textwidth}
        \centering
    \includegraphics[width=\textwidth]{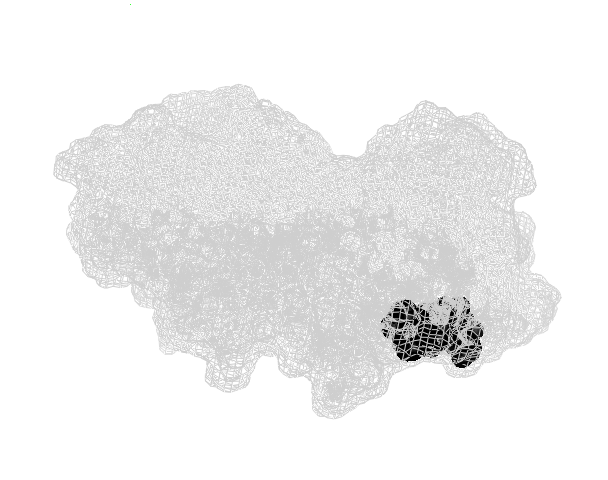}
        \caption{}\label{subfig:mv_below_perc}
    \end{subfigure}
    \begin{subfigure}{0.32\textwidth}
        \centering
    \includegraphics[width=\textwidth]{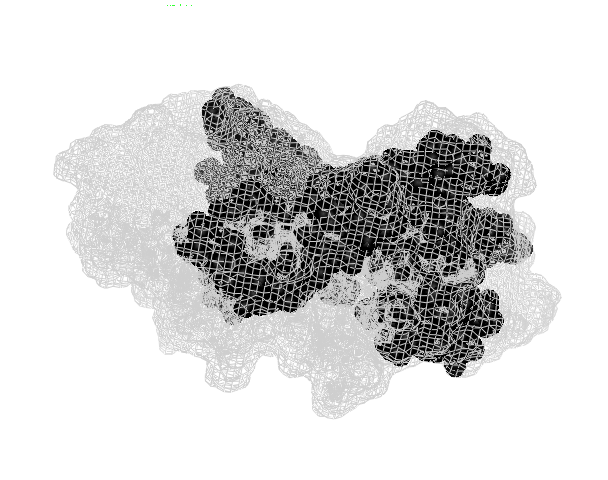}
        \caption{}\label{subfig:mv_at_perc}
    \end{subfigure}
    \begin{subfigure}{0.32\textwidth}
        \centering
    \includegraphics[width=\textwidth]{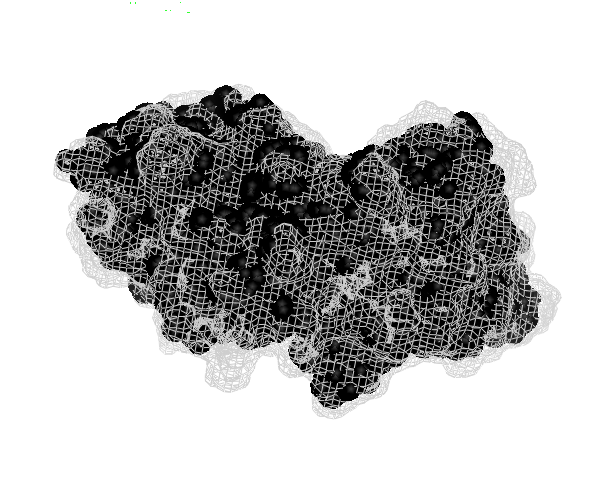}
        \caption{}\label{subfig:mv_above_perc}
    \end{subfigure}
\caption{Visualization of the largest microvoid cluster, shown in black, (a) below, (b) at, and (c) just above the percolation threshold.}
\label{fig:mv_cluster_vis}
\end{figure*}

\subsection{Microvoid percolation}\label{subsec:mv_perc}

Percolation theory provides a ubiquitous framework to understand clustering phenomena in a wide array of systems \citep{Jacobs1995,Geistlinger2015,Hassan2016,Iglauer2016,Huang2018,Roy2018,Stauffer1987} and has been previously used to characterize both protein volume and cavity volume within proteins \citep{Liang2001,Tang2017}. Here, we investigate the characteristics of microvoid percolation. Fig. \ref{subfig:volume_densities} shows that microvoid volume is a monotonically increasing function of probe radius. Fig. \ref{fig:mv_cluster_vis} illustrates how the largest cluster of microvoid volume in a protein changes as probe radius increases. For large probe sizes, the largest cluster of microvoid volume will span the entire protein, forming a percolating cluster. Note that for visualization purposes, the coordinates of all grid points, their volume types, and cluster labels are saved in memory as the HK method scans over a protein to identify specific clusters.  These coordinates and cluster labels are saved such that they can be visualized using PyMOL \citep{Schrodinger2015}. For example, in Fig. \ref{fig:mv_cluster_vis}, the protein volume is shown in light gray with black shading representing the largest microvoid cluster for probe sizes below, at, and just above the percolation threshold. The largest cluster at the threshold in Fig. \ref{subfig:mv_at_perc} has an irregular shape and does not span the protein in all directions simultaneously. Microvoid clusters are typically found to be non-spherical and they decrease in sphericity (as defined in \citep{Wadell1935}) with increasing microvoid cluster size.

\begin{figure*}
 \centering
    \begin{subfigure}{0.45\textwidth}
        \centering
    \includegraphics[width=\textwidth]{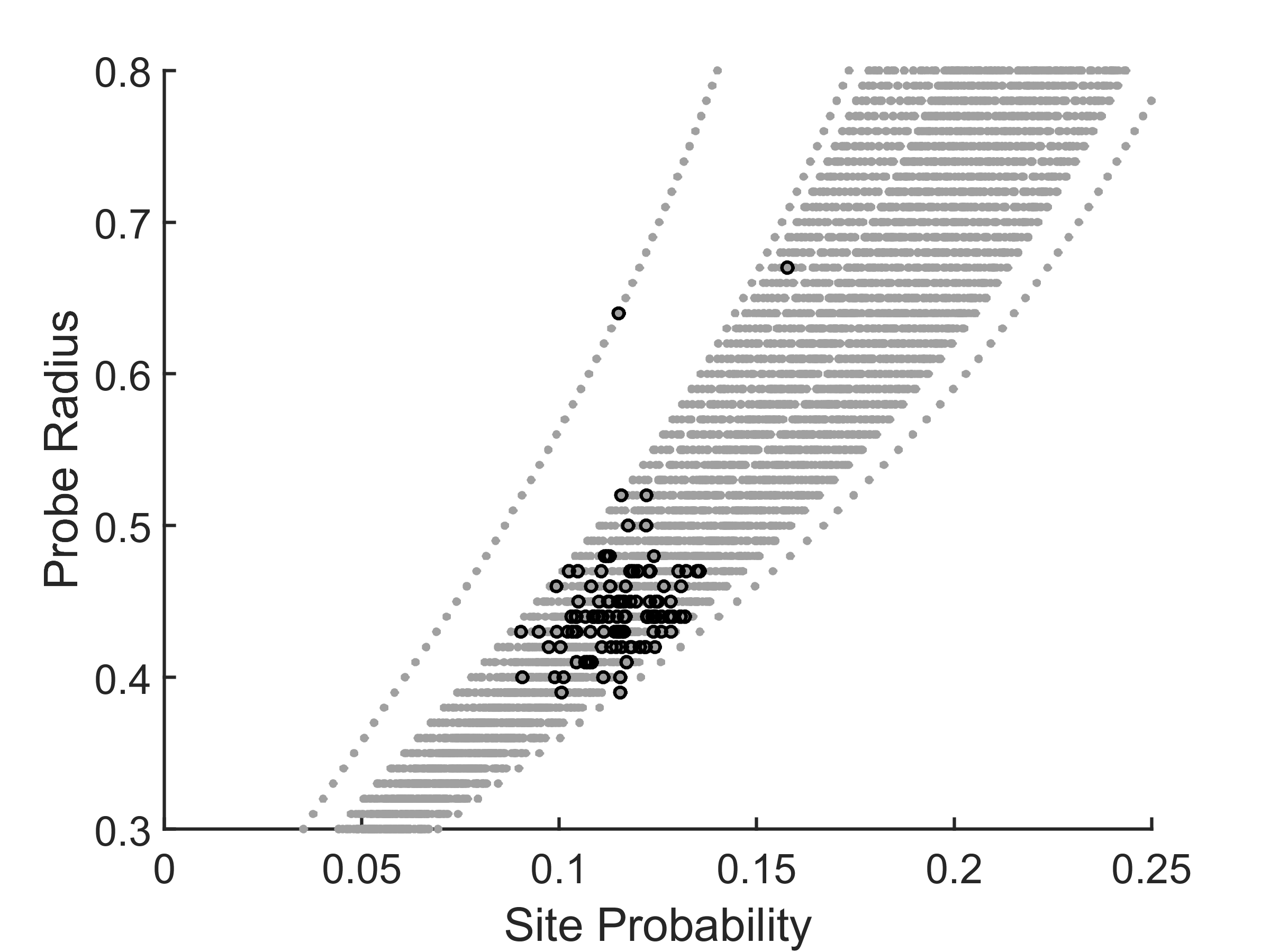}
        \caption{}\label{subfig:probe_rad_vs_prob}
    \end{subfigure}
    \begin{subfigure}{0.45\textwidth}
        \centering
    \includegraphics[width=\textwidth]{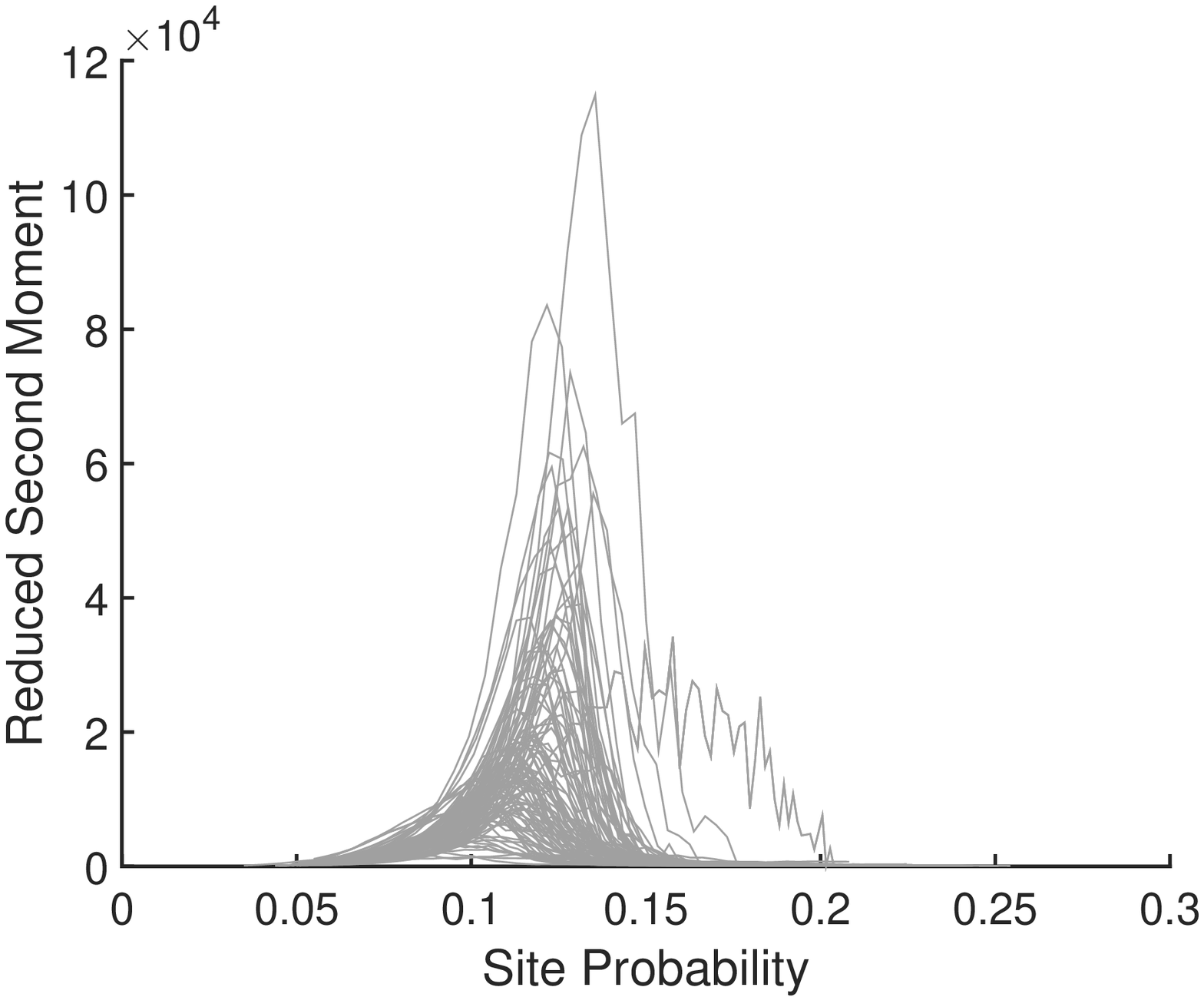}
        \caption{}\label{subfig:rsm_vs_prob}
    \end{subfigure}
    \begin{subfigure}{0.45\textwidth}
        \centering
    \includegraphics[width=\textwidth]{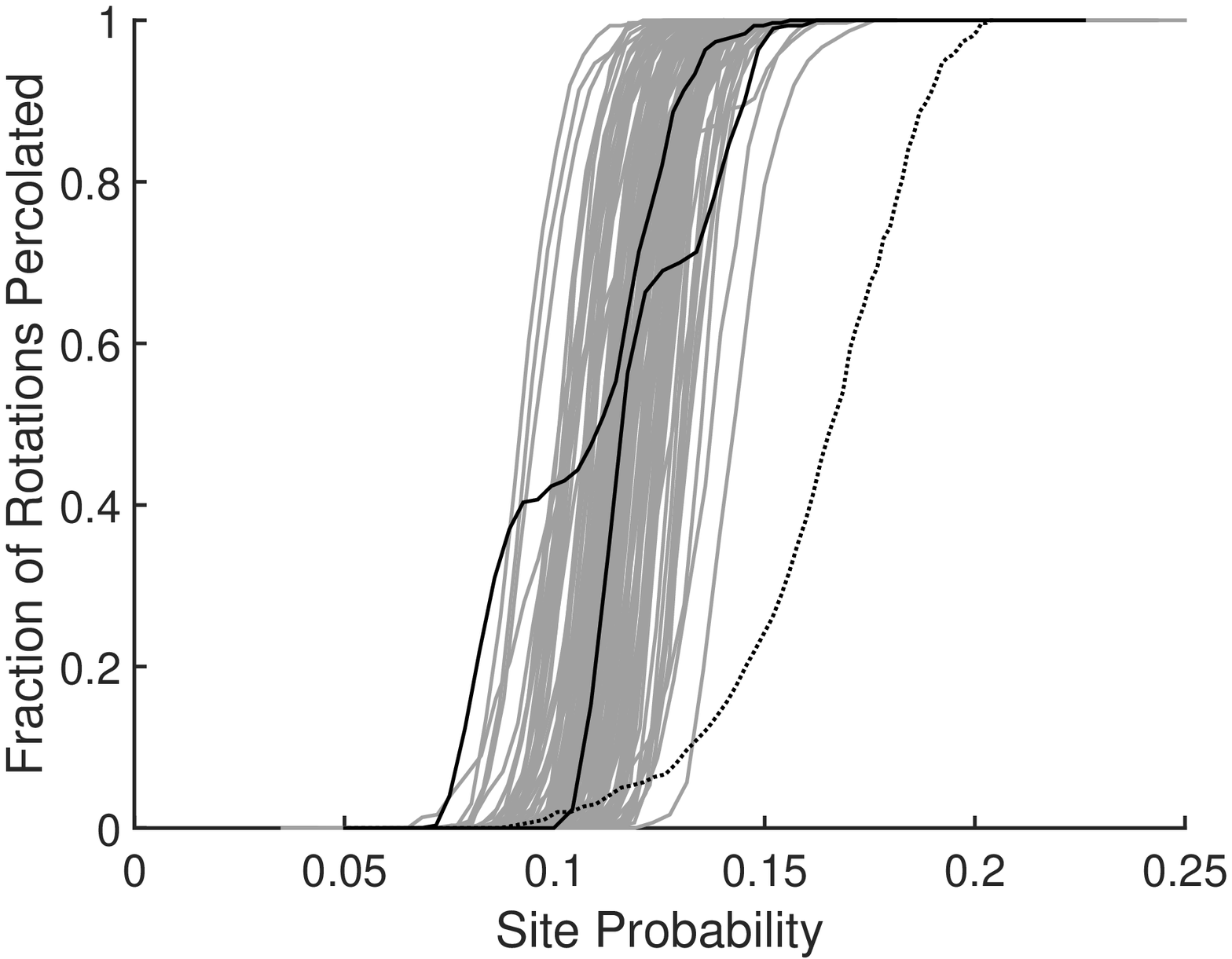}
        \caption{}\label{subfig:perc_frac_vs_prob}
    \end{subfigure}
    \begin{subfigure}{0.45\textwidth}
        \centering
    \includegraphics[width=\textwidth]{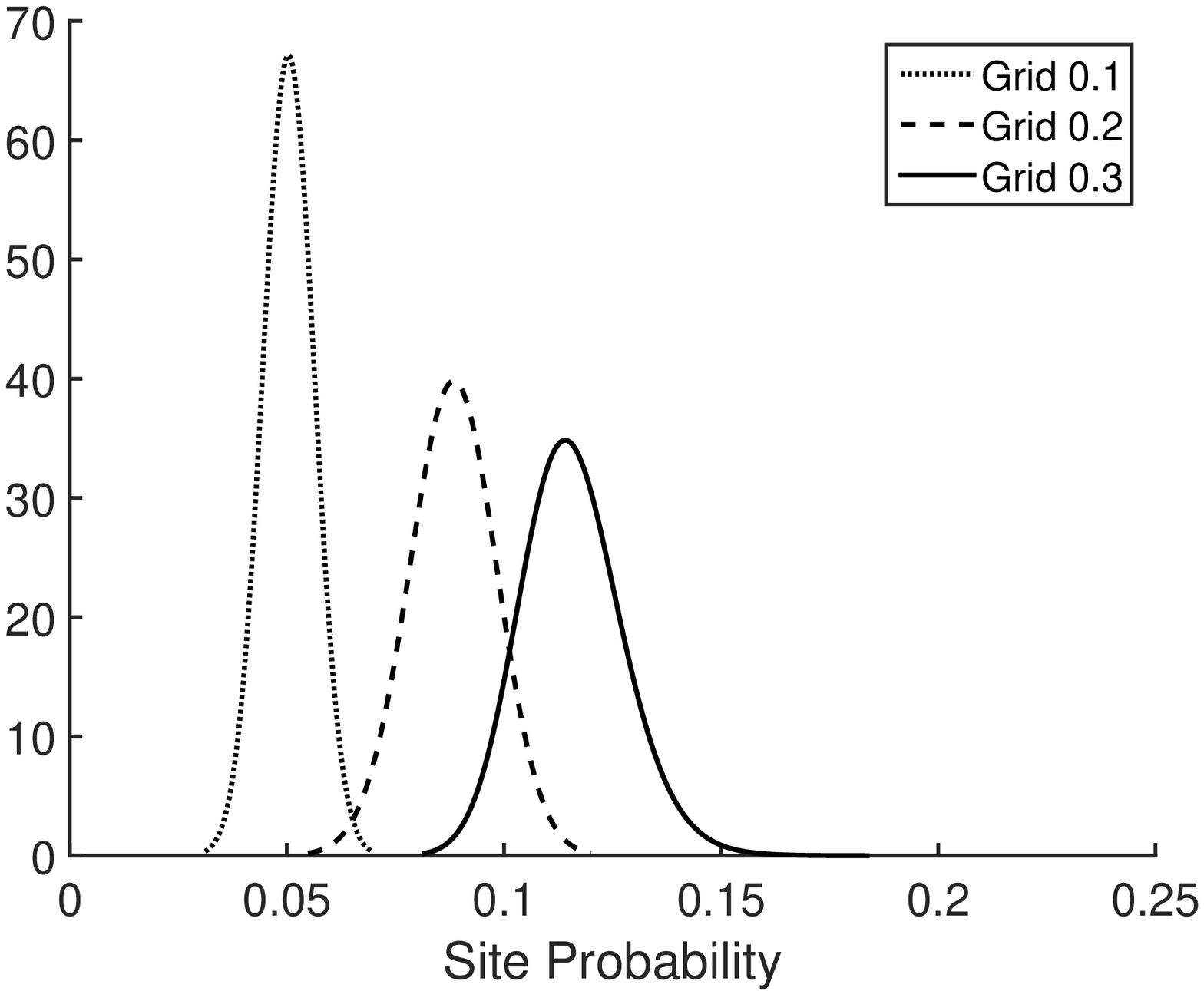}
        \caption{}\label{subfig:pdf_vs_prob}
    \end{subfigure}
\caption{Microvoid percolation characteristics as a function of site probability for 108 proteins (gray lines). (a) Mapping to probe radius, threshold probabilities indicated by black circles. (b) Reduced second moment. (c) Fraction of percolated realizations with exceptions to smooth transitions shown in black. (d) Lattice constant dependence.}
\label{fig:mv_perc}
\end{figure*}

To characterize microvoid volume in the context of percolation theory, we set $L_s$ to a fixed value to maintain a constant system volume as probe size is varied. At each probe size we run the calculation over 300 random orientations. We define $V_m$, $V_c$ and $V_b$ to be the microvoid, cavity, and boundary volumes, respectively. Probe size is then mapped to site probability using the relationship: 

\begin{equation}\label{eqn:site_prob}
    p=\frac{V_m}{V_m + V_c + V_{b}} .
\end{equation}
Note that vdW volume is omitted from consideration in Eq. \eqref{eqn:site_prob} because it can never be occupied by microvoid volume. The denominator in Eq. \eqref{eqn:site_prob} represents total void space, which is a fixed value as probe size changes when $L_s$ is held constant. The mapping of this site probability to probe size for all proteins is shown in Fig. \ref{subfig:probe_rad_vs_prob} for the case that $L_s=0.55$ \AA{}. We consider different values for $L_s$ ranging from 0 to 5.6 \AA{}, finding that the site percolation threshold monotonically decreases as a function of $L_s$. This is obvious because $V_b$ increases with $L_s$, which lowers the range of possible site probabilities. Furthermore, we consider a range of grid sizes from 0.1 \AA{} to 0.3 \AA{}. Although we arrive at the same conclusion regarding percolation properties for all choices of $L_s$ and grid size considered, the analysis becomes more difficult to perform using very small or large values of $L_s$. The choice of $L_s=0.55$ \AA{} provides a good working range of site probabilities for all proteins in our dataset.

For finite systems, an operational definition for the effective site percolation threshold $p_c$ is the site probability associated with the peak value of the reduced second moment (RSM) \citep{Hoshen1976,Dean1963}, which is defined as:

\begin{equation}
    \textrm{RSM} = \frac{\sum_s s^2 n_s}{\sum_s n_s} ,
\end{equation}
where $s$ is the size of a microvoid cluster taken as number of grid points, $n_s$ gives the number of clusters of size $s$, and the summation runs over non-percolating clusters only. The percolation threshold for each of the 108 proteins is calculated individually and is markedly similar for all proteins. To see this, the dark circles in Fig. \ref{subfig:probe_rad_vs_prob} mark the probe size and site probability at the percolation threshold for each individual protein. Fig. \ref{subfig:rsm_vs_prob} shows RSM as a function of site probability using the collective cluster size statistics across all 300 rotations. Similarly, Fig. \ref{subfig:perc_frac_vs_prob} plots the fraction of percolated realizations as a function of the site probability for each protein by using the peaks in RSMs that are calculated separately for each rotation. In almost all cases, the expected sigmoid-shaped curve describing a single sharp transition is present. 

Among the 108 proteins, only two exceptions were found that have multiple discrete transition jumps leading to plateaus in the percolation probability curves, shown as black solid lines in Fig. \ref{subfig:perc_frac_vs_prob}. Interestingly, the two exceptions are the largest protein in our dataset (839 residues) and one of the smallest proteins (61 residues).  A third exception, shown as the rightmost dashed black line in Fig. \ref{subfig:perc_frac_vs_prob}, is an example of a non-spherical protein with an unusually shallow percolation probability curve with a larger threshold value compared to typical cases. The large amount of variation in RSM characteristics, illustrated in Fig. \ref{subfig:pdf_vs_prob}, makes it challenging to apply percolation theory to microvoid volume because proteins of similar size are inherently inhomogeneous. Nonetheless, the site probabilities and percolation thresholds are remarkably similar across a diverse collection of globular proteins.

All of the results discussed thus far in Figs. \ref{subfig:probe_rad_vs_prob}, \ref{subfig:rsm_vs_prob}, and \ref{subfig:perc_frac_vs_prob} were produced using a lattice constant of 0.3 \AA{}.  Fig. \ref{subfig:pdf_vs_prob} shows the probability density for the percolation thresholds for three different lattice constants, which highlights a few interesting characteristics. A shift in $p_c$ towards smaller values occurs with decreasing grid size, while the spread in the distributions narrow. This qualitative trend is expected due to the greater resolution of spatial connectivity as the grid size decreases. Quantitatively, it is curious that the coefficient of variation for all three cases is approximately constant. Note that grid sizes greater than 0.4 \AA{} cannot be used to determine the microvoid percolation threshold because the grid spacing cannot be greater than the probe radius, otherwise the probe size would be smaller than the grid resolution. However, according to standard percolation theory, the values for critical thresholds do not have any significance regarding scaling laws. 

The remaining analysis regarding microvoid percolation uses a grid size of 0.3 \AA{}, which yields a mean percolation threshold of $p_c=0.114$ over the 108 proteins. Notice that this percolation threshold for site percolation is significantly smaller than that for the random cubic lattice, which is approximately 0.311 \citep{Stauffer1987,Wang2013}. Unlike a cubic lattice that has homogeneous site probability throughout the system, in proteins, the grid points representing vdW volume and the outer limits of the boundary volume are not accessible to microvoid. It is therefore clear that for large $L_s$, the microvoid volume will be a small percentage of total void volume, trivially decreasing the percolation threshold. More interestingly, the geometrical constraints that originate from the underlying molecular structure induce correlation between microvoid volume and protein conformation. In a general context, correlated site probabilities have been shown to be an important factor that can lower the percolation threshold compared to an uncorrelated random lattice \citep{Harter2005,Duckers1974,Duckers1978,Mueller-Krumbhaar1974}.

In a standard percolation problem, a system can be defined by a simulation box of linear dimension $L$, where it is possible to numerically estimate $p_c (L)$ with small statistical errors. The $p_c (L)$ estimates for different system sizes can be used subsequently in the context of finite-size scaling (FSS). In comparison, the application of percolation theory to microvoid volume within a protein is unconventional in three respects. First, the site percolation probability given in Eq. \eqref{eqn:site_prob} is correlated due to the covalently bonded molecular structure underlying the native conformation of the protein being analyzed. Second, protein structure is not homogeneous and it is therefore not possible to increase system size of a given protein by simply doubling its size, for example. Third, FSS involves combining different size proteins together, but the statistics per protein is very limited. In addition, our sample size of 108 proteins is not especially large in the context of percolation theory. Below, we present an analysis of microvoid percolation using FSS to extract critical exponents. When considering the quality and size of our dataset, the deviations that we find in the critical exponents relative to connectivity percolation in three dimensions are not sufficiently large to claim microvoid percolation lies within a different universality class. Said another way, the critical exponents that describe microvoid percolation in globular proteins are consistent with the standard connectivity percolation critical exponents in 3 dimensions.

We apply FSS to estimate a hypothetical percolation threshold corresponding to the limit of an infinite-sized globular protein (i.e. not repeat proteins). The threshold in this theoretical limit can be extracted from the effective probability $p_c (L)$, determined by the peak in the RSM, where $L$ gives the linear dimension of the system. We calculate the linear size $L$ of a protein as done previously in the literature \citep{Liang2001,Lorenz1993}, where $L$ is defined as the average of the protein's maximum extent along each coordinate axes:

\begin{equation}\label{eqn:linear_size}
    L=\frac{1}{6}\sum_{j=1}^{3}(x_{j,\textrm{max}}-x_{j,\textrm{min}}) .
\end{equation}
The most extreme coordinates of all the atoms in each protein are used in each of the $x$, $y$, and $z$ directions. We checked different variants of linear size with respect to rotation, such as by using the original orientation of the protein from the input PDB files, the average of this quantity over 300 random rotations, and the greatest value from the 300 random rotations. These variants have no effect on the subsequent analyses and conclusions. The results shown below are based on Eq. \eqref{eqn:linear_size} with the simplest implementation, namely with calculating $L$ from the initial orientation of the protein from the input PDB file. According to FSS, the percolation threshold for a finite-size system has the following form:

\begin{equation}\label{eqn:p_c_L}
    p_c (L)=aL^{-{\frac{1}{\nu}}}+p_c (\infty ),	
\end{equation}
where $p_c (L)$ is empirically determined from the RSM, $a$ is a fitting parameter and $p_c(\infty)$ is the percolation threshold for the hypothetical infinite-size globular protein. The finite-size scaling exponent $\nu$ is universal in the sense that it depends only on the dimension of the system.

The critical exponent $\beta$ can be determined by calculating the strength of the infinite cluster according to the equation

\begin{equation}\label{eqn:p_inf}
    P_\infty \sim |p-p_c|^\beta,	
\end{equation}
where $P_\infty$ is the probability that a random site is a member of the infinite cluster in a hypothetical infinitely large protein. Applying FSS to Eq. \eqref{eqn:p_inf} yields a scaling law \citep{Rintoul1997} of the form

\begin{equation}\label{eqn:p_inf_given_p_c}
    P_\infty \big(L|p=p_c(\infty)\big)\sim L^{-{\frac{\beta}{\nu}}},
\end{equation}
where $P_\infty$ is evaluated at the same site probability, equal to $p_c(\infty)$, for all systems regardless of their size. The exponent $\frac{\beta}{\nu}$ for this power law is calculated from the slope of the straight line that results when $\log\big(P_\infty (L)\big)$ versus $\log (L)$ for each protein at the estimated $p_c(\infty)$ (data not shown). This means the estimate for $\frac{\beta}{\nu}$ depends on the extrapolated value of $p_c(\infty)$. Unfortunately, it is difficult to accurately extract $p_c(\infty)$ from Eq. \eqref{eqn:p_c_L} due to the noise in the data. We therefore develop a systematic procedure to extract the critical exponents and $p_c(\infty)$. This procedure will work for any percolation problem and, to our knowledge, it is a novel approach.

\begin{figure*}
 \centering
    \begin{subfigure}{0.45\textwidth}
        \centering
    \includegraphics[width=\textwidth]{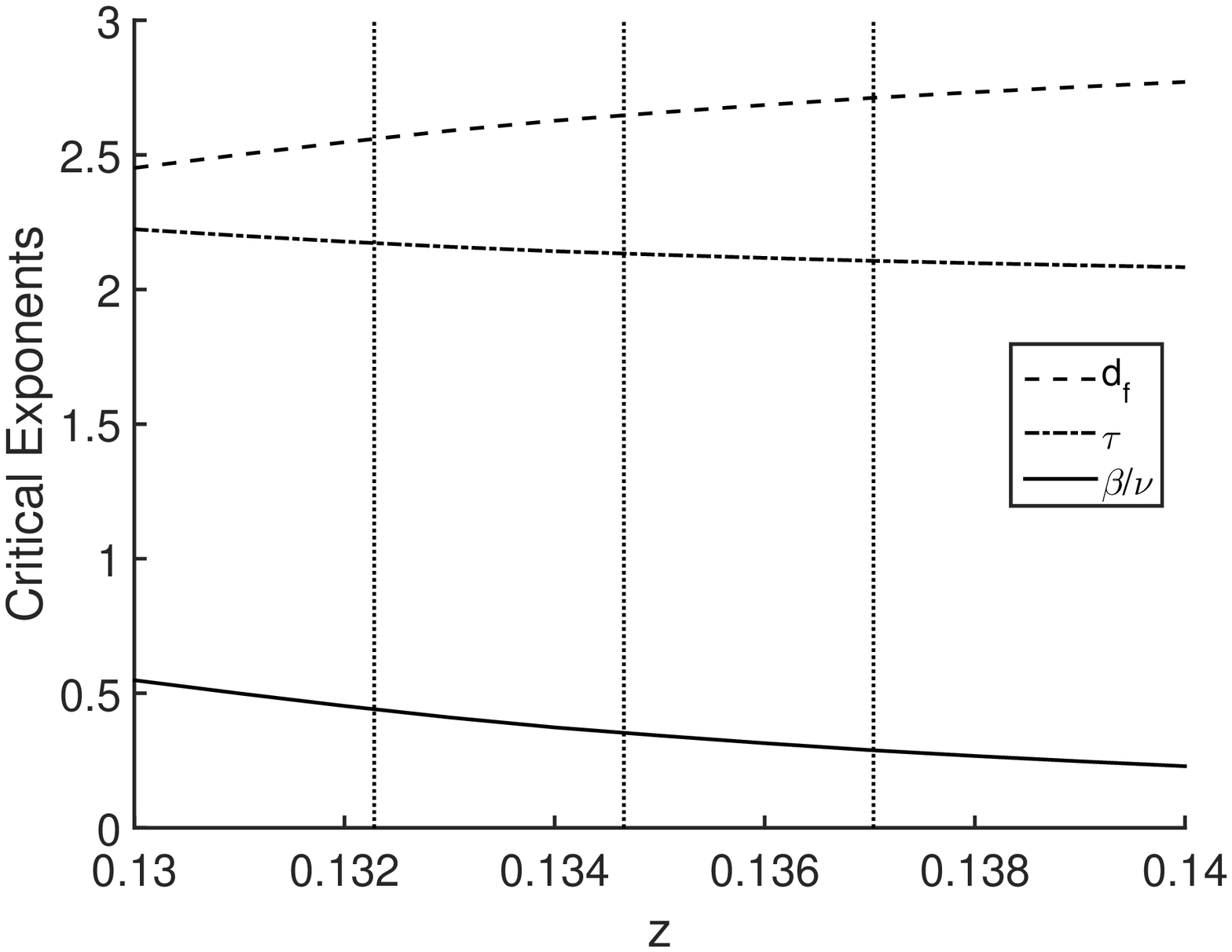}
        \caption{}\label{subfig:scaling_exps}
    \end{subfigure}
    \begin{subfigure}{0.45\textwidth}
        \centering
    \includegraphics[width=\textwidth]{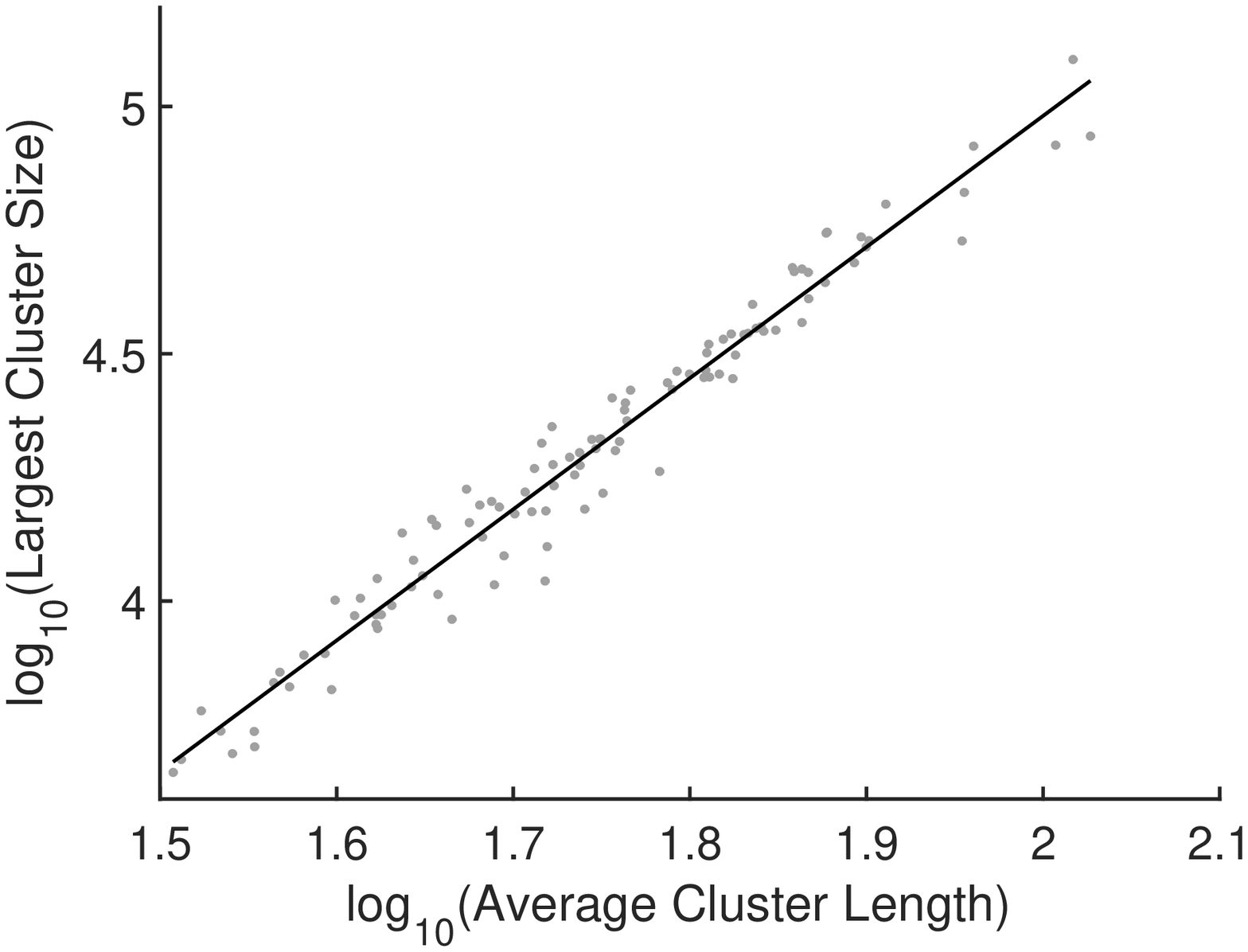}
        \caption{}\label{subfig:fractal}
    \end{subfigure}
    \begin{subfigure}{0.45\textwidth}
        \centering
    \includegraphics[width=\textwidth]{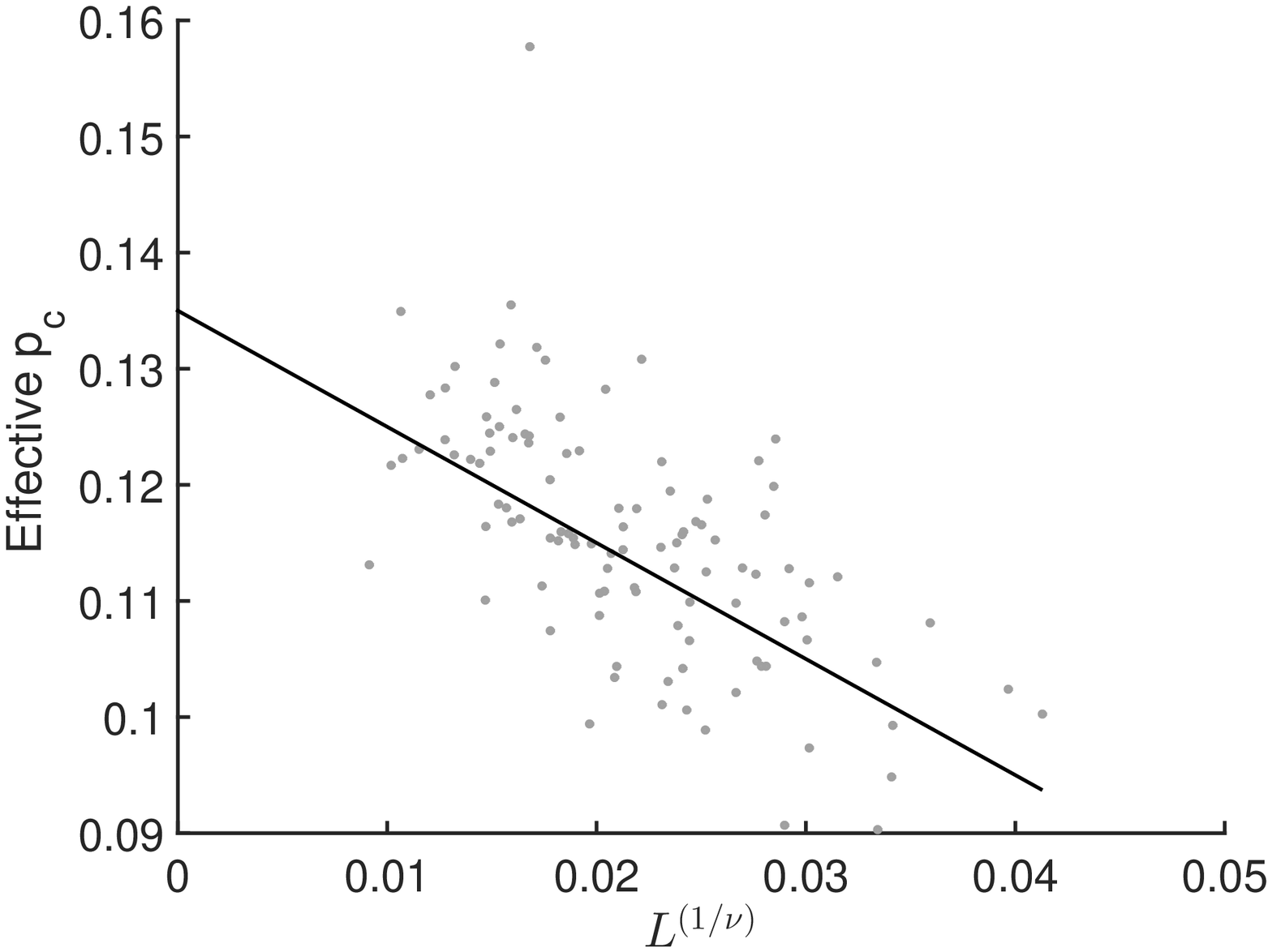}
        \caption{}\label{subfig:p_c}
    \end{subfigure}
    \begin{subfigure}{0.45\textwidth}
        \centering
    \includegraphics[width=\textwidth]{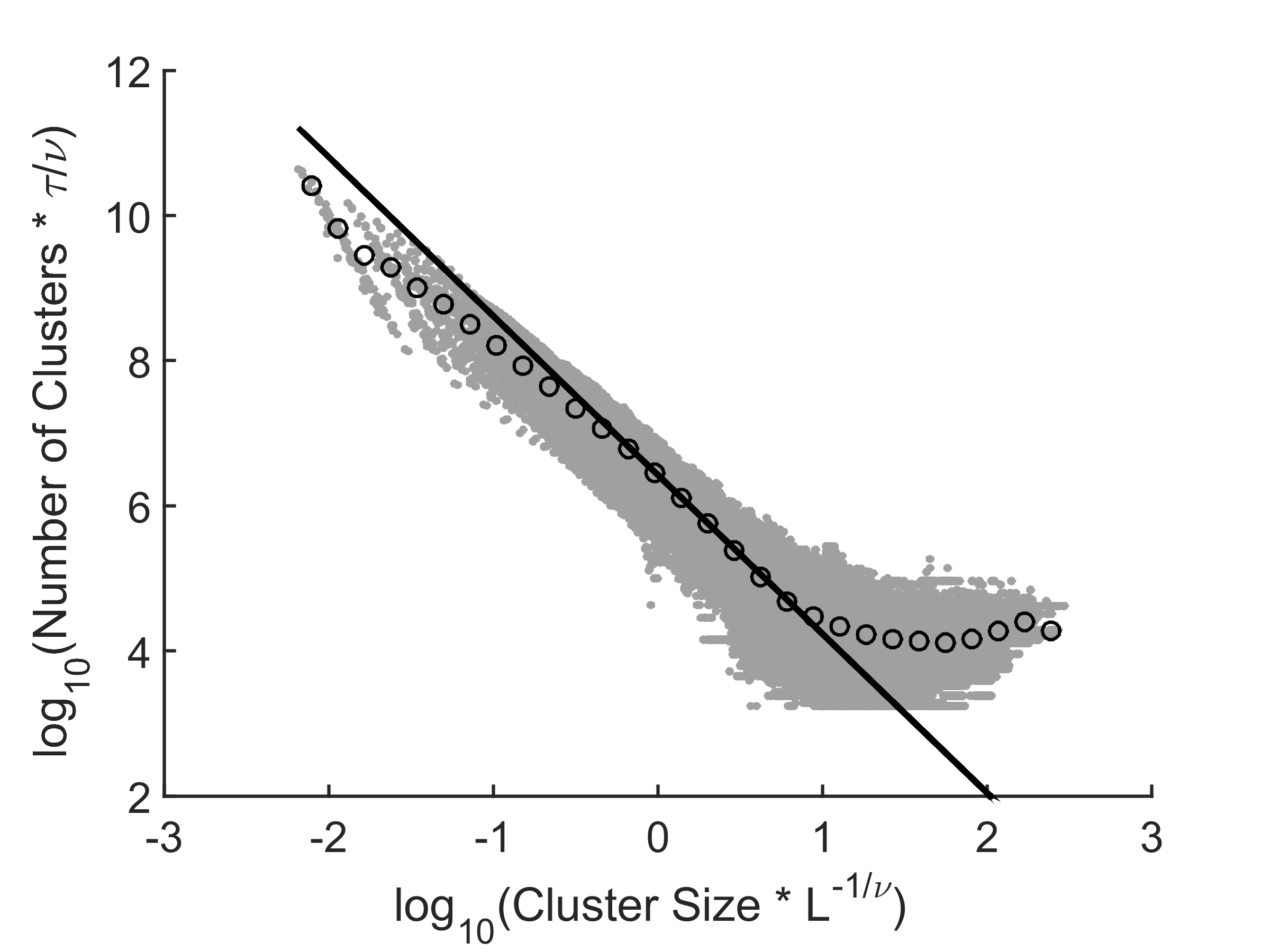}
        \caption{}\label{subfig:fss}
    \end{subfigure}
\caption{Percolation threshold properties. (a) Scaling exponents as a function of possible critical threshold. (b) Estimation of the fraction dimension. (c) Percolation threshold as a function of system size. (d) Finite-size scaling applied to the cluster number statistics.}
\label{fig:perc_thresh_props}
\end{figure*}

The first step is to treat $p_c(\infty)$ as an independent variable that we call $z$. Note that a conservative range of $z$ is considered such that $z_\textrm{min} \le p_c(\infty) \le z_\textrm{max}$ is true based on Eq. \eqref{eqn:p_c_L} and associated data, $p_c(L)$. We then determine the $\frac{\beta}{\nu}$ exponent from Eq. \eqref{eqn:p_inf_given_p_c} as explained above for different values of $z$ that represent the putative $p_c(\infty)$ within the range $[z_\textrm{min},z_\textrm{max}]$. After this scan is complete, we have $\frac{\beta}{\nu}$ as a function of $z$. Furthermore, other exponents can also be obtained based on universal scaling relationships. For example, the fractal dimension $d_f$ and critical exponent for cluster statistics $\tau$ can be calculated through the universal scaling relations

\begin{equation}\label{eqn:fractal_tau}
    d_f=d - \frac{\beta}{\nu} \quad \textrm{and} \quad \tau=1 + \frac{d}{d-{\frac{\beta}{\nu}}},
\end{equation}
where $d_f$ and $\tau$ are treated as dependent variables and the dimension of space $d=3$. Fig. \ref{subfig:scaling_exps} shows the fitted values of $\frac{\beta}{\nu}$ over a range of $z$ and from the relationships in Eq. \eqref{eqn:fractal_tau}, $d_f$ and $\tau$ are plotted on the same graph as a function of $z$.

As a second step, we estimate the fractal dimension. Fig. \ref{subfig:fractal} shows the volume of the largest microvoid cluster in a protein at the critical threshold (based on the peak in the RSM) as a function of the cluster's length, $\ell$. The volume of this largest cluster is expected to follow a power-law scaling that gives the fractal dimension $d_f$ according to

\begin{equation}
    V_\textrm{max}=b\ell^{d_f} .
\end{equation}
The length of a microvoid cluster is determined in the same way as done for the length of a protein based on Eq. \eqref{eqn:linear_size}. The linear regression on the log-log plot gives $d_f=2.651 \pm 0.061$, which is slightly above the accepted value of 2.52 in percolation theory. Using this empirical range of $d_f$ values, we add dotted vertical lines in Fig. \ref{subfig:scaling_exps}, where the middle line marks the $z$ that gives the best estimate for $p_c(\infty)$, while the flanking lines bracket the error in this estimate. Therefore, $p_c(\infty)=0.135\pm0.002$. These vertical lines also cross the curves for $\frac{\beta}{\nu}$ and $\tau$, which yield estimates for these exponents as $\frac{\beta}{\nu} =0.353 \pm 0.061$ and $\tau=2.134 \pm 0.033$. 

Using this estimate for $p_c(\infty)$, the third step is to fit Eq. \eqref{eqn:p_c_L} to the data shown in Fig. \ref{subfig:p_c}, where the slope of the line is fit over a wide range of protein sizes while the $y$-intercept $p_c(\infty)$ is fixed. This is done for the mean and extreme values of $p_c(\infty)$ to obtain an estimate of $\nu$ and bracket the error. From these best fits, we find that $\nu = 0.722 \pm 0.061$. Note that in the standard percolation problem, this step could include the averaging over thousands of random realizations per system size to reduce noise and the FSS could be applied to a very large range of system sizes to ultimately determine $\nu$ more accurately \citep{Roy2018,Wang2013,Harter2005,Rintoul1997,Ding2014}. In our case, the large scatter in Fig. \ref{subfig:p_c} reflects inhomogeneity within proteins and the fact that only one realization is available per protein. 

The fourth step is to apply FSS to cluster statistics by starting with the usual scaling form that $n(s,p)\rightarrow s^{-\tau} g[\frac{s}{\xi(p-p_c)}]$ where $n(s,p)$ is the number of clusters of size $s$ at a fixed site probability $p$. Assuming there is a single length scale in the problem, the correlation length is given by $\xi(p-p_c)= |p-p_c(L)|^{\frac{1}{\sigma}}$, where $\sigma$ is a scaling exponent that characterizes the rate of divergence in the correlation function near criticality. Moreover,  $g(x)$ is a scaling function that must appear if there is data collapse. Taken together, $n(s,p)\rightarrow s^{-\tau} g \big[s |p-p_c(L)|^{\frac{-1}{\sigma}} \big]$. 

The next step is to apply Eq. \eqref{eqn:p_c_L} to this last equation in similar fashion as was done for $P_\infty$. In addition, cluster statistics are collected at the same $p_c(\infty)$ for all system sizes as the data from 108 proteins is combined. This combined data is described by 
$$n_s\big(L|p=p_c(\infty)\big) \sim L^{-{\frac{\tau}{\sigma\nu}}}f\big(sL^{\frac{-1}{\sigma\nu}}\big),$$ where it must be that $f (x)\rightarrow x^{-\tau}$ for $x\gg1$. This form recovers the usual cluster size scaling when $L\rightarrow\infty$. It appears this scaling form will allow the cluster statistics to collapse onto a single curve for more than one set of exponents. This freedom allows us to ignore the $\sigma$ exponent dependence. 

As shown in Fig. \ref{subfig:fss}, we scale the cluster statistics data by introducing the scale factor in the $x$-direction as $L^{\frac{-1}{\nu}}$ and in the $y$-direction as $L^{\frac{\tau}{\nu}}$ to create a data collapse. The values of $\tau$ and $\nu$ used in these scaling factors come directly from the procedure above and are held fixed. Note that the quality of the data collapse shown in Fig. \ref{subfig:fss} is similar to any other data collapse we obtained by allowing the scaling factors to be optimized by treating $\tau$ and $\nu$ as free parameters. The characteristic leveling off at the large cluster size tail reflects poor sampling in the largest cluster sizes, with only one or two per bin \citep{Iglauer2016,Ding2014,Liu2011}. 

Within the scaling region, the slope of the resulting data-collapsed curve in a log-log plot should be $-\tau$, which we use as an independent check on the overall consistency. To get this slope, the cluster statistics are scaled and then binned logarithmically. The circles in Fig. \ref{subfig:fss} represent the binned averages and the gray dots include all of the raw data. The straight line shown has a slope of $-2.040 \pm 0.380$, which encloses the accepted value of 2.19 for 3D connectivity percolation as well as the value estimated above in the third step ($\tau=2.134 \pm 0.033$). Taken together, the exponents reported here are derived self-consistently and agree fairly well with the connectivity percolation exponents in 3D considering the inhomogeneous nature of the proteins being compared.

\section{Packing Density Characteristics}\label{sec:pd}

The long-standing accepted value of the mean packing fraction in the core region of globular proteins is approximately 0.75 \citep{Liang2001,Richards1977,Cuff2004}. In a recent study focused on a re-evaluation of vdW radii, a mean packing density of 0.56 within core regions of globular proteins was reported \citep{Gaines2017,Gaines2016}. In our work, the mean packing density in core regions agrees with the accepted value of 0.75 when Bondi vdW radii \citep{Bondi1964} are employed and 0.56 is reproduced when the vdW radii proposed by \citet{Gaines2016} is used. Clearly, this discrepancy is related to the modeling of the vdW interactions, which reinforces our view that the results presented above are more physically meaningful when vdW interactions are modeled as soft interactions in the presence of dynamical fluctuations. We use Bondi radii for all subsequent discussion.

For Voronoi methods, the local packing density $\phi$ is generally defined as

\begin{equation}\label{eqn:pd}
    \phi = \frac{V_\textrm{vdW}}{V_\textrm{cell}},
\end{equation}
where $V_\textrm{cell}$ is the Voronoi volume of a group of atoms of interest and $V_\textrm{vdW}$ is the total vdW volume for all atoms contained within $V_\textrm{cell}$. The packing density given by Eq. \eqref{eqn:pd} can be applied at different spatial resolutions that range from a single atom, to a residue, as well as up the level of all protein atoms to determine a mean packing density for the entire protein. Voronoi methods make calculating the packing density simple in principle, but as discussed in Section \ref{sec:intro}, the details of how to obtain the Voronoi volume of atoms on the surface of a protein within an implicit solvent model critically affects the accuracy of the density calculations. In a recent study focused on obtaining accurate packing densities in proteins \citep{Esque2010}, water molecules were treated explicitly to obtain a realistic solvent boundary at the protein surface, MD simulations were used to account for atomic fluctuations, and the Voronoi method was improved through empirically derived weighting parameters to accurately partition space among atoms. It is worth mentioning that using weight factors in the Voronoi method is critical for a realistic tessellation when dealing with differing atom sizes (i.e. vdW radii) found in proteins \citep{Richards1974,Esque2010}. Within this explicit-solvent framework, the volumes associated with residues buried in the core of a globular protein are approximately the same as those on the surface \citep{Esque2010}. Moreover, this result was shown to be consistent with experimental data. Thus, the implication is that packing density is indeed approximately a constant throughout a globular protein in its native state. Previous results suggested proteins are packed more densely in the core compared to the surface \citep{Liang2001,Fleming2000,Gerstein1995}. 

Considering the work of \citet{Esque2010}, we elucidate why differences in predictions for local packing density occur across various models.  Our partial volume calculations provide a means for studying intrinsic packing densities of core and surface residues separately.  We propose an initial definition for packing density in terms of partial volumes as

\begin{equation}\label{eqn:phi_0}
    \phi_0 = \frac{V_\textrm{vdW}}{V_\textrm{vdW}+V_m + V_c}.
\end{equation}
This frequently invoked definition calculates packing density as the ratio of vdW volume over molecular volume. Notice that partial boundary volume is not included in $\phi_0$. However, the partial boundary volume, as defined when $L_s=5.6$ \AA{}, is used to rank order how deeply buried a residue is within a protein according to the buried fraction $f_b$, given as

\begin{equation}\label{eqn:buried_frac}
    f_b = \frac{V_m}{V_m + V_b} .
\end{equation}
Based on Eq. \eqref{eqn:buried_frac}, it can be seen that completely buried residues have no boundary volume and hence $f_b=1$, whereas $f_b\rightarrow0$ for residues at the surface when $V_b\gg V_m$. All residues are ranked by $f_b$ for each protein and split into two equal groups based on the median $f_b$ of that protein. Reference to core or surface residues are those that are respectively above or below the median $f_b$ for a given protein. In this way, each protein contributes the same number of samples to both groups. We also classified residues into buried and exposed groups such that buried residues have no boundary volume, whereas the exposed residues have a nonzero amount of boundary volume (no matter how small). 

\begin{figure*}
 \centering
    \begin{subfigure}{0.45\textwidth}
        \centering
    \includegraphics[width=\textwidth]{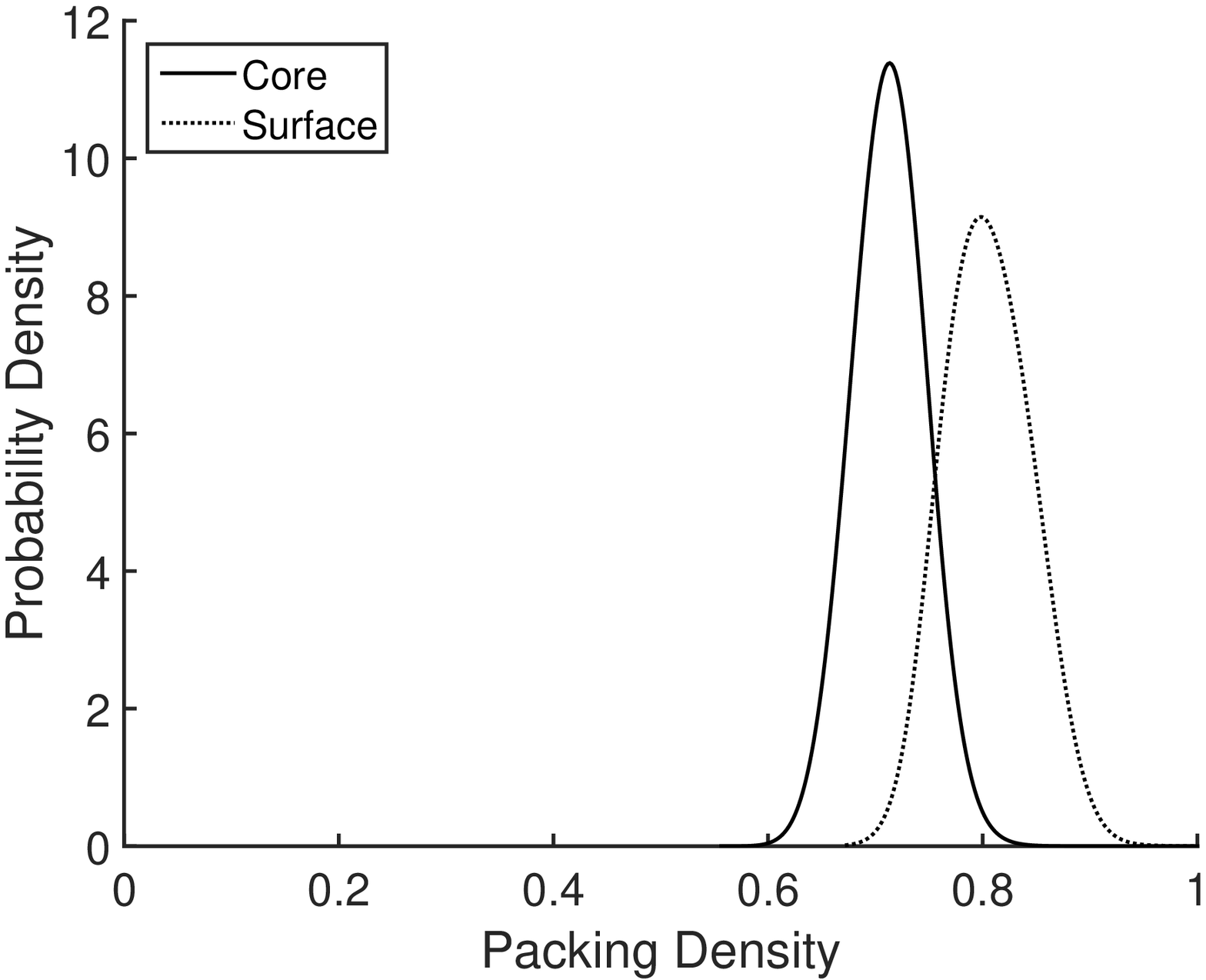}
        \caption{}\label{subfig:packing_no_bv}
    \end{subfigure}
    \begin{subfigure}{0.45\textwidth}
        \centering
    \includegraphics[width=\textwidth]{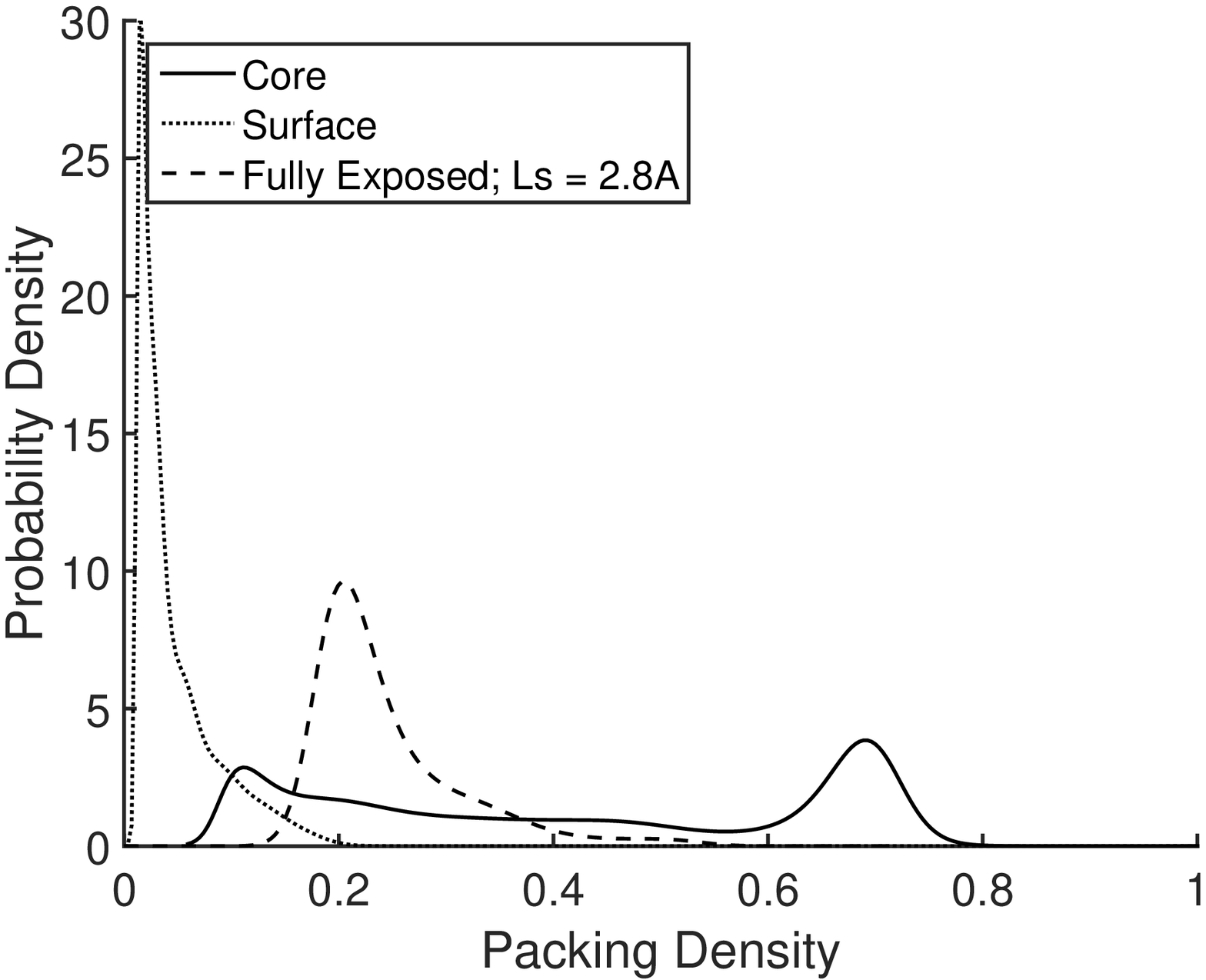}
        \caption{}\label{subfig:packing_large_bv}
    \end{subfigure}
    \begin{subfigure}{0.45\textwidth}
        \centering
    \includegraphics[width=\textwidth]{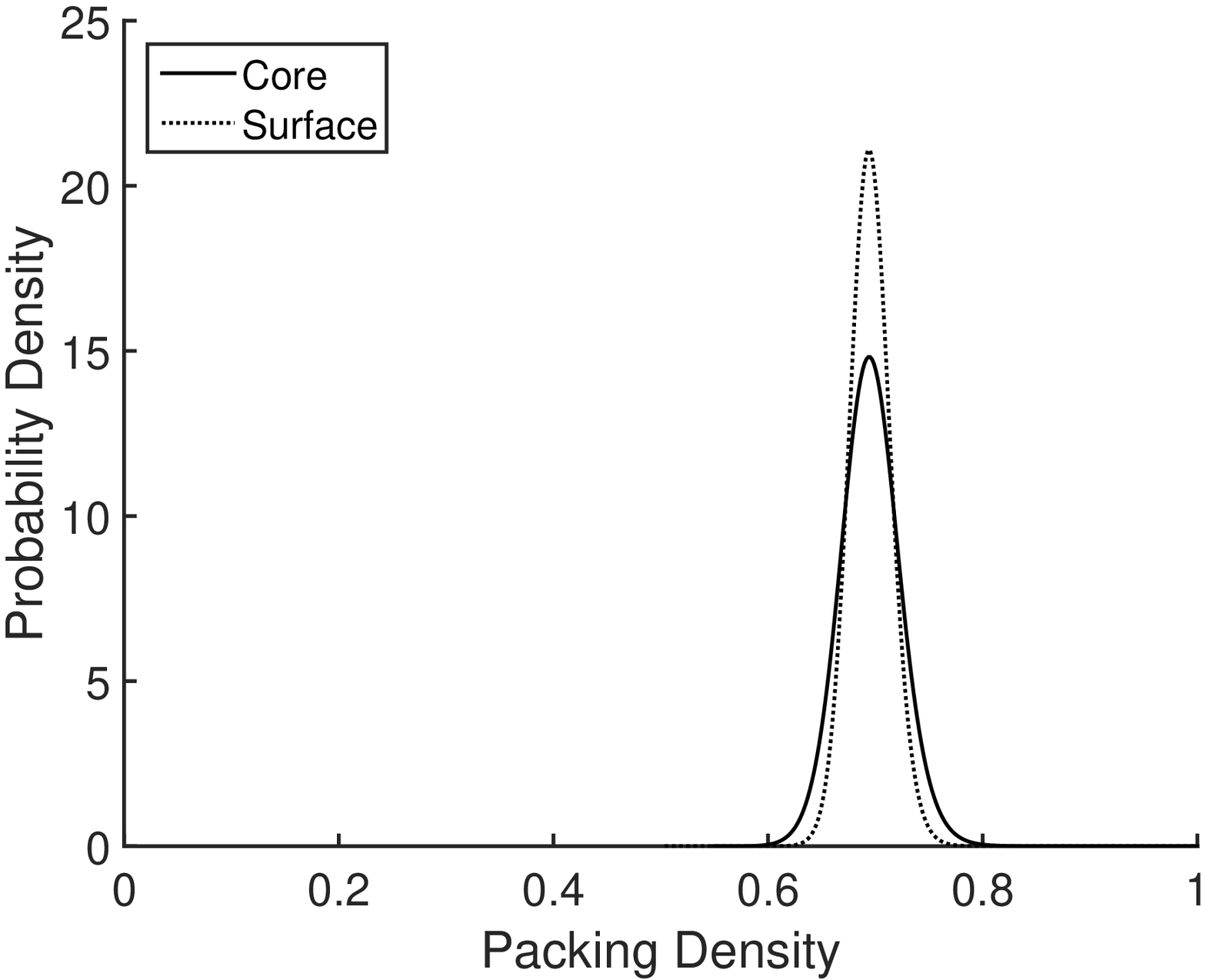}
        \caption{}\label{subfig:packing_uniform}
    \end{subfigure}
    \begin{subfigure}{0.45\textwidth}
        \centering
    \includegraphics[width=\textwidth]{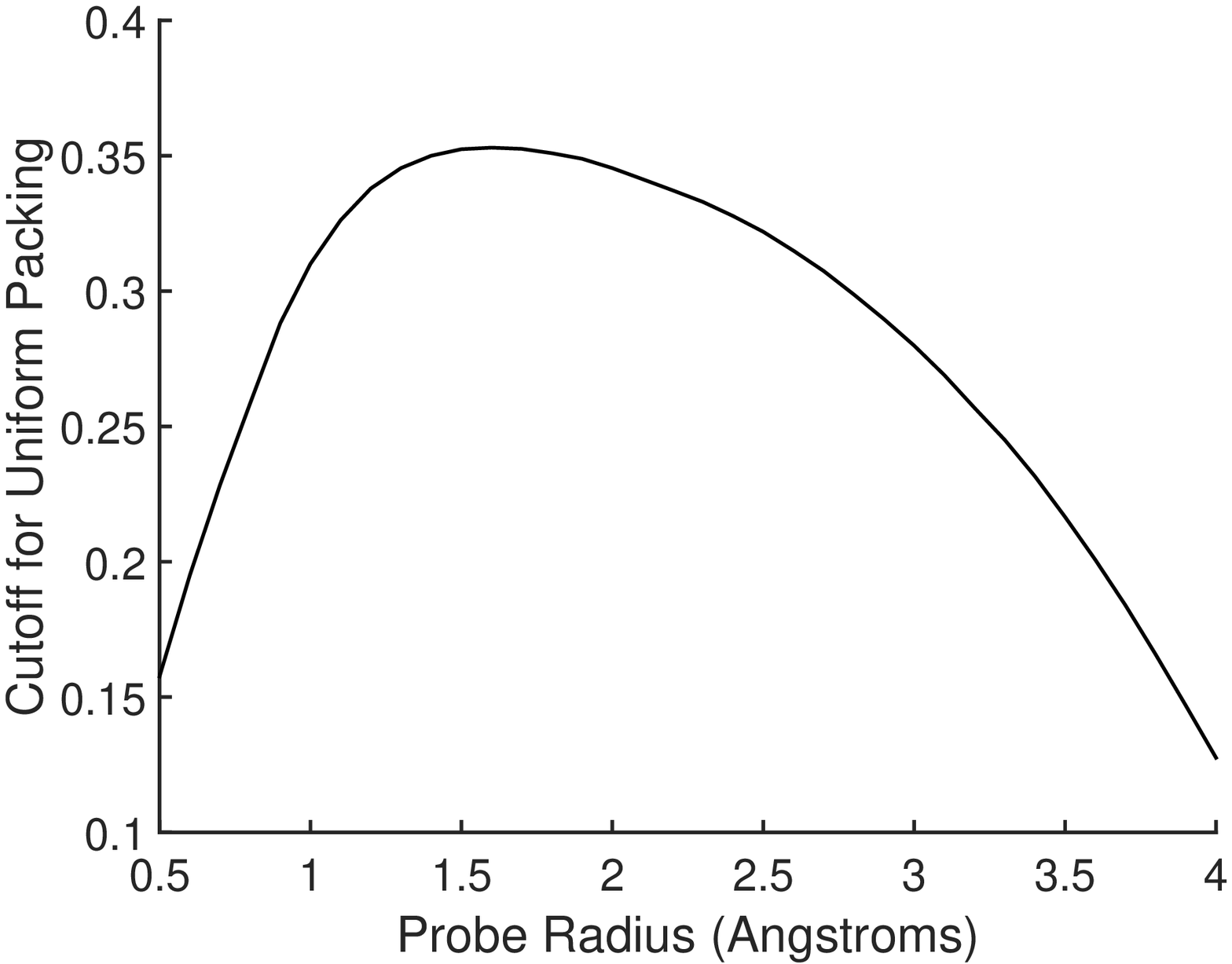}
        \caption{}\label{subfig:cutoff_vs_probe}
    \end{subfigure}
\caption{Distributions of packing density for residues classified either as core or surface depending on boundary volume influence. (a) No boundary volume component. (b) Large boundary volume component. (c) Boundary volume tuned to give uniform results for a probe radius of 1.4 \AA{}. (d) Boundary volume cutoff distance as a function of probe radius.}
\label{fig:packing_densities}
\end{figure*}

Distributions for packing density among the two groups of residues that are classified as buried or on the surface using a probe radius of 1.4 \AA{} and grid size of 0.5 \AA{} are shown in Figs. \ref{subfig:packing_no_bv}, \ref{subfig:packing_large_bv}, and \ref{subfig:packing_uniform}. In Fig. \ref{subfig:packing_no_bv}, the packing density distribution among core residues agrees well with Voronoi and OSP estimates. The packing density of residues near the surface, however, is slightly higher (c.f. Fig. \ref{subfig:packing_no_bv}), contrary to the predictions of many other methods \citep{Fleming2000,Gerstein1996}. In particular, surface packing densities from OSP are predicted to be lower than core densities, dipping as low as 0.2 to 0.4. Voronoi methods estimate different surface packing densities depending on how the solvent is modeled at the protein surface, but typically the packing density of surface residues is found to be slightly lower than that in the core.  To understand the reason for differences in packing density between the core and the surface of a protein, the definition for packing density must be generalized to included boundary volume.

An alternative definition for packing density is given as

\begin{equation}\label{eqn:phi_1}
    \phi_1 (R,L_s)=\frac{V_\textrm{vdW}}{V_\textrm{vdW}+V_m+V_c+V_b},
\end{equation}
where, for the moment, probe radius is $R=1.4$ \AA{} and $L_s=5.6$ \AA{}, as above. The distributions for the $\phi_1$ packing density for buried and surface residues is shown in Fig. \ref{subfig:packing_large_bv}. Not surprisingly, the most buried residues have the same average packing density compared with $\phi_0$ because these residues do not carry partial boundary volume at all. However, there is a decrease in packing density for partially buried residues and a dramatic decrease for surface residues. This large effect is due to the generous 5.6 \AA{} layer of boundary surrounding the protein. The dashed line in Fig. \ref{subfig:packing_large_bv} shows the distribution of packing density for exposed residues when $L_s$ is set to 2.8 \AA{}, which gives a 1-water molecule boundary layer that is also employed in the OSP method. The criteria produce surface packing density ranges in good agreement with OSP calculations.

When modeling water molecules explicitly within a simulation box, no water molecules count as void space, although there is microvoid volume between water molecules. In an implicit solvent model, void space in the extended volume layer represents vdW, cavity and microvoid volumes lumped together. The question then becomes how much of this extended volume should be attributed to microvoid volume and subsequently assigned as partial volume to surface atoms. The answer can be obtained by generalizing the definition of packing density to include a free parameter $r_c$ to control the amount of boundary volume to retain:

\begin{equation}\label{eqn:phi_gen}
    \phi (R,r_c)=\frac{V_\textrm{vdW}}{V_\textrm{vdW}+V_m+V_c+V_b(r_c)}.
\end{equation}
Specifically, we count only those boundary grid points that extend beyond the molecular surface up to a cutoff distance of $r_c$. Note that Eq. \eqref{eqn:phi_0} and Eq. \eqref{eqn:phi_1} are special cases of Eq. \eqref{eqn:phi_gen}. We then find the optimal $r_c$ to satisfy the condition that the average packing density for core residues is equal to that for surface residues. As shown in Fig. \ref{subfig:packing_uniform}, for a probe radius of 1.4 \AA{}, a cutoff distance of 0.35 \AA{} results in uniform packing density throughout globular proteins, matching quantitatively to the finely-tuned Voronoi tessellation method \citep{Esque2010}.

It is clear from Eq. \eqref{eqn:pd} that if an explicit water model is employed, packing density will be independent of probe radius. To achieve uniform packing density that is independent of probe size, it is of interest to determine the $r_c$ that keeps packing density uniform as a function of probe radius $R$. The dependence of $r_c$ on $R$ is shown in Fig. \ref{subfig:cutoff_vs_probe}. This curve exhibits a maximum. Interestingly, the maximum cutoff distance occurs near the probe size consistent with a water molecule. Additionally, this maximum is relatively flat, meaning a cutoff distance of 0.35 \AA{} is sufficient to yield roughly uniform packing density over a range of probe sizes between 1.3 \AA{} to 2.1 \AA{}. The lower cutoff values for smaller probe sizes is somewhat intuitive as we expect less microvoid to be present when a smaller probe is used. The reason for a decreasing cutoff for larger probes is less obvious but suggests a geometric relationship between the relative sizes of the protein residues and the solvent molecules at the surface. It is worth noting that similar results were obtained using the vdW radii of \citet{Gaines2016}. We also replaced all the vdW radii of different atom types with a single value and varied this single value to remove possible frustration effects caused by different atomic sizes not fitting regularly.  Similarly-shaped cutoff curves were found in all cases, albeit with different $r_c$ values and peak radii. This extended investigation of packing fractions finds no particular fine-tuned dependence on the size of water. 

\begin{figure}
  \centering
    \includegraphics[width=0.48\textwidth]{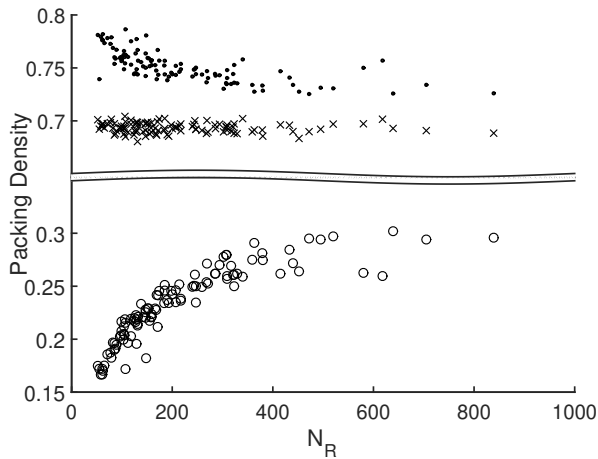}
\caption{Packing densities as a function of number of protein residues for 108 proteins, for three different definitions of packing density.  Points, crosses, and circles represent Eqs. \eqref{eqn:phi_0}, \eqref{eqn:phi_gen}, and \eqref{eqn:phi_1}, respectively}\label{fig:packing_vs_length}
\end{figure}

Protein length is another factor that affects packing density, but the correct dependence has not yet been determined in the literature \citep{Liang2001,Fleming2000,Gaines2017,Zhang2003}. Applying Eq. \eqref{eqn:phi_0} to the total protein volumes for all structures in our dataset, we find that packing density is smaller for larger proteins as shown in Fig. \ref{fig:packing_vs_length}. If we include a generous portion of boundary volume as in Eq. \eqref{eqn:phi_1}, the length dependence reverses, in agreement with OSP.  Finally, by applying Eq. \eqref{eqn:phi_gen} with a cutoff of 0.35 \AA{}, we find spatially-uniform packing densities that are independent of protein length. Thus, our unified model with a single adjustable parameter can be tuned to explain the differences in packing density between various models that have been reported in the literature.

\section{Discussion}\label{sec:discussion}

We implement a fast and accurate grid-based method with a low memory requirement for calculating the volume properties of molecular systems. The method is based on a generalization of the Hoshen-Kopelman cluster-labeling technique for identifying and uniquely labeling all contiguous clusters of the same volume type. This approach classifies microvoid, cavity, and boundary volume as separate quantities. Microvoid represents void space that is too small to accommodate a spherical probe of a specified radius. Grid points are identified on the fly as either cavity or microvoid by employing a compressive energy model that mitigates discrete lattice effects by allowing a tolerance for atomic clashing. Furthermore, a narrow distribution of different probe sizes is sampled in order to account for dynamical atomic motion when static structures are analyzed.  

A representative sample of globular proteins has been selected to test and benchmark this method as well as to collect statistics on typical properties.  The clustering of microvoid volume is analyzed using percolation theory. We show that microvoid clustering is well described by standard connectivity percolation. Furthermore, partial volumes for each volume type are assigned to each atom to give detailed maps of how volume and packing density are distributed throughout a protein. Our implicit-solvent model for packing density provides a unified framework with a single adjustable parameter to control the amount of microvoid volume outside the molecular volume of a protein that is attributed to the local volume of a protein atom. After parameterizing the model by matching to the results from explicit-solvent models that are consistent with experimental data, the model predicts packing density to be uniform within globular proteins, independent of the protein size. By leaving the adjustable parameter for packing density free, we recover the predictions of previous implicit-solvent models that lead to disparate results, thus explaining from a unified perspective the discrepancies in predictions for a variety of packing density properties found in the literature. 

As an implicit-solvent method, our approach has many applications to high-throughput analysis of protein structures. The solvation layer length and probe radius are key parameters that can be adjusted when investigating the distribution of volume in proteins. For example, microvoid volume characteristics have a universal character in the native state across diverse globular proteins. In ongoing work, the universal cluster statistics is being employed to place confidence levels on protein structure predictions to better identify decoys. Identifying cavities at different length scales as the test probe radius is varied also helps glean insight into protein function. In future work, the method to obtain partial volumes will be used to quantify differences in van der Waal interaction parameterization. The method can be applied to molecular dynamics trajectories where individual protein conformations provide statistical sampling, removing the need to sample over different probe sizes (along with random orientations). The improved statistics per protein should greatly improve the finite size scaling analysis given here. Simply by considering period boundary conditions, the method can analyze proteins in a simulation box with explicit solvent. Moreover, this method will generalize to any molecular system, and thus can support many applications in condensed matter systems with explicit modeling of all atoms.  

\begin{acknowledgments}
This work was partly supported by NIH grant GM101570. SBG was partly supported by the NOAA Ernest F. Hollings Scholarship during this work. The authors wish to thank Dennis Livesay for stimulating discussions in the early stages of this project.
\end{acknowledgments}

\bibliography{bibliography}

\begin{thebibliography}{94}%
\makeatletter
\providecommand \@ifxundefined [1]{%
 \@ifx{#1\undefined}
}%
\providecommand \@ifnum [1]{%
 \ifnum #1\expandafter \@firstoftwo
 \else \expandafter \@secondoftwo
 \fi
}%
\providecommand \@ifx [1]{%
 \ifx #1\expandafter \@firstoftwo
 \else \expandafter \@secondoftwo
 \fi
}%
\providecommand \natexlab [1]{#1}%
\providecommand \enquote  [1]{``#1''}%
\providecommand \bibnamefont  [1]{#1}%
\providecommand \bibfnamefont [1]{#1}%
\providecommand \citenamefont [1]{#1}%
\providecommand \href@noop [0]{\@secondoftwo}%
\providecommand \href [0]{\begingroup \@sanitize@url \@href}%
\providecommand \@href[1]{\@@startlink{#1}\@@href}%
\providecommand \@@href[1]{\endgroup#1\@@endlink}%
\providecommand \@sanitize@url [0]{\catcode `\\12\catcode `\$12\catcode
  `\&12\catcode `\#12\catcode `\^12\catcode `\_12\catcode `\%12\relax}%
\providecommand \@@startlink[1]{}%
\providecommand \@@endlink[0]{}%
\providecommand \url  [0]{\begingroup\@sanitize@url \@url }%
\providecommand \@url [1]{\endgroup\@href {#1}{\urlprefix }}%
\providecommand \urlprefix  [0]{URL }%
\providecommand \Eprint [0]{\href }%
\providecommand \doibase [0]{http://dx.doi.org/}%
\providecommand \selectlanguage [0]{\@gobble}%
\providecommand \bibinfo  [0]{\@secondoftwo}%
\providecommand \bibfield  [0]{\@secondoftwo}%
\providecommand \translation [1]{[#1]}%
\providecommand \BibitemOpen [0]{}%
\providecommand \bibitemStop [0]{}%
\providecommand \bibitemNoStop [0]{.\EOS\space}%
\providecommand \EOS [0]{\spacefactor3000\relax}%
\providecommand \BibitemShut  [1]{\csname bibitem#1\endcsname}%
\let\auto@bib@innerbib\@empty
\bibitem [{\citenamefont {Liang}\ and\ \citenamefont {Dill}(2001)}]{Liang2001}%
  \BibitemOpen
  \bibfield  {author} {\bibinfo {author} {\bibfnamefont {J.}~\bibnamefont
  {Liang}}\ and\ \bibinfo {author} {\bibfnamefont {K.~A.}\ \bibnamefont
  {Dill}},\ }\href {\doibase 10.1016/S0006-3495(01)75739-6} {\bibfield
  {journal} {\bibinfo  {journal} {Biophys. J.}\ }\textbf {\bibinfo {volume}
  {81}},\ \bibinfo {pages} {751} (\bibinfo {year} {2001})}\BibitemShut
  {NoStop}%
\bibitem [{\citenamefont {Richards}(1974)}]{Richards1974}%
  \BibitemOpen
  \bibfield  {author} {\bibinfo {author} {\bibfnamefont {F.~M.}\ \bibnamefont
  {Richards}},\ }\href {\doibase 10.1016/0022-2836(74)90570-1} {\bibfield
  {journal} {\bibinfo  {journal} {J. Mol. Biol.}\ }\textbf {\bibinfo {volume}
  {82}},\ \bibinfo {pages} {1} (\bibinfo {year} {1974})}\BibitemShut {NoStop}%
\bibitem [{\citenamefont {Tsai}\ \emph {et~al.}(1999)\citenamefont {Tsai},
  \citenamefont {Taylor}, \citenamefont {Chothia},\ and\ \citenamefont
  {Gerstein}}]{Tsai1999}%
  \BibitemOpen
  \bibfield  {author} {\bibinfo {author} {\bibfnamefont {J.}~\bibnamefont
  {Tsai}}, \bibinfo {author} {\bibfnamefont {R.}~\bibnamefont {Taylor}},
  \bibinfo {author} {\bibfnamefont {C.}~\bibnamefont {Chothia}}, \ and\
  \bibinfo {author} {\bibfnamefont {M.}~\bibnamefont {Gerstein}},\ }\href
  {\doibase 10.1006/jmbi.1999.2829} {\bibfield  {journal} {\bibinfo  {journal}
  {J. Mol. Biol.}\ }\textbf {\bibinfo {volume} {290}},\ \bibinfo {pages} {253}
  (\bibinfo {year} {1999})}\BibitemShut {NoStop}%
\bibitem [{\citenamefont {Cioni}(2006)}]{Cioni2006}%
  \BibitemOpen
  \bibfield  {author} {\bibinfo {author} {\bibfnamefont {P.}~\bibnamefont
  {Cioni}},\ }\href {\doibase 10.1529/biophysj.106.085670} {\bibfield
  {journal} {\bibinfo  {journal} {Biophys. J.}\ }\textbf {\bibinfo {volume}
  {91}},\ \bibinfo {pages} {3390} (\bibinfo {year} {2006})}\BibitemShut
  {NoStop}%
\bibitem [{\citenamefont {Frye}\ and\ \citenamefont {Royer}(1998)}]{Frye1998}%
  \BibitemOpen
  \bibfield  {author} {\bibinfo {author} {\bibfnamefont {K.~J.}\ \bibnamefont
  {Frye}}\ and\ \bibinfo {author} {\bibfnamefont {C.~A.}\ \bibnamefont
  {Royer}},\ }\href {\doibase 10.1002/pro.5560071020} {\bibfield  {journal}
  {\bibinfo  {journal} {Protein Sci.}\ }\textbf {\bibinfo {volume} {7}},\
  \bibinfo {pages} {2217} (\bibinfo {year} {1998})}\BibitemShut {NoStop}%
\bibitem [{\citenamefont {Roche}\ \emph {et~al.}(2012)\citenamefont {Roche},
  \citenamefont {Caro}, \citenamefont {Norberto}, \citenamefont {Barthe},
  \citenamefont {Roumestand}, \citenamefont {Schlessman}, \citenamefont
  {Garcia}, \citenamefont {Garcia-Moreno~E},\ and\ \citenamefont
  {Royer}}]{Roche2012}%
  \BibitemOpen
  \bibfield  {author} {\bibinfo {author} {\bibfnamefont {J.}~\bibnamefont
  {Roche}}, \bibinfo {author} {\bibfnamefont {J.~A.}\ \bibnamefont {Caro}},
  \bibinfo {author} {\bibfnamefont {D.~R.}\ \bibnamefont {Norberto}}, \bibinfo
  {author} {\bibfnamefont {P.}~\bibnamefont {Barthe}}, \bibinfo {author}
  {\bibfnamefont {C.}~\bibnamefont {Roumestand}}, \bibinfo {author}
  {\bibfnamefont {J.~L.}\ \bibnamefont {Schlessman}}, \bibinfo {author}
  {\bibfnamefont {A.~E.}\ \bibnamefont {Garcia}}, \bibinfo {author}
  {\bibfnamefont {B.}~\bibnamefont {Garcia-Moreno~E}}, \ and\ \bibinfo {author}
  {\bibfnamefont {C.~A.}\ \bibnamefont {Royer}},\ }\href {\doibase
  10.1073/pnas.1200915109} {\bibfield  {journal} {\bibinfo  {journal} {Proc.
  Natl. Acad. Sci. U.S.A.}\ }\textbf {\bibinfo {volume} {109}},\ \bibinfo
  {pages} {6945} (\bibinfo {year} {2012})}\BibitemShut {NoStop}%
\bibitem [{\citenamefont {P{\'e}rot}\ \emph {et~al.}(2010)\citenamefont
  {P{\'e}rot}, \citenamefont {Sperandio}, \citenamefont {Miteva}, \citenamefont
  {Camproux},\ and\ \citenamefont {Villoutreix}}]{Perot2010}%
  \BibitemOpen
  \bibfield  {author} {\bibinfo {author} {\bibfnamefont {S.}~\bibnamefont
  {P{\'e}rot}}, \bibinfo {author} {\bibfnamefont {O.}~\bibnamefont
  {Sperandio}}, \bibinfo {author} {\bibfnamefont {M.~A.}\ \bibnamefont
  {Miteva}}, \bibinfo {author} {\bibfnamefont {A.-C.}\ \bibnamefont
  {Camproux}}, \ and\ \bibinfo {author} {\bibfnamefont {B.~O.}\ \bibnamefont
  {Villoutreix}},\ }\href {\doibase 10.1016/j.drudis.2010.05.015} {\bibfield
  {journal} {\bibinfo  {journal} {Drug Discov. Today}\ }\textbf {\bibinfo
  {volume} {15}},\ \bibinfo {pages} {656} (\bibinfo {year} {2010})}\BibitemShut
  {NoStop}%
\bibitem [{\citenamefont {Raunest}\ and\ \citenamefont
  {Kandt}(2011)}]{Raunest2011}%
  \BibitemOpen
  \bibfield  {author} {\bibinfo {author} {\bibfnamefont {M.}~\bibnamefont
  {Raunest}}\ and\ \bibinfo {author} {\bibfnamefont {C.}~\bibnamefont
  {Kandt}},\ }\href {\doibase 10.1016/j.jmgm.2011.02.003} {\bibfield  {journal}
  {\bibinfo  {journal} {J. Mol. Graph. Model.}\ }\textbf {\bibinfo {volume}
  {29}},\ \bibinfo {pages} {895} (\bibinfo {year} {2011})}\BibitemShut
  {NoStop}%
\bibitem [{\citenamefont {Pet{\v r}ek}\ \emph {et~al.}(2006)\citenamefont
  {Pet{\v r}ek}, \citenamefont {Otyepka}, \citenamefont {Ban{\'a}{\v s}},
  \citenamefont {Ko{\v s}inov{\'a}}, \citenamefont {Ko{\v c}a},\ and\
  \citenamefont {Damborsk{\'y}}}]{Petrek2006}%
  \BibitemOpen
  \bibfield  {author} {\bibinfo {author} {\bibfnamefont {M.}~\bibnamefont
  {Pet{\v r}ek}}, \bibinfo {author} {\bibfnamefont {M.}~\bibnamefont
  {Otyepka}}, \bibinfo {author} {\bibfnamefont {P.}~\bibnamefont {Ban{\'a}{\v
  s}}}, \bibinfo {author} {\bibfnamefont {P.}~\bibnamefont {Ko{\v
  s}inov{\'a}}}, \bibinfo {author} {\bibfnamefont {J.}~\bibnamefont {Ko{\v
  c}a}}, \ and\ \bibinfo {author} {\bibfnamefont {J.}~\bibnamefont
  {Damborsk{\'y}}},\ }\href {\doibase 10.1186/1471-2105-7-316} {\bibfield
  {journal} {\bibinfo  {journal} {BMC Bioinf.}\ }\textbf {\bibinfo {volume}
  {7}},\ \bibinfo {pages} {1} (\bibinfo {year} {2006})}\BibitemShut {NoStop}%
\bibitem [{\citenamefont {Brezovsky}\ \emph {et~al.}(2013)\citenamefont
  {Brezovsky}, \citenamefont {Chovancova}, \citenamefont {Gora}, \citenamefont
  {Pavelka}, \citenamefont {Biedermannova},\ and\ \citenamefont
  {Damborsky}}]{Brezovsky2013}%
  \BibitemOpen
  \bibfield  {author} {\bibinfo {author} {\bibfnamefont {J.}~\bibnamefont
  {Brezovsky}}, \bibinfo {author} {\bibfnamefont {E.}~\bibnamefont
  {Chovancova}}, \bibinfo {author} {\bibfnamefont {A.}~\bibnamefont {Gora}},
  \bibinfo {author} {\bibfnamefont {A.}~\bibnamefont {Pavelka}}, \bibinfo
  {author} {\bibfnamefont {L.}~\bibnamefont {Biedermannova}}, \ and\ \bibinfo
  {author} {\bibfnamefont {J.}~\bibnamefont {Damborsky}},\ }\href {\doibase
  10.1016/j.biotechadv.2012.02.002} {\bibfield  {journal} {\bibinfo  {journal}
  {Biotechnol. Adv.}\ }\textbf {\bibinfo {volume} {31}},\ \bibinfo {pages} {38}
  (\bibinfo {year} {2013})}\BibitemShut {NoStop}%
\bibitem [{\citenamefont {Baldwin}(2013)}]{Baldwin2013}%
  \BibitemOpen
  \bibfield  {author} {\bibinfo {author} {\bibfnamefont {R.~L.}\ \bibnamefont
  {Baldwin}},\ }\href {\doibase 10.1016/j.febslet.2013.01.006} {\bibfield
  {journal} {\bibinfo  {journal} {FEBS Lett.}\ }\textbf {\bibinfo {volume}
  {587}},\ \bibinfo {pages} {1062} (\bibinfo {year} {2013})}\BibitemShut
  {NoStop}%
\bibitem [{\citenamefont {Chandler}(2005)}]{Chandler2005}%
  \BibitemOpen
  \bibfield  {author} {\bibinfo {author} {\bibfnamefont {D.}~\bibnamefont
  {Chandler}},\ }\href {\doibase 10.1038/nature04162} {\bibfield  {journal}
  {\bibinfo  {journal} {Nature}\ }\textbf {\bibinfo {volume} {437}},\ \bibinfo
  {pages} {640} (\bibinfo {year} {2005})}\BibitemShut {NoStop}%
\bibitem [{\citenamefont {Reynolds}\ \emph {et~al.}(1974)\citenamefont
  {Reynolds}, \citenamefont {Gilbert},\ and\ \citenamefont
  {Tanford}}]{Reynolds1974}%
  \BibitemOpen
  \bibfield  {author} {\bibinfo {author} {\bibfnamefont {J.~A.}\ \bibnamefont
  {Reynolds}}, \bibinfo {author} {\bibfnamefont {D.~B.}\ \bibnamefont
  {Gilbert}}, \ and\ \bibinfo {author} {\bibfnamefont {C.}~\bibnamefont
  {Tanford}},\ }\href {\doibase 10.1073/pnas.71.8.2925} {\bibfield  {journal}
  {\bibinfo  {journal} {Proc. Natl. Acad. Sci. U.S.A.}\ }\textbf {\bibinfo
  {volume} {71}},\ \bibinfo {pages} {2925} (\bibinfo {year}
  {1974})}\BibitemShut {NoStop}%
\bibitem [{\citenamefont {Winter}\ \emph {et~al.}(2011)\citenamefont {Winter},
  \citenamefont {Herzik}, \citenamefont {Kuriyan},\ and\ \citenamefont
  {Marletta}}]{Winter2011}%
  \BibitemOpen
  \bibfield  {author} {\bibinfo {author} {\bibfnamefont {M.~B.}\ \bibnamefont
  {Winter}}, \bibinfo {author} {\bibfnamefont {J.~M.~A.}\ \bibnamefont
  {Herzik}}, \bibinfo {author} {\bibfnamefont {J.}~\bibnamefont {Kuriyan}}, \
  and\ \bibinfo {author} {\bibfnamefont {M.~A.}\ \bibnamefont {Marletta}},\
  }\href {\doibase 10.1073/pnas.1114038108} {\bibfield  {journal} {\bibinfo
  {journal} {Proc. Natl. Acad. Sci. U.S.A.}\ }\textbf {\bibinfo {volume} {108}}
  (\bibinfo {year} {2011}),\ 10.1073/pnas.1114038108}\BibitemShut {NoStop}%
\bibitem [{\citenamefont {McCallum}\ \emph {et~al.}(2000)\citenamefont
  {McCallum}, \citenamefont {Hitchens}, \citenamefont {Torborg},\ and\
  \citenamefont {Rule}}]{McCallum2000}%
  \BibitemOpen
  \bibfield  {author} {\bibinfo {author} {\bibfnamefont {S.~A.}\ \bibnamefont
  {McCallum}}, \bibinfo {author} {\bibfnamefont {T.~K.}\ \bibnamefont
  {Hitchens}}, \bibinfo {author} {\bibfnamefont {C.}~\bibnamefont {Torborg}}, \
  and\ \bibinfo {author} {\bibfnamefont {G.~S.}\ \bibnamefont {Rule}},\ }\href
  {\doibase 10.1021/bi992767d} {\bibfield  {journal} {\bibinfo  {journal}
  {Biochemistry (Mosc.)}\ }\textbf {\bibinfo {volume} {39}},\ \bibinfo {pages}
  {7343} (\bibinfo {year} {2000})}\BibitemShut {NoStop}%
\bibitem [{\citenamefont {Liang}\ \emph {et~al.}(1998)\citenamefont {Liang},
  \citenamefont {Edelsbrunner},\ and\ \citenamefont {Woodward}}]{Liang1998}%
  \BibitemOpen
  \bibfield  {author} {\bibinfo {author} {\bibfnamefont {J.}~\bibnamefont
  {Liang}}, \bibinfo {author} {\bibfnamefont {H.}~\bibnamefont {Edelsbrunner}},
  \ and\ \bibinfo {author} {\bibfnamefont {C.}~\bibnamefont {Woodward}},\
  }\href {\doibase 10.1002/pro.5560070905} {\bibfield  {journal} {\bibinfo
  {journal} {Protein Sci.}\ }\textbf {\bibinfo {volume} {7}},\ \bibinfo {pages}
  {1884} (\bibinfo {year} {1998})}\BibitemShut {NoStop}%
\bibitem [{\citenamefont {Weisel}\ \emph {et~al.}(2007)\citenamefont {Weisel},
  \citenamefont {Proschak},\ and\ \citenamefont {Schneider}}]{Weisel2007}%
  \BibitemOpen
  \bibfield  {author} {\bibinfo {author} {\bibfnamefont {M.}~\bibnamefont
  {Weisel}}, \bibinfo {author} {\bibfnamefont {E.}~\bibnamefont {Proschak}}, \
  and\ \bibinfo {author} {\bibfnamefont {G.}~\bibnamefont {Schneider}},\ }\href
  {\doibase 10.1186/1752-153x-1-7} {\bibfield  {journal} {\bibinfo  {journal}
  {Chem. Cent. J.}\ }\textbf {\bibinfo {volume} {1}},\ \bibinfo {pages} {7}
  (\bibinfo {year} {2007})}\BibitemShut {NoStop}%
\bibitem [{\citenamefont {Schmidtke}\ \emph {et~al.}(2010)\citenamefont
  {Schmidtke}, \citenamefont {Souaille}, \citenamefont {Estienne},
  \citenamefont {Baurin},\ and\ \citenamefont {Kroemer}}]{Schmidtke2010}%
  \BibitemOpen
  \bibfield  {author} {\bibinfo {author} {\bibfnamefont {P.}~\bibnamefont
  {Schmidtke}}, \bibinfo {author} {\bibfnamefont {C.}~\bibnamefont {Souaille}},
  \bibinfo {author} {\bibfnamefont {F.}~\bibnamefont {Estienne}}, \bibinfo
  {author} {\bibfnamefont {N.}~\bibnamefont {Baurin}}, \ and\ \bibinfo {author}
  {\bibfnamefont {R.}~\bibnamefont {Kroemer}},\ }\href {\doibase
  10.1021/ci1000289} {\bibfield  {journal} {\bibinfo  {journal} {J. Chem. Inf.
  Model.}\ }\textbf {\bibinfo {volume} {50}},\ \bibinfo {pages} {2191}
  (\bibinfo {year} {2010})}\BibitemShut {NoStop}%
\bibitem [{\citenamefont {Chen}\ and\ \citenamefont
  {Makhatadze}(2015)}]{Chen2015}%
  \BibitemOpen
  \bibfield  {author} {\bibinfo {author} {\bibfnamefont {C.~R.}\ \bibnamefont
  {Chen}}\ and\ \bibinfo {author} {\bibfnamefont {G.~I.}\ \bibnamefont
  {Makhatadze}},\ }\href {\doibase 10.1186/s12859-015-0531-2} {\bibfield
  {journal} {\bibinfo  {journal} {BMC Bioinf.}\ }\textbf {\bibinfo {volume}
  {16}},\ \bibinfo {pages} {101} (\bibinfo {year} {2015})}\BibitemShut
  {NoStop}%
\bibitem [{\citenamefont {Voloshin}\ \emph {et~al.}(2011)\citenamefont
  {Voloshin}, \citenamefont {Anikeenko}, \citenamefont {Medvedev},\ and\
  \citenamefont {Geiger}}]{Voloshin2011}%
  \BibitemOpen
  \bibfield  {author} {\bibinfo {author} {\bibfnamefont {V.~P.}\ \bibnamefont
  {Voloshin}}, \bibinfo {author} {\bibfnamefont {A.~V.}\ \bibnamefont
  {Anikeenko}}, \bibinfo {author} {\bibfnamefont {N.~N.}\ \bibnamefont
  {Medvedev}}, \ and\ \bibinfo {author} {\bibfnamefont {A.}~\bibnamefont
  {Geiger}},\ }\href {\doibase 10.1109/ISVD.2011.30} {\bibfield  {journal}
  {\bibinfo  {journal} {2011 Eighth International Symposium on Voronoi Diagrams
  in Science and Engineering}\ ,\ \bibinfo {pages} {170}} (\bibinfo {year}
  {2011})}\BibitemShut {NoStop}%
\bibitem [{\citenamefont {Gaines}\ \emph {et~al.}(2018)\citenamefont {Gaines},
  \citenamefont {Acebes}, \citenamefont {Virrueta}, \citenamefont {Butler},
  \citenamefont {Regan},\ and\ \citenamefont {O'Hern}}]{Gaines2018}%
  \BibitemOpen
  \bibfield  {author} {\bibinfo {author} {\bibfnamefont {J.~C.}\ \bibnamefont
  {Gaines}}, \bibinfo {author} {\bibfnamefont {S.}~\bibnamefont {Acebes}},
  \bibinfo {author} {\bibfnamefont {A.}~\bibnamefont {Virrueta}}, \bibinfo
  {author} {\bibfnamefont {M.}~\bibnamefont {Butler}}, \bibinfo {author}
  {\bibfnamefont {L.}~\bibnamefont {Regan}}, \ and\ \bibinfo {author}
  {\bibfnamefont {C.~S.}\ \bibnamefont {O'Hern}},\ }\href {\doibase
  10.1002/prot.25479} {\bibfield  {journal} {\bibinfo  {journal} {Proteins}\
  }\textbf {\bibinfo {volume} {86}},\ \bibinfo {pages} {581} (\bibinfo {year}
  {2018})}\BibitemShut {NoStop}%
\bibitem [{\citenamefont {Shaytan}\ \emph {et~al.}(2009)\citenamefont
  {Shaytan}, \citenamefont {Shaitan},\ and\ \citenamefont
  {Khokhlov}}]{Shaytan2009}%
  \BibitemOpen
  \bibfield  {author} {\bibinfo {author} {\bibfnamefont {A.~K.}\ \bibnamefont
  {Shaytan}}, \bibinfo {author} {\bibfnamefont {K.~V.}\ \bibnamefont
  {Shaitan}}, \ and\ \bibinfo {author} {\bibfnamefont {A.~R.}\ \bibnamefont
  {Khokhlov}},\ }\href {\doibase 10.1021/bm8015169} {\bibfield  {journal}
  {\bibinfo  {journal} {Biomacromolecules}\ }\textbf {\bibinfo {volume} {10}},\
  \bibinfo {pages} {1224} (\bibinfo {year} {2009})}\BibitemShut {NoStop}%
\bibitem [{\citenamefont {Richards}(1977)}]{Richards1977}%
  \BibitemOpen
  \bibfield  {author} {\bibinfo {author} {\bibfnamefont {F.~M.}\ \bibnamefont
  {Richards}},\ }\href {\doibase 10.1146/annurev.bb.06.060177.001055}
  {\bibfield  {journal} {\bibinfo  {journal} {Ann. Rev. Biophys.}\ }\textbf
  {\bibinfo {volume} {6}},\ \bibinfo {pages} {151} (\bibinfo {year}
  {1977})}\BibitemShut {NoStop}%
\bibitem [{\citenamefont {Fleming}\ and\ \citenamefont
  {Richards}(2000)}]{Fleming2000}%
  \BibitemOpen
  \bibfield  {author} {\bibinfo {author} {\bibfnamefont {P.~J.}\ \bibnamefont
  {Fleming}}\ and\ \bibinfo {author} {\bibfnamefont {F.~M.}\ \bibnamefont
  {Richards}},\ }\href {\doibase 10.1006/jmbi.2000.3750} {\bibfield  {journal}
  {\bibinfo  {journal} {J. Mol. Biol.}\ }\textbf {\bibinfo {volume} {299}},\
  \bibinfo {pages} {487} (\bibinfo {year} {2000})}\BibitemShut {NoStop}%
\bibitem [{\citenamefont {Voss}\ and\ \citenamefont
  {Gerstein}(2010)}]{Voss2010}%
  \BibitemOpen
  \bibfield  {author} {\bibinfo {author} {\bibfnamefont {N.~R.}\ \bibnamefont
  {Voss}}\ and\ \bibinfo {author} {\bibfnamefont {M.}~\bibnamefont
  {Gerstein}},\ }\href {\doibase 10.1093/nar/gkq395} {\bibfield  {journal}
  {\bibinfo  {journal} {Nucleic Acids Res.}\ }\textbf {\bibinfo {volume}
  {38}},\ \bibinfo {pages} {W555} (\bibinfo {year} {2010})}\BibitemShut
  {NoStop}%
\bibitem [{\citenamefont {Ho}\ and\ \citenamefont {Marshall}(1990)}]{Ho1990}%
  \BibitemOpen
  \bibfield  {author} {\bibinfo {author} {\bibfnamefont {C.~M.~W.}\
  \bibnamefont {Ho}}\ and\ \bibinfo {author} {\bibfnamefont {G.~R.}\
  \bibnamefont {Marshall}},\ }\href {\doibase 10.1007/BF00117400} {\bibfield
  {journal} {\bibinfo  {journal} {J. Comput.-Aided Mol. Des.}\ }\textbf
  {\bibinfo {volume} {4}},\ \bibinfo {pages} {337} (\bibinfo {year}
  {1990})}\BibitemShut {NoStop}%
\bibitem [{\citenamefont {Kleywegt}\ and\ \citenamefont
  {Jones}(1994)}]{Kleywegt1994}%
  \BibitemOpen
  \bibfield  {author} {\bibinfo {author} {\bibfnamefont {G.~J.}\ \bibnamefont
  {Kleywegt}}\ and\ \bibinfo {author} {\bibfnamefont {T.~A.}\ \bibnamefont
  {Jones}},\ }\href {\doibase 10.1107/s0907444993011333} {\bibfield  {journal}
  {\bibinfo  {journal} {Acta Crystallogr. D}\ }\textbf {\bibinfo {volume}
  {50}},\ \bibinfo {pages} {178} (\bibinfo {year} {1994})}\BibitemShut
  {NoStop}%
\bibitem [{\citenamefont {Levitt}\ and\ \citenamefont
  {Banaszak}(1992)}]{Levitt1992}%
  \BibitemOpen
  \bibfield  {author} {\bibinfo {author} {\bibfnamefont {D.~G.}\ \bibnamefont
  {Levitt}}\ and\ \bibinfo {author} {\bibfnamefont {L.~J.}\ \bibnamefont
  {Banaszak}},\ }\href {\doibase 10.1016/0263-7855(92)80074-N} {\bibfield
  {journal} {\bibinfo  {journal} {J. Mol. Graphics}\ }\textbf {\bibinfo
  {volume} {10}},\ \bibinfo {pages} {229} (\bibinfo {year} {1992})}\BibitemShut
  {NoStop}%
\bibitem [{\citenamefont {Laskowski}(1995)}]{Laskowski1995}%
  \BibitemOpen
  \bibfield  {author} {\bibinfo {author} {\bibfnamefont {R.~A.}\ \bibnamefont
  {Laskowski}},\ }\href {\doibase 10.1016/0263-7855(95)00073-9} {\bibfield
  {journal} {\bibinfo  {journal} {J. Mol. Graphics}\ }\textbf {\bibinfo
  {volume} {13}},\ \bibinfo {pages} {323} (\bibinfo {year} {1995})}\BibitemShut
  {NoStop}%
\bibitem [{\citenamefont {Voronoi}(1908)}]{Voronoi1908}%
  \BibitemOpen
  \bibfield  {author} {\bibinfo {author} {\bibfnamefont {G.}~\bibnamefont
  {Voronoi}},\ }\href {https://eudml.org/doc/149291} {\bibfield  {journal}
  {\bibinfo  {journal} {Journal f{\"u}r die Reine und Angewandte Mathematik}\
  }\textbf {\bibinfo {volume} {1908}},\ \bibinfo {pages} {198} (\bibinfo {year}
  {1908})}\BibitemShut {NoStop}%
\bibitem [{\citenamefont {Finney}(1970)}]{Finney1970}%
  \BibitemOpen
  \bibfield  {author} {\bibinfo {author} {\bibfnamefont {J.~L.}\ \bibnamefont
  {Finney}},\ }\href {\doibase 10.1098/rspa.1970.0190} {\bibfield  {journal}
  {\bibinfo  {journal} {Proc. Royal Soc. A}\ }\textbf {\bibinfo {volume}
  {319}},\ \bibinfo {pages} {495} (\bibinfo {year} {1970})}\BibitemShut
  {NoStop}%
\bibitem [{\citenamefont {Edelsbrunner}\ \emph {et~al.}(1983)\citenamefont
  {Edelsbrunner}, \citenamefont {Kirkpatrick},\ and\ \citenamefont
  {Seidel}}]{Edelsbrunner1983}%
  \BibitemOpen
  \bibfield  {author} {\bibinfo {author} {\bibfnamefont {H.}~\bibnamefont
  {Edelsbrunner}}, \bibinfo {author} {\bibfnamefont {D.~G.}\ \bibnamefont
  {Kirkpatrick}}, \ and\ \bibinfo {author} {\bibfnamefont {R.}~\bibnamefont
  {Seidel}},\ }\href {\doibase 10.1109/TIT.1983.1056714} {\bibfield  {journal}
  {\bibinfo  {journal} {IEEE Trans. Inf. Theory}\ }\textbf {\bibinfo {volume}
  {29}},\ \bibinfo {pages} {551} (\bibinfo {year} {1983})}\BibitemShut
  {NoStop}%
\bibitem [{\citenamefont {Delaunay}(1934)}]{Delaunay1934}%
  \BibitemOpen
  \bibfield  {author} {\bibinfo {author} {\bibfnamefont {B.}~\bibnamefont
  {Delaunay}},\ }\href {http://www.mathnet.ru/eng/im4937} {\bibfield  {journal}
  {\bibinfo  {journal} {Bulletin de l'Acad{\'e}mie des Sciences de l'URSS
  Classe des sciences math{\'e}matiques et naturelles}\ }\textbf {\bibinfo
  {volume} {6}},\ \bibinfo {pages} {793} (\bibinfo {year} {1934})}\BibitemShut
  {NoStop}%
\bibitem [{\citenamefont {Li}\ \emph {et~al.}(2013)\citenamefont {Li},
  \citenamefont {Mach},\ and\ \citenamefont {Koehl}}]{Li2013}%
  \BibitemOpen
  \bibfield  {author} {\bibinfo {author} {\bibfnamefont {J.}~\bibnamefont
  {Li}}, \bibinfo {author} {\bibfnamefont {P.}~\bibnamefont {Mach}}, \ and\
  \bibinfo {author} {\bibfnamefont {P.}~\bibnamefont {Koehl}},\ }\href
  {\doibase 10.5936/csbj.201309001} {\bibfield  {journal} {\bibinfo  {journal}
  {Comput. Struct. Biotechnol. J.}\ }\textbf {\bibinfo {volume} {8}},\ \bibinfo
  {pages} {e201309001} (\bibinfo {year} {2013})}\BibitemShut {NoStop}%
\bibitem [{\citenamefont {Albou}\ \emph {et~al.}(2009)\citenamefont {Albou},
  \citenamefont {Schwarz}, \citenamefont {Poch}, \citenamefont {Marie~Wurtz},\
  and\ \citenamefont {Moras}}]{Albou2009}%
  \BibitemOpen
  \bibfield  {author} {\bibinfo {author} {\bibfnamefont {L.-P.}\ \bibnamefont
  {Albou}}, \bibinfo {author} {\bibfnamefont {B.}~\bibnamefont {Schwarz}},
  \bibinfo {author} {\bibfnamefont {O.}~\bibnamefont {Poch}}, \bibinfo {author}
  {\bibfnamefont {J.}~\bibnamefont {Marie~Wurtz}}, \ and\ \bibinfo {author}
  {\bibfnamefont {D.}~\bibnamefont {Moras}},\ }\href {\doibase
  10.1002/prot.22301} {\bibfield  {journal} {\bibinfo  {journal} {Proteins}\
  }\textbf {\bibinfo {volume} {76}},\ \bibinfo {pages} {1} (\bibinfo {year}
  {2009})}\BibitemShut {NoStop}%
\bibitem [{\citenamefont {Zhou}\ and\ \citenamefont {Yan}(2014)}]{Zhou2014}%
  \BibitemOpen
  \bibfield  {author} {\bibinfo {author} {\bibfnamefont {W.}~\bibnamefont
  {Zhou}}\ and\ \bibinfo {author} {\bibfnamefont {H.}~\bibnamefont {Yan}},\
  }\href {\doibase 10.1093/bib/bbs077} {\bibfield  {journal} {\bibinfo
  {journal} {Briefings Bioinf.}\ }\textbf {\bibinfo {volume} {15}},\ \bibinfo
  {pages} {54} (\bibinfo {year} {2014})}\BibitemShut {NoStop}%
\bibitem [{\citenamefont {Mach}\ and\ \citenamefont {Koehl}(2011)}]{Mach2011}%
  \BibitemOpen
  \bibfield  {author} {\bibinfo {author} {\bibfnamefont {P.}~\bibnamefont
  {Mach}}\ and\ \bibinfo {author} {\bibfnamefont {P.}~\bibnamefont {Koehl}},\
  }\href {\doibase 10.1002/jcc.21884} {\bibfield  {journal} {\bibinfo
  {journal} {J. Comput. Chem.}\ }\textbf {\bibinfo {volume} {32}},\ \bibinfo
  {pages} {3023} (\bibinfo {year} {2011})}\BibitemShut {NoStop}%
\bibitem [{\citenamefont {Hoshen}\ and\ \citenamefont
  {Kopelman}(1976)}]{Hoshen1976}%
  \BibitemOpen
  \bibfield  {author} {\bibinfo {author} {\bibfnamefont {J.}~\bibnamefont
  {Hoshen}}\ and\ \bibinfo {author} {\bibfnamefont {R.}~\bibnamefont
  {Kopelman}},\ }\href {\doibase 10.1103/physrevb.14.3438} {\bibfield
  {journal} {\bibinfo  {journal} {Phys. Rev. B}\ }\textbf {\bibinfo {volume}
  {14}},\ \bibinfo {pages} {3438} (\bibinfo {year} {1976})}\BibitemShut
  {NoStop}%
\bibitem [{\citenamefont {Gerstein}\ \emph {et~al.}(1995)\citenamefont
  {Gerstein}, \citenamefont {Tsai},\ and\ \citenamefont
  {Levitt}}]{Gerstein1995}%
  \BibitemOpen
  \bibfield  {author} {\bibinfo {author} {\bibfnamefont {M.}~\bibnamefont
  {Gerstein}}, \bibinfo {author} {\bibfnamefont {J.}~\bibnamefont {Tsai}}, \
  and\ \bibinfo {author} {\bibfnamefont {M.}~\bibnamefont {Levitt}},\ }\href
  {\doibase 10.1006/jmbi.1995.0351} {\bibfield  {journal} {\bibinfo  {journal}
  {J. Mol. Biol.}\ }\textbf {\bibinfo {volume} {249}},\ \bibinfo {pages} {955}
  (\bibinfo {year} {1995})}\BibitemShut {NoStop}%
\bibitem [{\citenamefont {Esque}\ \emph {et~al.}(2010)\citenamefont {Esque},
  \citenamefont {Oguey},\ and\ \citenamefont {de~Brevern}}]{Esque2010}%
  \BibitemOpen
  \bibfield  {author} {\bibinfo {author} {\bibfnamefont {J.}~\bibnamefont
  {Esque}}, \bibinfo {author} {\bibfnamefont {C.}~\bibnamefont {Oguey}}, \ and\
  \bibinfo {author} {\bibfnamefont {A.~G.}\ \bibnamefont {de~Brevern}},\ }\href
  {\doibase 10.1021/ci9004892} {\bibfield  {journal} {\bibinfo  {journal} {J.
  Chem. Inf. Model.}\ }\textbf {\bibinfo {volume} {50}},\ \bibinfo {pages}
  {947} (\bibinfo {year} {2010})}\BibitemShut {NoStop}%
\bibitem [{\citenamefont {Dill}(1990)}]{Dill1990}%
  \BibitemOpen
  \bibfield  {author} {\bibinfo {author} {\bibfnamefont {K.~A.}\ \bibnamefont
  {Dill}},\ }\href {\doibase 10.1021/bi00483a001} {\bibfield  {journal}
  {\bibinfo  {journal} {Biochemistry (Mosc.)}\ }\textbf {\bibinfo {volume}
  {29}},\ \bibinfo {pages} {7133} (\bibinfo {year} {1990})}\BibitemShut
  {NoStop}%
\bibitem [{\citenamefont {Rother}\ \emph {et~al.}(2003)\citenamefont {Rother},
  \citenamefont {Preissner}, \citenamefont {Goede},\ and\ \citenamefont
  {Frmmel}}]{Rother2003}%
  \BibitemOpen
  \bibfield  {author} {\bibinfo {author} {\bibfnamefont {K.}~\bibnamefont
  {Rother}}, \bibinfo {author} {\bibfnamefont {R.}~\bibnamefont {Preissner}},
  \bibinfo {author} {\bibfnamefont {A.}~\bibnamefont {Goede}}, \ and\ \bibinfo
  {author} {\bibfnamefont {C.}~\bibnamefont {Frmmel}},\ }\href {\doibase
  10.1093/bioinformatics/btg292} {\bibfield  {journal} {\bibinfo  {journal}
  {Bioinformatics}\ }\textbf {\bibinfo {volume} {19}},\ \bibinfo {pages} {2112}
  (\bibinfo {year} {2003})}\BibitemShut {NoStop}%
\bibitem [{\citenamefont {Rycroft}(2009)}]{Rycroft2009}%
  \BibitemOpen
  \bibfield  {author} {\bibinfo {author} {\bibfnamefont {C.~H.}\ \bibnamefont
  {Rycroft}},\ }\href {\doibase 10.1063/1.3215722} {\bibfield  {journal}
  {\bibinfo  {journal} {Chaos}\ }\textbf {\bibinfo {volume} {19}} (\bibinfo
  {year} {2009}),\ 10.1063/1.3215722}\BibitemShut {NoStop}%
\bibitem [{\citenamefont {Andersson}\ and\ \citenamefont
  {Hovmoller}(1998)}]{Andersson1998}%
  \BibitemOpen
  \bibfield  {author} {\bibinfo {author} {\bibfnamefont {K.}~\bibnamefont
  {Andersson}}\ and\ \bibinfo {author} {\bibfnamefont {S.}~\bibnamefont
  {Hovmoller}},\ }\href {\doibase 10.1524/zkri.1998.213.7-8.369} {\bibfield
  {journal} {\bibinfo  {journal} {Z. Kristallogr. Cryst. Mater.}\ }\textbf
  {\bibinfo {volume} {213}},\ \bibinfo {pages} {369} (\bibinfo {year}
  {1998})}\BibitemShut {NoStop}%
\bibitem [{\citenamefont {Gerstein}\ and\ \citenamefont
  {Chothia}(1996)}]{Gerstein1996}%
  \BibitemOpen
  \bibfield  {author} {\bibinfo {author} {\bibfnamefont {M.}~\bibnamefont
  {Gerstein}}\ and\ \bibinfo {author} {\bibfnamefont {C.}~\bibnamefont
  {Chothia}},\ }\href {\doibase 10.1073/pnas.93.19.10167} {\bibfield  {journal}
  {\bibinfo  {journal} {Proc. Natl. Acad. Sci. U.S.A.}\ }\textbf {\bibinfo
  {volume} {93}},\ \bibinfo {pages} {10167} (\bibinfo {year}
  {1996})}\BibitemShut {NoStop}%
\bibitem [{\citenamefont {Chakravarty}\ \emph {et~al.}(2002)\citenamefont
  {Chakravarty}, \citenamefont {Bhinge},\ and\ \citenamefont
  {Varadarajan}}]{Chakravarty2002}%
  \BibitemOpen
  \bibfield  {author} {\bibinfo {author} {\bibfnamefont {S.}~\bibnamefont
  {Chakravarty}}, \bibinfo {author} {\bibfnamefont {A.}~\bibnamefont {Bhinge}},
  \ and\ \bibinfo {author} {\bibfnamefont {R.}~\bibnamefont {Varadarajan}},\
  }\href {\doibase 10.1074/jbc.M201373200} {\bibfield  {journal} {\bibinfo
  {journal} {J. Biol. Chem.}\ }\textbf {\bibinfo {volume} {277}},\ \bibinfo
  {pages} {31345} (\bibinfo {year} {2002})}\BibitemShut {NoStop}%
\bibitem [{\citenamefont {Angelov}\ \emph {et~al.}(2002)\citenamefont
  {Angelov}, \citenamefont {Sadoc}, \citenamefont {Jullien}, \citenamefont
  {Soyer}, \citenamefont {Mornon},\ and\ \citenamefont
  {Chomilier}}]{Angelov2002}%
  \BibitemOpen
  \bibfield  {author} {\bibinfo {author} {\bibfnamefont {B.}~\bibnamefont
  {Angelov}}, \bibinfo {author} {\bibfnamefont {J.}~\bibnamefont {Sadoc}},
  \bibinfo {author} {\bibfnamefont {R.}~\bibnamefont {Jullien}}, \bibinfo
  {author} {\bibfnamefont {A.}~\bibnamefont {Soyer}}, \bibinfo {author}
  {\bibfnamefont {J.}~\bibnamefont {Mornon}}, \ and\ \bibinfo {author}
  {\bibfnamefont {J.}~\bibnamefont {Chomilier}},\ }\href {\doibase
  10.1002/prot.10220} {\bibfield  {journal} {\bibinfo  {journal} {Proteins}\
  }\textbf {\bibinfo {volume} {49}},\ \bibinfo {pages} {446} (\bibinfo {year}
  {2002})}\BibitemShut {NoStop}%
\bibitem [{\citenamefont {Richmond}(1984)}]{Richmond1984}%
  \BibitemOpen
  \bibfield  {author} {\bibinfo {author} {\bibfnamefont {T.~J.}\ \bibnamefont
  {Richmond}},\ }\href {\doibase 10.1016/0022-2836(84)90231-6} {\bibfield
  {journal} {\bibinfo  {journal} {J. Mol. Biol.}\ }\textbf {\bibinfo {volume}
  {178}},\ \bibinfo {pages} {63} (\bibinfo {year} {1984})}\BibitemShut
  {NoStop}%
\bibitem [{\citenamefont {Connolly}(1983)}]{Connolly1983}%
  \BibitemOpen
  \bibfield  {author} {\bibinfo {author} {\bibfnamefont {M.~L.}\ \bibnamefont
  {Connolly}},\ }\href {\doibase 10.1107/S0021889883010985} {\bibfield
  {journal} {\bibinfo  {journal} {J. Appl. Crystallogr.}\ }\textbf {\bibinfo
  {volume} {16}},\ \bibinfo {pages} {548} (\bibinfo {year} {1983})}\BibitemShut
  {NoStop}%
\bibitem [{\citenamefont {Connolly}(1984)}]{Connolly1984}%
  \BibitemOpen
  \bibfield  {author} {\bibinfo {author} {\bibfnamefont {M.~L.}\ \bibnamefont
  {Connolly}},\ }\href {\doibase 10.1021/ja00291a006} {\bibfield  {journal}
  {\bibinfo  {journal} {J. Am. Chem. Soc.}\ }\textbf {\bibinfo {volume}
  {107}},\ \bibinfo {pages} {1118} (\bibinfo {year} {1984})}\BibitemShut
  {NoStop}%
\bibitem [{\citenamefont {Lee}\ and\ \citenamefont {Richards}(1971)}]{Lee1971}%
  \BibitemOpen
  \bibfield  {author} {\bibinfo {author} {\bibfnamefont {B.}~\bibnamefont
  {Lee}}\ and\ \bibinfo {author} {\bibfnamefont {F.~M.}\ \bibnamefont
  {Richards}},\ }\href {\doibase 10.1016/0022-2836(71)90324-X} {\bibfield
  {journal} {\bibinfo  {journal} {J. Mol. Biol.}\ }\textbf {\bibinfo {volume}
  {55}},\ \bibinfo {pages} {379,IN3} (\bibinfo {year} {1971})}\BibitemShut
  {NoStop}%
\bibitem [{\citenamefont {Teuler}\ and\ \citenamefont
  {Gimel}(2000)}]{Teuler2000}%
  \BibitemOpen
  \bibfield  {author} {\bibinfo {author} {\bibfnamefont {J.~M.}\ \bibnamefont
  {Teuler}}\ and\ \bibinfo {author} {\bibfnamefont {J.~C.}\ \bibnamefont
  {Gimel}},\ }\href {\doibase 10.1016/S0010-4655(00)00046-1} {\bibfield
  {journal} {\bibinfo  {journal} {Comput. Phys. Commun.}\ }\textbf {\bibinfo
  {volume} {130}},\ \bibinfo {pages} {118} (\bibinfo {year}
  {2000})}\BibitemShut {NoStop}%
\bibitem [{\citenamefont {Al-Futaisi}\ and\ \citenamefont
  {Patzek}(2003)}]{Al-Futaisi2003}%
  \BibitemOpen
  \bibfield  {author} {\bibinfo {author} {\bibfnamefont {A.}~\bibnamefont
  {Al-Futaisi}}\ and\ \bibinfo {author} {\bibfnamefont {T.~W.}\ \bibnamefont
  {Patzek}},\ }\href {\doibase 10.1016/S0378-4371(02)01586-8} {\bibfield
  {journal} {\bibinfo  {journal} {Physica A}\ }\textbf {\bibinfo {volume}
  {321}} (\bibinfo {year} {2003}),\ 10.1016/S0378-4371(02)01586-8}\BibitemShut
  {NoStop}%
\bibitem [{\citenamefont {Hoshen}(1998)}]{Hoshen1998}%
  \BibitemOpen
  \bibfield  {author} {\bibinfo {author} {\bibfnamefont {J.}~\bibnamefont
  {Hoshen}},\ }\href {\doibase 10.1016/S0167-8655(98)00018-X} {\bibfield
  {journal} {\bibinfo  {journal} {Pattern Recognit. Lett.}\ }\textbf {\bibinfo
  {volume} {19}},\ \bibinfo {pages} {575} (\bibinfo {year} {1998})}\BibitemShut
  {NoStop}%
\bibitem [{\citenamefont {Frijters}\ \emph {et~al.}(2015)\citenamefont
  {Frijters}, \citenamefont {Kr{\"u}ger},\ and\ \citenamefont
  {Harting}}]{Frijters2015}%
  \BibitemOpen
  \bibfield  {author} {\bibinfo {author} {\bibfnamefont {S.}~\bibnamefont
  {Frijters}}, \bibinfo {author} {\bibfnamefont {T.}~\bibnamefont
  {Kr{\"u}ger}}, \ and\ \bibinfo {author} {\bibfnamefont {J.}~\bibnamefont
  {Harting}},\ }\href {\doibase 10.1016/j.cpc.2014.12.014} {\bibfield
  {journal} {\bibinfo  {journal} {Comput. Phys. Commun.}\ }\textbf {\bibinfo
  {volume} {189}},\ \bibinfo {pages} {92} (\bibinfo {year} {2015})}\BibitemShut
  {NoStop}%
\bibitem [{\citenamefont {Hoshen}\ \emph {et~al.}(1997)\citenamefont {Hoshen},
  \citenamefont {Berry},\ and\ \citenamefont {Minser}}]{Hoshen1997}%
  \BibitemOpen
  \bibfield  {author} {\bibinfo {author} {\bibfnamefont {J.}~\bibnamefont
  {Hoshen}}, \bibinfo {author} {\bibfnamefont {M.~W.}\ \bibnamefont {Berry}}, \
  and\ \bibinfo {author} {\bibfnamefont {K.~S.}\ \bibnamefont {Minser}},\
  }\href {\doibase 10.1103/PhysRevE.56.1455} {\bibfield  {journal} {\bibinfo
  {journal} {Phys. Rev. E}\ }\textbf {\bibinfo {volume} {56}},\ \bibinfo
  {pages} {1455} (\bibinfo {year} {1997})}\BibitemShut {NoStop}%
\bibitem [{\citenamefont {Khaimov}\ \emph {et~al.}(1989)\citenamefont
  {Khaimov}, \citenamefont {Mikhalev}, \citenamefont {Savchenko},\ and\
  \citenamefont {Iordanov}}]{Khaimov1989}%
  \BibitemOpen
  \bibfield  {author} {\bibinfo {author} {\bibfnamefont {S.}~\bibnamefont
  {Khaimov}}, \bibinfo {author} {\bibfnamefont {M.}~\bibnamefont {Mikhalev}},
  \bibinfo {author} {\bibfnamefont {A.}~\bibnamefont {Savchenko}}, \ and\
  \bibinfo {author} {\bibfnamefont {O.}~\bibnamefont {Iordanov}},\ }\href
  {\doibase 10.1109/TGRS.1989.35943} {\bibfield  {journal} {\bibinfo  {journal}
  {IEEE Trans. Geosci. Remote Sens.}\ }\textbf {\bibinfo {volume} {27}},\
  \bibinfo {pages} {606} (\bibinfo {year} {1989})}\BibitemShut {NoStop}%
\bibitem [{\citenamefont {Moreira}\ and\ \citenamefont
  {Barrufet}(1996)}]{Moreira1996}%
  \BibitemOpen
  \bibfield  {author} {\bibinfo {author} {\bibfnamefont {R.~G.}\ \bibnamefont
  {Moreira}}\ and\ \bibinfo {author} {\bibfnamefont {M.~A.}\ \bibnamefont
  {Barrufet}},\ }\href {\doibase 10.1016/0260-8774(95)00010-0} {\bibfield
  {journal} {\bibinfo  {journal} {J. Food Eng.}\ }\textbf {\bibinfo {volume}
  {27}},\ \bibinfo {pages} {279} (\bibinfo {year} {1996})}\BibitemShut
  {NoStop}%
\bibitem [{\citenamefont {Touw}\ \emph {et~al.}(2015)\citenamefont {Touw},
  \citenamefont {Baakman}, \citenamefont {Black}, \citenamefont {Beek},
  \citenamefont {Krieger}, \citenamefont {Joosten},\ and\ \citenamefont
  {Vriend}}]{Touw2015}%
  \BibitemOpen
  \bibfield  {author} {\bibinfo {author} {\bibfnamefont {W.~G.}\ \bibnamefont
  {Touw}}, \bibinfo {author} {\bibfnamefont {C.~A.~B.}\ \bibnamefont
  {Baakman}}, \bibinfo {author} {\bibfnamefont {J.}~\bibnamefont {Black}},
  \bibinfo {author} {\bibfnamefont {T.~A.~v.}\ \bibnamefont {Beek}}, \bibinfo
  {author} {\bibfnamefont {E.}~\bibnamefont {Krieger}}, \bibinfo {author}
  {\bibfnamefont {R.~P.}\ \bibnamefont {Joosten}}, \ and\ \bibinfo {author}
  {\bibfnamefont {G.}~\bibnamefont {Vriend}},\ }\href {\doibase
  10.1093/nar/gku1028} {\bibfield  {journal} {\bibinfo  {journal} {Nucleic
  Acids Res.}\ }\textbf {\bibinfo {volume} {43}},\ \bibinfo {pages} {D364}
  (\bibinfo {year} {2015})}\BibitemShut {NoStop}%
\bibitem [{\citenamefont {Hubbard}\ and\ \citenamefont
  {Thornton}(1993)}]{Hubbard1993}%
  \BibitemOpen
  \bibfield  {author} {\bibinfo {author} {\bibfnamefont {S.}~\bibnamefont
  {Hubbard}}\ and\ \bibinfo {author} {\bibfnamefont {J.}~\bibnamefont
  {Thornton}},\ }\href {http://wolf.bms.umist.ac.uk/naccess/} {\enquote
  {\bibinfo {title} {{NACCESS}},}\ } (\bibinfo {year} {1993})\BibitemShut
  {NoStop}%
\bibitem [{\citenamefont {Auton}\ and\ \citenamefont
  {Bolen}(2005)}]{Auton2005}%
  \BibitemOpen
  \bibfield  {author} {\bibinfo {author} {\bibfnamefont {M.}~\bibnamefont
  {Auton}}\ and\ \bibinfo {author} {\bibfnamefont {D.~W.}\ \bibnamefont
  {Bolen}},\ }\href {\doibase 10.1073/pnas.0507053102} {\bibfield  {journal}
  {\bibinfo  {journal} {Proc. Natl. Acad. Sci. U.S.A.}\ }\textbf {\bibinfo
  {volume} {102}},\ \bibinfo {pages} {15065} (\bibinfo {year}
  {2005})}\BibitemShut {NoStop}%
\bibitem [{\citenamefont {Hobohm}\ and\ \citenamefont
  {Sander}(1994)}]{Hobohm1994}%
  \BibitemOpen
  \bibfield  {author} {\bibinfo {author} {\bibfnamefont {U.}~\bibnamefont
  {Hobohm}}\ and\ \bibinfo {author} {\bibfnamefont {C.}~\bibnamefont
  {Sander}},\ }\href {\doibase 10.1002/pro.5560030317} {\bibfield  {journal}
  {\bibinfo  {journal} {Protein Sci.}\ }\textbf {\bibinfo {volume} {3}},\
  \bibinfo {pages} {522} (\bibinfo {year} {1994})}\BibitemShut {NoStop}%
\bibitem [{\citenamefont {Hubbard}\ \emph {et~al.}(1994)\citenamefont
  {Hubbard}, \citenamefont {Gross},\ and\ \citenamefont {Argos}}]{Hubbard1994}%
  \BibitemOpen
  \bibfield  {author} {\bibinfo {author} {\bibfnamefont {S.~J.}\ \bibnamefont
  {Hubbard}}, \bibinfo {author} {\bibfnamefont {K.-H.}\ \bibnamefont {Gross}},
  \ and\ \bibinfo {author} {\bibfnamefont {P.}~\bibnamefont {Argos}},\ }\href
  {\doibase 10.1093/protein/7.5.613} {\bibfield  {journal} {\bibinfo  {journal}
  {Protein Eng. Des. Sel.}\ }\textbf {\bibinfo {volume} {7}},\ \bibinfo {pages}
  {613} (\bibinfo {year} {1994})}\BibitemShut {NoStop}%
\bibitem [{\citenamefont {Eriksson}\ \emph {et~al.}(1992)\citenamefont
  {Eriksson}, \citenamefont {Baase}, \citenamefont {Zhang}, \citenamefont
  {Heinz}, \citenamefont {Blaber}, \citenamefont {Baldwin},\ and\ \citenamefont
  {Matthews}}]{Eriksson1992}%
  \BibitemOpen
  \bibfield  {author} {\bibinfo {author} {\bibfnamefont {A.~E.}\ \bibnamefont
  {Eriksson}}, \bibinfo {author} {\bibfnamefont {W.~A.}\ \bibnamefont {Baase}},
  \bibinfo {author} {\bibfnamefont {X.~J.}\ \bibnamefont {Zhang}}, \bibinfo
  {author} {\bibfnamefont {D.~W.}\ \bibnamefont {Heinz}}, \bibinfo {author}
  {\bibfnamefont {M.}~\bibnamefont {Blaber}}, \bibinfo {author} {\bibfnamefont
  {E.~P.}\ \bibnamefont {Baldwin}}, \ and\ \bibinfo {author} {\bibfnamefont
  {B.~W.}\ \bibnamefont {Matthews}},\ }\href {\doibase 10.1126/science.1553543}
  {\bibfield  {journal} {\bibinfo  {journal} {Science}\ }\textbf {\bibinfo
  {volume} {255}},\ \bibinfo {pages} {178} (\bibinfo {year}
  {1992})}\BibitemShut {NoStop}%
\bibitem [{\citenamefont {Spyrakis}\ \emph {et~al.}(2011)\citenamefont
  {Spyrakis}, \citenamefont {Faggiano}, \citenamefont {Abbruzzetti},
  \citenamefont {Dominici}, \citenamefont {Cacciatori}, \citenamefont
  {Astegno}, \citenamefont {Droghetti}, \citenamefont {Feis}, \citenamefont
  {Smulevich}, \citenamefont {Bruno}, \citenamefont {Mozzarelli}, \citenamefont
  {Cozzini}, \citenamefont {Viappiani}, \citenamefont {Bidon-Chanal},\ and\
  \citenamefont {Luque}}]{Spyrakis2011}%
  \BibitemOpen
  \bibfield  {author} {\bibinfo {author} {\bibfnamefont {F.}~\bibnamefont
  {Spyrakis}}, \bibinfo {author} {\bibfnamefont {S.}~\bibnamefont {Faggiano}},
  \bibinfo {author} {\bibfnamefont {S.}~\bibnamefont {Abbruzzetti}}, \bibinfo
  {author} {\bibfnamefont {P.}~\bibnamefont {Dominici}}, \bibinfo {author}
  {\bibfnamefont {E.}~\bibnamefont {Cacciatori}}, \bibinfo {author}
  {\bibfnamefont {A.}~\bibnamefont {Astegno}}, \bibinfo {author} {\bibfnamefont
  {E.}~\bibnamefont {Droghetti}}, \bibinfo {author} {\bibfnamefont
  {A.}~\bibnamefont {Feis}}, \bibinfo {author} {\bibfnamefont {G.}~\bibnamefont
  {Smulevich}}, \bibinfo {author} {\bibfnamefont {S.}~\bibnamefont {Bruno}},
  \bibinfo {author} {\bibfnamefont {A.}~\bibnamefont {Mozzarelli}}, \bibinfo
  {author} {\bibfnamefont {P.}~\bibnamefont {Cozzini}}, \bibinfo {author}
  {\bibfnamefont {C.}~\bibnamefont {Viappiani}}, \bibinfo {author}
  {\bibfnamefont {A.}~\bibnamefont {Bidon-Chanal}}, \ and\ \bibinfo {author}
  {\bibfnamefont {F.~J.}\ \bibnamefont {Luque}},\ }\href {\doibase
  10.1021/jp110816h} {\bibfield  {journal} {\bibinfo  {journal} {J. Phys. Chem.
  B}\ }\textbf {\bibinfo {volume} {115}},\ \bibinfo {pages} {4138} (\bibinfo
  {year} {2011})}\BibitemShut {NoStop}%
\bibitem [{\citenamefont {Voss}\ \emph {et~al.}(2006)\citenamefont {Voss},
  \citenamefont {Gerstein}, \citenamefont {Steitz},\ and\ \citenamefont
  {Moore}}]{Voss2006}%
  \BibitemOpen
  \bibfield  {author} {\bibinfo {author} {\bibfnamefont {N.~R.}\ \bibnamefont
  {Voss}}, \bibinfo {author} {\bibfnamefont {M.}~\bibnamefont {Gerstein}},
  \bibinfo {author} {\bibfnamefont {T.~A.}\ \bibnamefont {Steitz}}, \ and\
  \bibinfo {author} {\bibfnamefont {P.~B.}\ \bibnamefont {Moore}},\ }\href
  {\doibase 10.1016/j.jmb.2006.05.023} {\bibfield  {journal} {\bibinfo
  {journal} {J. Mol. Biol.}\ }\textbf {\bibinfo {volume} {360}},\ \bibinfo
  {pages} {893} (\bibinfo {year} {2006})}\BibitemShut {NoStop}%
\bibitem [{\citenamefont {Bui}\ \emph {et~al.}(2004)\citenamefont {Bui},
  \citenamefont {Tai},\ and\ \citenamefont {McCammon}}]{Bui2004}%
  \BibitemOpen
  \bibfield  {author} {\bibinfo {author} {\bibfnamefont {J.~M.}\ \bibnamefont
  {Bui}}, \bibinfo {author} {\bibfnamefont {K.}~\bibnamefont {Tai}}, \ and\
  \bibinfo {author} {\bibfnamefont {J.~A.}\ \bibnamefont {McCammon}},\ }\href
  {\doibase 10.1021/ja0485715} {\bibfield  {journal} {\bibinfo  {journal} {J.
  Am. Chem. Soc.}\ }\textbf {\bibinfo {volume} {126}},\ \bibinfo {pages} {7198}
  (\bibinfo {year} {2004})}\BibitemShut {NoStop}%
\bibitem [{\citenamefont {Farmer}\ and\ \citenamefont
  {Jacobs}(2018)}]{Farmer2018}%
  \BibitemOpen
  \bibfield  {author} {\bibinfo {author} {\bibfnamefont {J.}~\bibnamefont
  {Farmer}}\ and\ \bibinfo {author} {\bibfnamefont {D.}~\bibnamefont
  {Jacobs}},\ }\href {\doibase 10.1371/journal.pone.0196937} {\bibfield
  {journal} {\bibinfo  {journal} {PLoS One}\ }\textbf {\bibinfo {volume}
  {13}},\ \bibinfo {pages} {e0196937} (\bibinfo {year} {2018})}\BibitemShut
  {NoStop}%
\bibitem [{\citenamefont {F{\"u}l{\"o}p}\ \emph {et~al.}(1998)\citenamefont
  {F{\"u}l{\"o}p}, \citenamefont {B{\"o}cskei},\ and\ \citenamefont
  {Polg{\'a}r}}]{Fulop1998}%
  \BibitemOpen
  \bibfield  {author} {\bibinfo {author} {\bibfnamefont {V.}~\bibnamefont
  {F{\"u}l{\"o}p}}, \bibinfo {author} {\bibfnamefont {Z.}~\bibnamefont
  {B{\"o}cskei}}, \ and\ \bibinfo {author} {\bibfnamefont {L.}~\bibnamefont
  {Polg{\'a}r}},\ }\href {\doibase 10.1016/S0092-8674(00)81416-6} {\bibfield
  {journal} {\bibinfo  {journal} {Cell}\ }\textbf {\bibinfo {volume} {94}},\
  \bibinfo {pages} {161} (\bibinfo {year} {1998})}\BibitemShut {NoStop}%
\bibitem [{\citenamefont {Jacobs}\ and\ \citenamefont
  {Thorpe}(1995)}]{Jacobs1995}%
  \BibitemOpen
  \bibfield  {author} {\bibinfo {author} {\bibfnamefont {D.~J.}\ \bibnamefont
  {Jacobs}}\ and\ \bibinfo {author} {\bibfnamefont {M.~F.}\ \bibnamefont
  {Thorpe}},\ }\href {\doibase 10.1103/PhysRevLett.75.4051} {\bibfield
  {journal} {\bibinfo  {journal} {Phys. Rev. Lett.}\ }\textbf {\bibinfo
  {volume} {75}},\ \bibinfo {pages} {4051} (\bibinfo {year}
  {1995})}\BibitemShut {NoStop}%
\bibitem [{\citenamefont {Geistlinger}\ and\ \citenamefont
  {Mohammadian}(2015)}]{Geistlinger2015}%
  \BibitemOpen
  \bibfield  {author} {\bibinfo {author} {\bibfnamefont {H.}~\bibnamefont
  {Geistlinger}}\ and\ \bibinfo {author} {\bibfnamefont {S.}~\bibnamefont
  {Mohammadian}},\ }\href {\doibase 10.1016/j.advwatres.2015.02.010} {\bibfield
   {journal} {\bibinfo  {journal} {Adv. Water Resour.}\ }\textbf {\bibinfo
  {volume} {79}},\ \bibinfo {pages} {35} (\bibinfo {year} {2015})}\BibitemShut
  {NoStop}%
\bibitem [{\citenamefont {Hassan}\ and\ \citenamefont
  {Rahman}(2016)}]{Hassan2016}%
  \BibitemOpen
  \bibfield  {author} {\bibinfo {author} {\bibfnamefont {M.~K.}\ \bibnamefont
  {Hassan}}\ and\ \bibinfo {author} {\bibfnamefont {M.~M.}\ \bibnamefont
  {Rahman}},\ }\href {\doibase 10.1103/PhysRevE.94.042109} {\bibfield
  {journal} {\bibinfo  {journal} {Phys. Rev. E}\ }\textbf {\bibinfo {volume}
  {94}},\ \bibinfo {pages} {4} (\bibinfo {year} {2016})}\BibitemShut {NoStop}%
\bibitem [{\citenamefont {Iglauer}\ and\ \citenamefont
  {W{\"u}lling}(2016)}]{Iglauer2016}%
  \BibitemOpen
  \bibfield  {author} {\bibinfo {author} {\bibfnamefont {S.}~\bibnamefont
  {Iglauer}}\ and\ \bibinfo {author} {\bibfnamefont {W.}~\bibnamefont
  {W{\"u}lling}},\ }\href {\doibase 10.1002/2016GL071298} {\bibfield  {journal}
  {\bibinfo  {journal} {Geophys. Res. Lett.}\ }\textbf {\bibinfo {volume}
  {43}},\ \bibinfo {pages} {11,253} (\bibinfo {year} {2016})}\BibitemShut
  {NoStop}%
\bibitem [{\citenamefont {Huang}\ \emph {et~al.}(2018)\citenamefont {Huang},
  \citenamefont {Hou}, \citenamefont {Wang}, \citenamefont {Ziff},\ and\
  \citenamefont {Deng}}]{Huang2018}%
  \BibitemOpen
  \bibfield  {author} {\bibinfo {author} {\bibfnamefont {W.}~\bibnamefont
  {Huang}}, \bibinfo {author} {\bibfnamefont {P.}~\bibnamefont {Hou}}, \bibinfo
  {author} {\bibfnamefont {J.}~\bibnamefont {Wang}}, \bibinfo {author}
  {\bibfnamefont {R.~M.}\ \bibnamefont {Ziff}}, \ and\ \bibinfo {author}
  {\bibfnamefont {Y.}~\bibnamefont {Deng}},\ }\href {\doibase
  10.1103/PhysRevE.97.022107} {\bibfield  {journal} {\bibinfo  {journal} {Phys.
  Rev. E}\ }\textbf {\bibinfo {volume} {97}} (\bibinfo {year} {2018}),\
  10.1103/PhysRevE.97.022107}\BibitemShut {NoStop}%
\bibitem [{\citenamefont {Roy}\ and\ \citenamefont {Santra}(2018)}]{Roy2018}%
  \BibitemOpen
  \bibfield  {author} {\bibinfo {author} {\bibfnamefont {B.}~\bibnamefont
  {Roy}}\ and\ \bibinfo {author} {\bibfnamefont {S.~B.}\ \bibnamefont
  {Santra}},\ }\href {\doibase 10.1016/j.physa.2017.11.028} {\bibfield
  {journal} {\bibinfo  {journal} {Physica A}\ }\textbf {\bibinfo {volume}
  {492}},\ \bibinfo {pages} {969} (\bibinfo {year} {2018})}\BibitemShut
  {NoStop}%
\bibitem [{\citenamefont {Stauffer}\ and\ \citenamefont
  {Aharony}(1992)}]{Stauffer1987}%
  \BibitemOpen
  \bibfield  {author} {\bibinfo {author} {\bibfnamefont {D.}~\bibnamefont
  {Stauffer}}\ and\ \bibinfo {author} {\bibfnamefont {A.}~\bibnamefont
  {Aharony}},\ }\href@noop {} {\emph {\bibinfo {title} {Introduction to
  percolation theory}}},\ \bibinfo {edition} {2nd}\ ed.\ (\bibinfo  {publisher}
  {Taylor \& Francis London},\ \bibinfo {year} {1992})\BibitemShut {NoStop}%
\bibitem [{\citenamefont {Tang}\ \emph {et~al.}(2017)\citenamefont {Tang},
  \citenamefont {Zhang}, \citenamefont {Wang}, \citenamefont {Wang},\ and\
  \citenamefont {Chialvo}}]{Tang2017}%
  \BibitemOpen
  \bibfield  {author} {\bibinfo {author} {\bibfnamefont {Q.~Y.}\ \bibnamefont
  {Tang}}, \bibinfo {author} {\bibfnamefont {Y.~Y.}\ \bibnamefont {Zhang}},
  \bibinfo {author} {\bibfnamefont {J.}~\bibnamefont {Wang}}, \bibinfo {author}
  {\bibfnamefont {W.}~\bibnamefont {Wang}}, \ and\ \bibinfo {author}
  {\bibfnamefont {D.~R.}\ \bibnamefont {Chialvo}},\ }\href {\doibase
  10.1103/PhysRevLett.118.088102} {\bibfield  {journal} {\bibinfo  {journal}
  {Phys. Rev. Lett.}\ }\textbf {\bibinfo {volume} {118}} (\bibinfo {year}
  {2017}),\ 10.1103/PhysRevLett.118.088102}\BibitemShut {NoStop}%
\bibitem [{\citenamefont {{Schr{\" o}dinger, LLC}}(2015)}]{Schrodinger2015}%
  \BibitemOpen
  \bibfield  {author} {\bibinfo {author} {\bibnamefont {{Schr{\" o}dinger,
  LLC}}},\ }\href {https://pymol.org/} {\enquote {\bibinfo {title} {{The PyMOL
  Molecular Graphics Development System, Version 1.8}},}\ } (\bibinfo {year}
  {2015})\BibitemShut {NoStop}%
\bibitem [{\citenamefont {Wadell}(1935)}]{Wadell1935}%
  \BibitemOpen
  \bibfield  {author} {\bibinfo {author} {\bibfnamefont {H.}~\bibnamefont
  {Wadell}},\ }\href {\doibase 10.1086/624298} {\bibfield  {journal} {\bibinfo
  {journal} {J. Geol.}\ }\textbf {\bibinfo {volume} {43}},\ \bibinfo {pages}
  {250} (\bibinfo {year} {1935})}\BibitemShut {NoStop}%
\bibitem [{\citenamefont {Dean}(1963)}]{Dean1963}%
  \BibitemOpen
  \bibfield  {author} {\bibinfo {author} {\bibfnamefont {P.}~\bibnamefont
  {Dean}},\ }\href {\doibase 10.1017/S0305004100037026} {\bibfield  {journal}
  {\bibinfo  {journal} {Math. Proc. Camb. Philos. Soc.}\ }\textbf {\bibinfo
  {volume} {59}},\ \bibinfo {pages} {397} (\bibinfo {year} {1963})}\BibitemShut
  {NoStop}%
\bibitem [{\citenamefont {Wang}\ \emph {et~al.}(2013)\citenamefont {Wang},
  \citenamefont {Zhou}, \citenamefont {Zhang}, \citenamefont {Garoni},\ and\
  \citenamefont {Deng}}]{Wang2013}%
  \BibitemOpen
  \bibfield  {author} {\bibinfo {author} {\bibfnamefont {J.}~\bibnamefont
  {Wang}}, \bibinfo {author} {\bibfnamefont {Z.}~\bibnamefont {Zhou}}, \bibinfo
  {author} {\bibfnamefont {W.}~\bibnamefont {Zhang}}, \bibinfo {author}
  {\bibfnamefont {T.~M.}\ \bibnamefont {Garoni}}, \ and\ \bibinfo {author}
  {\bibfnamefont {Y.}~\bibnamefont {Deng}},\ }\href {\doibase
  10.1103/PhysRevE.87.052107} {\bibfield  {journal} {\bibinfo  {journal} {Phys.
  Rev. E}\ }\textbf {\bibinfo {volume} {87}},\ \bibinfo {pages} {052107}
  (\bibinfo {year} {2013})}\BibitemShut {NoStop}%
\bibitem [{\citenamefont {Harter}(2005)}]{Harter2005}%
  \BibitemOpen
  \bibfield  {author} {\bibinfo {author} {\bibfnamefont {T.}~\bibnamefont
  {Harter}},\ }\href {\doibase 10.1103/PhysRevE.72.026120} {\bibfield
  {journal} {\bibinfo  {journal} {Phys. Rev. E}\ }\textbf {\bibinfo {volume}
  {72}},\ \bibinfo {pages} {026120} (\bibinfo {year} {2005})}\BibitemShut
  {NoStop}%
\bibitem [{\citenamefont {Duckers}\ and\ \citenamefont
  {Ross}(1974)}]{Duckers1974}%
  \BibitemOpen
  \bibfield  {author} {\bibinfo {author} {\bibfnamefont {L.~J.}\ \bibnamefont
  {Duckers}}\ and\ \bibinfo {author} {\bibfnamefont {R.~G.}\ \bibnamefont
  {Ross}},\ }\href {\doibase 10.1016/0375-9601(74)90270-9} {\bibfield
  {journal} {\bibinfo  {journal} {Phys. Lett. A}\ }\textbf {\bibinfo {volume}
  {49}},\ \bibinfo {pages} {361} (\bibinfo {year} {1974})}\BibitemShut
  {NoStop}%
\bibitem [{\citenamefont {Duckers}(1978)}]{Duckers1978}%
  \BibitemOpen
  \bibfield  {author} {\bibinfo {author} {\bibfnamefont {L.~J.}\ \bibnamefont
  {Duckers}},\ }\href {\doibase 10.1016/0375-9601(78)90029-4} {\bibfield
  {journal} {\bibinfo  {journal} {Phys. Lett. A}\ }\textbf {\bibinfo {volume}
  {67}},\ \bibinfo {pages} {93} (\bibinfo {year} {1978})}\BibitemShut {NoStop}%
\bibitem [{\citenamefont {M{\"u}ller-Krumbhaar}(1974)}]{Mueller-Krumbhaar1974}%
  \BibitemOpen
  \bibfield  {author} {\bibinfo {author} {\bibfnamefont {H.}~\bibnamefont
  {M{\"u}ller-Krumbhaar}},\ }\href {\doibase 10.1016/0375-9601(74)90337-5}
  {\bibfield  {journal} {\bibinfo  {journal} {Phys. Lett. A}\ }\textbf
  {\bibinfo {volume} {50}},\ \bibinfo {pages} {27} (\bibinfo {year}
  {1974})}\BibitemShut {NoStop}%
\bibitem [{\citenamefont {Lorenz}\ \emph {et~al.}(1993)\citenamefont {Lorenz},
  \citenamefont {Orgzall},\ and\ \citenamefont {Heuer}}]{Lorenz1993}%
  \BibitemOpen
  \bibfield  {author} {\bibinfo {author} {\bibfnamefont {B.}~\bibnamefont
  {Lorenz}}, \bibinfo {author} {\bibfnamefont {I.}~\bibnamefont {Orgzall}}, \
  and\ \bibinfo {author} {\bibfnamefont {H.~O.}\ \bibnamefont {Heuer}},\ }\href
  {\doibase 10.1088/0305-4470/26/18/032} {\bibfield  {journal} {\bibinfo
  {journal} {J. Phys. A}\ }\textbf {\bibinfo {volume} {26}},\ \bibinfo {pages}
  {4711} (\bibinfo {year} {1993})}\BibitemShut {NoStop}%
\bibitem [{\citenamefont {Rintoul}(1997)}]{Rintoul1997}%
  \BibitemOpen
  \bibfield  {author} {\bibinfo {author} {\bibfnamefont {M.~D.}\ \bibnamefont
  {Rintoul}},\ }\href {\doibase 10.1088/0305-4470/30/16/005} {\bibfield
  {journal} {\bibinfo  {journal} {J. Phys. A}\ }\textbf {\bibinfo {volume}
  {30}},\ \bibinfo {pages} {L585} (\bibinfo {year} {1997})}\BibitemShut
  {NoStop}%
\bibitem [{\citenamefont {Ding}\ \emph {et~al.}(2014)\citenamefont {Ding},
  \citenamefont {Li}, \citenamefont {Zhang}, \citenamefont {Lu},\ and\
  \citenamefont {Ji}}]{Ding2014}%
  \BibitemOpen
  \bibfield  {author} {\bibinfo {author} {\bibfnamefont {B.}~\bibnamefont
  {Ding}}, \bibinfo {author} {\bibfnamefont {C.}~\bibnamefont {Li}}, \bibinfo
  {author} {\bibfnamefont {M.}~\bibnamefont {Zhang}}, \bibinfo {author}
  {\bibfnamefont {G.}~\bibnamefont {Lu}}, \ and\ \bibinfo {author}
  {\bibfnamefont {F.}~\bibnamefont {Ji}},\ }\href {\doibase
  10.1140/epjb/e2014-40996-4} {\bibfield  {journal} {\bibinfo  {journal} {Eur.
  Phys. J. B}\ }\textbf {\bibinfo {volume} {87}},\ \bibinfo {pages} {1}
  (\bibinfo {year} {2014})}\BibitemShut {NoStop}%
\bibitem [{\citenamefont {Liu}\ and\ \citenamefont
  {Regenauer-Lieb}(2011)}]{Liu2011}%
  \BibitemOpen
  \bibfield  {author} {\bibinfo {author} {\bibfnamefont {J.}~\bibnamefont
  {Liu}}\ and\ \bibinfo {author} {\bibfnamefont {K.}~\bibnamefont
  {Regenauer-Lieb}},\ }\href {\doibase 10.1103/PhysRevE.83.016106} {\bibfield
  {journal} {\bibinfo  {journal} {Phys. Rev. E}\ }\textbf {\bibinfo {volume}
  {83}} (\bibinfo {year} {2011}),\ 10.1103/PhysRevE.83.016106}\BibitemShut
  {NoStop}%
\bibitem [{\citenamefont {Cuff}\ and\ \citenamefont {Martin}(2004)}]{Cuff2004}%
  \BibitemOpen
  \bibfield  {author} {\bibinfo {author} {\bibfnamefont {A.~L.}\ \bibnamefont
  {Cuff}}\ and\ \bibinfo {author} {\bibfnamefont {A.~C.~R.}\ \bibnamefont
  {Martin}},\ }\href {\doibase 10.1016/j.jmb.2004.10.015} {\bibfield  {journal}
  {\bibinfo  {journal} {J. Mol. Biol.}\ }\textbf {\bibinfo {volume} {344}},\
  \bibinfo {pages} {1199} (\bibinfo {year} {2004})}\BibitemShut {NoStop}%
\bibitem [{\citenamefont {Gaines}\ \emph {et~al.}(2017)\citenamefont {Gaines},
  \citenamefont {Clark}, \citenamefont {Regan},\ and\ \citenamefont
  {O’Hern}}]{Gaines2017}%
  \BibitemOpen
  \bibfield  {author} {\bibinfo {author} {\bibfnamefont {J.~C.}\ \bibnamefont
  {Gaines}}, \bibinfo {author} {\bibfnamefont {A.~H.}\ \bibnamefont {Clark}},
  \bibinfo {author} {\bibfnamefont {L.}~\bibnamefont {Regan}}, \ and\ \bibinfo
  {author} {\bibfnamefont {C.~S.}\ \bibnamefont {O’Hern}},\ }\href {\doibase
  10.1088/1361-648X/aa75c2} {\bibfield  {journal} {\bibinfo  {journal} {J.
  Phys. Condens. Matter}\ }\textbf {\bibinfo {volume} {29}},\ \bibinfo {pages}
  {293001} (\bibinfo {year} {2017})}\BibitemShut {NoStop}%
\bibitem [{\citenamefont {Gaines}\ \emph {et~al.}(2016)\citenamefont {Gaines},
  \citenamefont {Smith}, \citenamefont {Regan},\ and\ \citenamefont
  {O'Hern}}]{Gaines2016}%
  \BibitemOpen
  \bibfield  {author} {\bibinfo {author} {\bibfnamefont {J.~C.}\ \bibnamefont
  {Gaines}}, \bibinfo {author} {\bibfnamefont {W.~W.}\ \bibnamefont {Smith}},
  \bibinfo {author} {\bibfnamefont {L.}~\bibnamefont {Regan}}, \ and\ \bibinfo
  {author} {\bibfnamefont {C.~S.}\ \bibnamefont {O'Hern}},\ }\href {\doibase
  10.1103/PhysRevE.93.032415} {\bibfield  {journal} {\bibinfo  {journal} {Phys.
  Rev. E}\ }\textbf {\bibinfo {volume} {93}},\ \bibinfo {pages} {032415}
  (\bibinfo {year} {2016})}\BibitemShut {NoStop}%
\bibitem [{\citenamefont {Bondi}(1964)}]{Bondi1964}%
  \BibitemOpen
  \bibfield  {author} {\bibinfo {author} {\bibfnamefont {A.}~\bibnamefont
  {Bondi}},\ }\href {\doibase 10.1021/j100785a001} {\bibfield  {journal}
  {\bibinfo  {journal} {J. Phys. Chem.}\ }\textbf {\bibinfo {volume} {68}},\
  \bibinfo {pages} {441} (\bibinfo {year} {1964})}\BibitemShut {NoStop}%
\bibitem [{\citenamefont {Zhang}\ \emph {et~al.}(2003)\citenamefont {Zhang},
  \citenamefont {Chen}, \citenamefont {Tang},\ and\ \citenamefont
  {Liang}}]{Zhang2003}%
  \BibitemOpen
  \bibfield  {author} {\bibinfo {author} {\bibfnamefont {J.}~\bibnamefont
  {Zhang}}, \bibinfo {author} {\bibfnamefont {R.}~\bibnamefont {Chen}},
  \bibinfo {author} {\bibfnamefont {C.}~\bibnamefont {Tang}}, \ and\ \bibinfo
  {author} {\bibfnamefont {J.}~\bibnamefont {Liang}},\ }\href {\doibase
  10.1063/1.1554395} {\bibfield  {journal} {\bibinfo  {journal} {J. Chem.
  Phys.}\ }\textbf {\bibinfo {volume} {118}},\ \bibinfo {pages} {6102}
  (\bibinfo {year} {2003})}\BibitemShut {NoStop}%
\end{thebibliography}%
\end{document}